\documentclass[lettersize,journal, onecolumn]{IEEEtran}
\usepackage{amsmath,amsfonts}
\usepackage{algorithmic}
\usepackage{algorithm}
\usepackage{array}
\usepackage[caption=false,font=normalsize,labelfont=sf,textfont=sf]{subfig}
\usepackage{textcomp}
\usepackage{stfloats}
\usepackage{url}
\usepackage{verbatim}
\usepackage{graphicx}
\usepackage{cite}
\usepackage{listings}
\usepackage{braket}
\usepackage{xcolor}
\usepackage{booktabs}

\begin{document}

\title{CMOS-Compatible Electrostatic SWAP Gate in Silicon Quantum Dots: Tight-Binding Model and Beyond}




\author{\IEEEauthorblockN{1\textsuperscript{st} Krzysztof Pomorski$^{1,2}$ } \\
\IEEEauthorblockA{\textit{1: University of New Mexico} \\
\textit{Center for High Technological Materials, Albuquerque, USA}\\
\textit{2: Quantum Hardware Systems, Lodz, Poland} \\
email: kdvpomorski@gmail.com \\
$ $
} 
\and \\ 
\IEEEauthorblockN{2\textsuperscript{nd} Eryk Halubek$^{3,4}$} \\
\IEEEauthorblockA{\textit{3: Lodz University of Technology} \\
\textit{ Institute of Physics, Lodz, Poland}\\
\textit{4: Quantum Hardware Systems, Lodz, Poland}\\
}
}


\maketitle

\begin{abstract}
We present a generalized electrostatic SWAP gate realized in a chain of two double quantum dots  operated in the single-electron regime. Using a minimalist tight-binding model, we derive analytical results and corroborate them with numerical simulations. We exploit the charge anticorrelation arising from Coulomb repulsion and quantify the resulting entanglement generation. We contrast classical and quantum descriptions and show how device geometry and coupling strengths govern entanglement dynamics and gate performance. The results are relevant to cryogenic, CMOS-compatible quantum technologies and suggest a route toward large-scale semiconductor implementations of quantum logic. Finally, we outline a~systematic procedure for translating classical electrostatic logic gates into single-electron quantum gates.
\end{abstract}

\begin{IEEEkeywords}
 Wannier position-based qubits, semiconductor single-electron devices, electrostatic quantum swap gates, semiconductor quantum dots
\end{IEEEkeywords}

\newpage
\tableofcontents
\newpage 

\section{Introduction}

\section{Current status of quantum technologies}

More than twenty hardware platforms are being explored across quantum computing, communication, and sensing. From a scalability point of view, two families currently show a credible path towards wafer-scale integration: (i) silicon spin/charge qubits compatible with CMOS process flows \cite{SiSpinCMOS,ChatterjeeNRP2021} and (ii) superconducting transmons supported by a mature cryogenic control stack \cite{IBMQexp}. Silicon benefits from 300\,mm fabrication and the prospect of dense 2D arrays with purely electrostatic control, which motivates our focus on electrostatically defined single-electron devices. Major system-level constraints remain error-correction overheads and the wiring/thermal budget at the mK stage, which strongly favor low-power, high-density solutions. In silicon quantum dots, progress in isotopic purification, gate-oxide quality, and cryo-CMOS control has reduced charge noise and improved uniformity, enabling reproducible exchange couplings across small arrays \cite{ChatterjeeNRP2021}. Against this backdrop, electrostatically defined single-electron devices offer a fabrication-aligned path to native two-qubit primitives (SWAP/$\sqrt{\text{SWAP}}$) realized with detuning and barrier pulses, making them natural candidates for scalable, all-electrostatic architectures.

\begin{figure}[h]
  \centering
  \includegraphics[width=0.86\columnwidth]{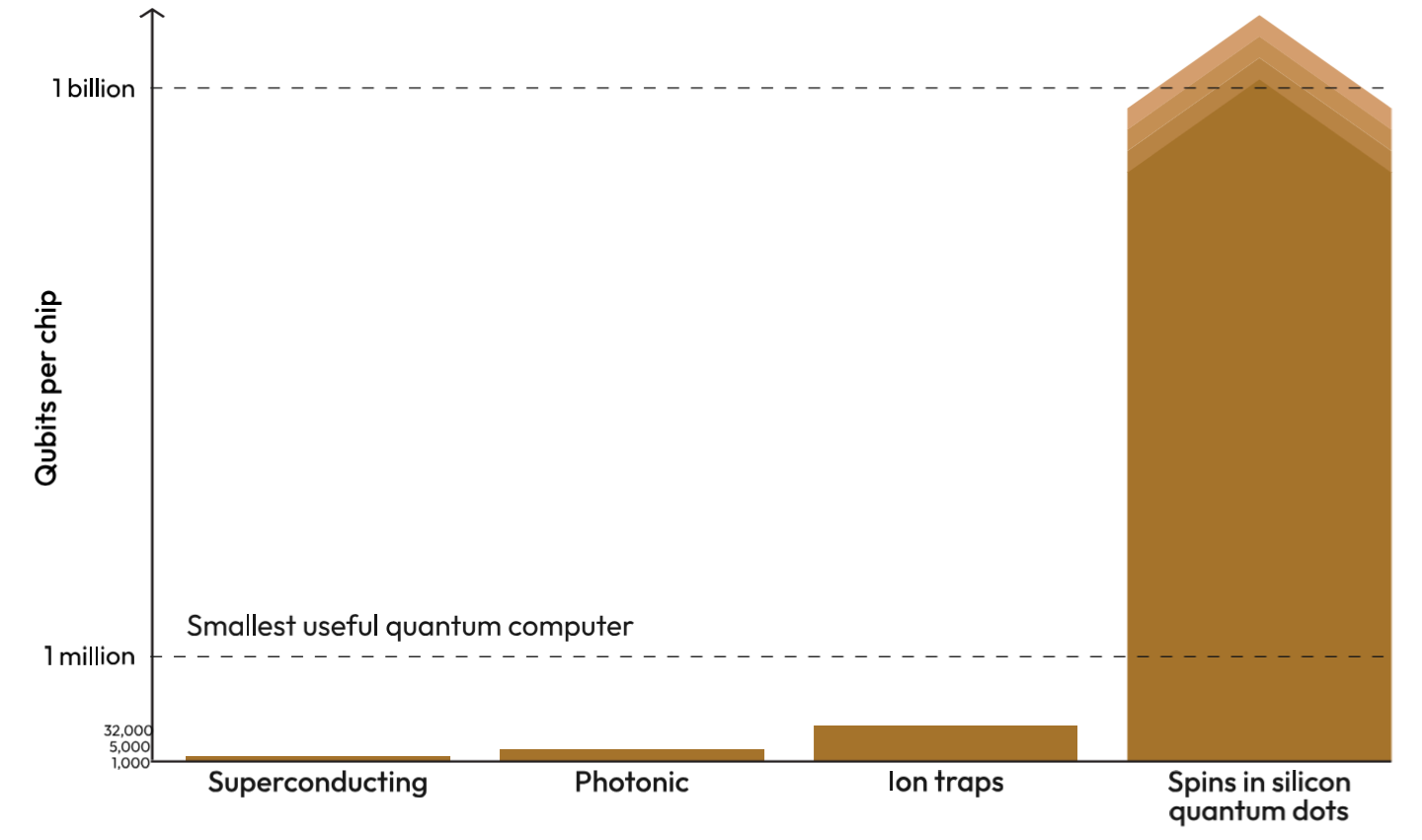}
  \caption{Indicative scalability of selected qubit architectures. Stylized comparison of estimated qubits-per-die across superconducting, photonic, ion-trap and silicon spin platforms (based on public roadmaps from Diraq company https://diraq.com/).}
  \label{fig:scalability}
\end{figure}


As summarized in Fig.~\ref{fig:scalability}, platform roadmaps consistently indicate that silicon and superconducting approaches currently offer the steepest scaling trajectories. In silicon, the decisive levers are 300\,mm manufacturing, isotopically purified $^{28}$Si with long coherence, and the possibility of co-integrating cryo-CMOS control within a tight thermal budget. Superconducting circuits, by contrast, benefit from a mature microwave toolchain and rapid gate speeds, yet face wiring density and footprint constraints at the mK stage. These system-level considerations motivate a device-centric look at CMOS-compatible, \emph{electrostatically} defined qubits.

\begin{figure}[!t]
  \centering
  \includegraphics[width=0.98\columnwidth]{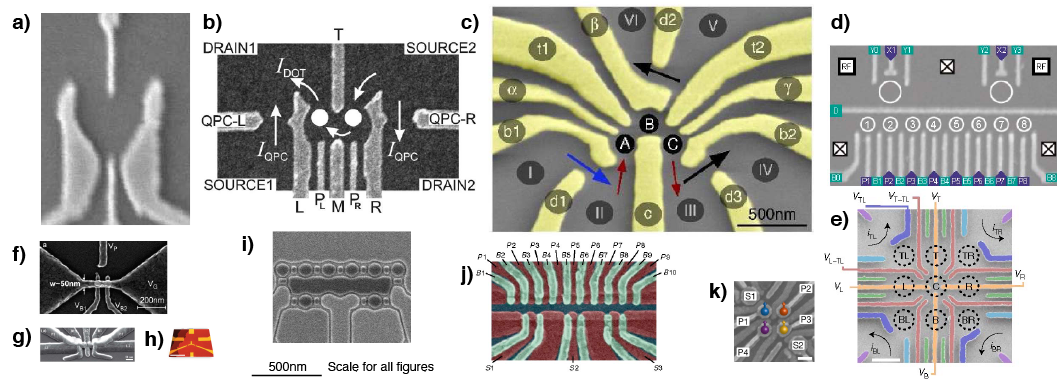}
  \caption{Representative implementations of silicon quantum dots and arrays (gate-defined MOS/SiGe, nanowires/holes, crossbar concepts, and 2D lattices). After Burkard \emph{et\,al.}, \emph{Rev.\ Mod.\ Phys. 95, 025003} (2023) \cite{BurkardRMP}.}
  \label{fig:rmp_devices}
\end{figure}

\begin{figure}[!t]
  \centering
  \includegraphics[width=0.88\columnwidth]{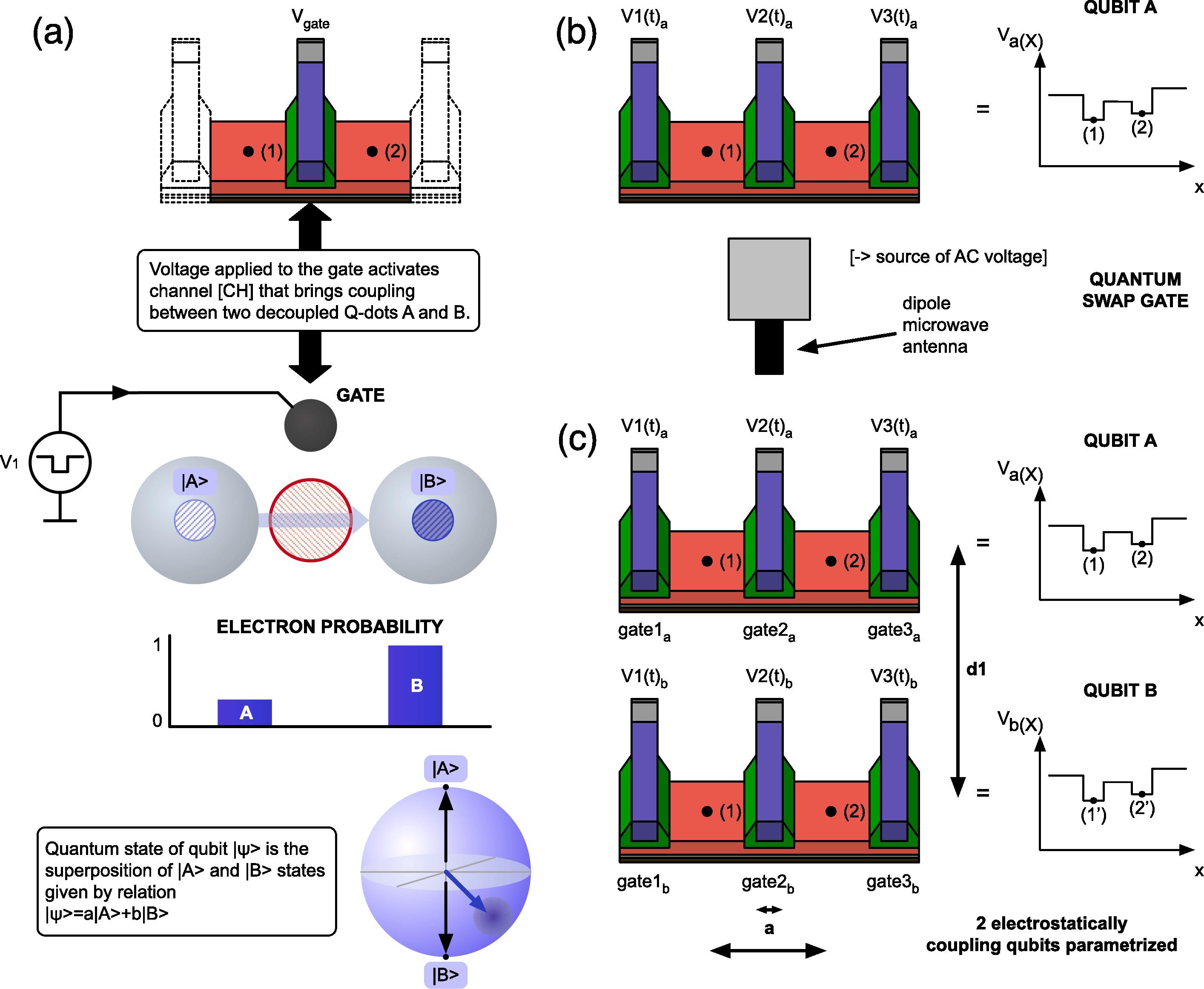}
  \caption{Electrostatic, position-based qubit and tunable SWAP in a CMOS-compatible DQD (Double Quantum Dot) network; analytical treatment and numerical corroboration as by Pomorski \cite{Cryogenics}.}
  \label{fig:swap_concept}
\end{figure}

Zooming in, Fig.~\ref{fig:rmp_devices} maps the concrete device landscape: gate-defined MOS and Si/SiGe quantum dots, hole-spin nanowires (Ge/Si), as well as crossbar and 2D-array layouts that enable capacitive and exchange coupling without permanent magnets. Two control knobs dominate across these realizations—detuning $\Delta$ (on-site energies) and barrier control (tunnel coupling $t_c$)—which together set the exchange interaction $J(\Delta,t_c)$ and hence the entangling dynamics. This architecture aligns with standard CMOS stacks and provides native access to SWAP/$\sqrt{\mathrm{SWAP}}$ primitives via time-shaped detuning and barrier pulses, paving the way for the tunable electrostatic SWAP operation analyzed next (Fig.~\ref{fig:swap_concept}).

Following \cite{Cryogenics}, we use a spinless tight-binding model in which on-site detuning $\Delta$ and barrier-controlled tunnel coupling $t_c$ are the only electrostatic knobs. In a DQD link, a SWAP (resp.\ $\sqrt{\mathrm{SWAP}}$) is realized by a pulse area of $\pi$ (resp.\ $\pi/2$) with residual phases removed by short detuning bursts; \cite{Cryogenics} provides closed-form propagators for such piecewise-constant control and explicit gate-phase relations. Coulomb-driven charge anticorrelation enforces single occupancy and supplies the resource for entanglement while keeping control fully capacitive—no micromagnets—thus aligning the device stack with CMOS (gate metals/oxide, mK operation, sub-$10\,\mu\mathrm{W}$/channel cryo-CMOS). Our numerics reproduce the analytic gate-time conditions and indicate tolerance to moderate dispersion in $(\Delta,t_c)$ along short chains, supporting the scalable SWAP primitive depicted in Fig.~\ref{fig:swap_concept}. Various milestones to be achieved by position-based qubits are specified by Table.1.

The purpose of this work is to describe position-based semiconductor qubits arranged in a chain of electrostatically controlled quantum dots, using single electrons and neglecting spin, which is discussed in the next section

\begin{table}[!t]
\centering
\caption{Selected milestones relevant to CMOS-compatible qubits to be achieved in near future. 
}
\label{tab:cmos_milestones}
\begin{tabular}{@{}p{0.38\linewidth}p{0.54\linewidth}@{}}
\toprule
\textbf{Milestone} & \textbf{Brief note} \\
\midrule
Cryo-CMOS control $<10\,\mu\mathrm{W}$/channel & mK operation with negligible coherence penalty. \\
Single-qubit fidelity $99.9\%$ & Demonstrated on 300\,mm lines; CMOS-compatible. \\
Two-qubit fidelity $>99\%$ & Silicon entangling gates beyond $99\%$. \\
Benchmarking consortia & Toward standardized metrics (e.g.\ DARPA/QBI). \\
\bottomrule
\end{tabular}
\end{table}

\section{Description of position based-qubit in tight-binding model}
\begin{figure}[htb]
\centering
	\includegraphics[scale=0.6]{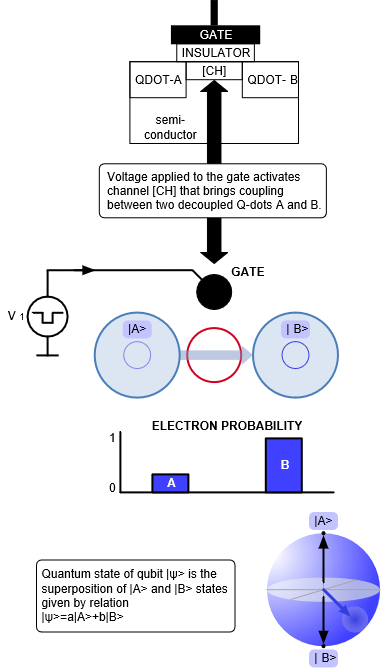} 
	\caption{Scheme of electrostatically controlled qubit made from 2 neighbouring quantum dots and referring to the previous figure showing implementation of the single qubit. } 
	\label{PositionQubit} 
\end{figure}
\begin{figure}
\centering
	\includegraphics[scale=0.6]{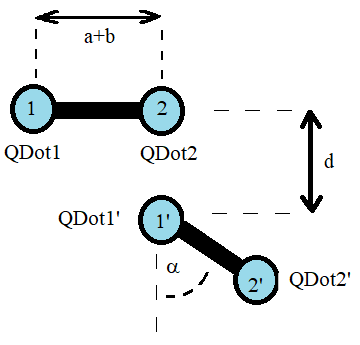} 
	\caption{Scheme of electrostatic quantum gate as two interacting qubits (\cite{Pomorski_spie, SEL, b3, Cryogenics, b5}) with geometrical parametrization of generalized electrostatic quantum gate.} 
	\label{GQSwapGate} 
\end{figure}
We refer to the physical situation from Fig.\ref{PositionQubit} as given by \cite{Dirk}, \cite{SEL}, \cite{Pomorski_spie}, \cite{Fujisawa}, \cite{Petta}  
and we consider position based-qubit in tight-binding model \cite{SEL} and its  Hamiltonian is given by
\begin{eqnarray}
\label{simplematrix}
\hat{H}(t)=
\begin{pmatrix}
E_{p1}(t) & t_{s12}(t) \\
t_{s12}^{\dag}(t) & E_{p2}(t)
\end{pmatrix}_{[x=(x_1,x_2)]}= \nonumber \\
(E_1(t)\ket{E_1}_t \bra{E_1}_t+E_2(t)\ket{E_2}\bra{E_2})_{[E=(E_1,E_2)]}.
\end{eqnarray}
The Hamiltonian $\hat{H}(t)$ eigenenergies $E_1(t)$ and $E_2(t)$ with $E_2(t)>E_1(t)$ are given as:
\begin{eqnarray}
E_1(t)= \left(-\sqrt{\frac{(E_{p1}(t)-E_{p2}(t))^2}{4}+|t_{s12}(t)|^2}+\frac{E_{p1}(t)+E_{p2}(t)}{2}\right), \nonumber \\
E_2(t)= \left(+\sqrt{\frac{(E_{p1}(t)-E_{p2}(t))^2}{4}+|t_{s12}(t)|^2}+\frac{E_{p1}(t)+E_{p2}(t)}{2}\right),
\end{eqnarray}
and energy eigenstates $\ket{E_1(t)}$ and $\ket{E_2(t)}$ have the following form
\begin{eqnarray}
\ket{E_1,t}=
\begin{pmatrix}
\frac{(E_{p2}(t)-E_{p1}(t))+\sqrt{\frac{(E_{p2}(t)-E_{p1}(t))^2}{2}+|t_{s12}(t)|^2}}{-i t_{sr}(t)+t_{si}(t)} \\
-1
\end{pmatrix},  \nonumber \\
\ket{E_2,t}=
\begin{pmatrix}
\frac{-(E_{p2}(t)-E_{p1}(t))+\sqrt{\frac{(E_{p2}(t)-E_{p1}(t))^2}{2}+|t_{s12}(t)|^2}}{t_{sr}(t) - i t_{si}(t)} \\
1
\end{pmatrix}.
\end{eqnarray}
This Hamiltonian gives a description of two coupled quantum wells as depicted in Fig.4.
In such situation we have real-valued functions $E_{p1}(t)$, $E_{p2}(t)$ and complex-valued functions $t_{s12}(t)=t_s(t)=t_{sr}(t)+i t_{si}(t)$ and $t_{s21}(t)=t_{s12}^{*}(t)$, which is equivalent to the knowledge of four real valued time-dependent continuous or discontinues functions $E_{p1}(t)$, $E_{p1}(2)$ , $t_{sr}(t)$ and $t_{si}(t)$. The quantum state is a superposition of state localized at node 1 and 2 and therefore is given as
\begin{equation}
\ket{\psi}_{[x]}=\alpha(t)\ket{1,0}_x+\beta(t)\ket{0,1}_x=
\alpha(t)
\begin{pmatrix}
1 \\
0 \\
\end{pmatrix}
+
\beta(t)
\begin{pmatrix}
0 \\
1 \\
\end{pmatrix} ,
\end{equation}
where $|\alpha(t)|^2$ ($|\beta(t)|^2$) is probability of finding particle at node 1(2) respectively, which brings $|\alpha(t)|^2+|\beta(t)|^2=1$ and obviously $\bra{1,0}_x\ket{|1,0}_x=1=\bra{0,1}_x\ket{|0,1}_x$ and $\bra{1,0}_x\ket{|0,1}_x=0=\bra{0,1}_x\ket{|1,0}_x$. In Schr\"odinger formalism, states $\ket{1,0}_x$ and $\ket{0,1}_x$ are Wannier functions that are parameterized by position $x$. We work in tight-binding approximation and quantum state evolution with time as given by
\begin{equation}
 i \hbar \frac{d}{dt}\ket{\psi(t)}=\hat{H}(t)\ket{\psi(t)}=E(t)\ket{\psi(t)}.
\end{equation}
The last equation has an analytic solution
\begin{equation}
\ket{\psi(t)}=e^{\frac{1}{i \hbar}\int_{t_0}^{t}\hat{H}(t_1)dt_1}\ket{\psi(t_0)}=e^{\frac{1}{i \hbar}\int_{t_0}^{t}\hat{H}(t_1)dt_1}
\begin{pmatrix}
\alpha(0) \\
\beta(0) \\
\end{pmatrix}
\end{equation}
and in quantum density matrix theory we obtain
\begin{eqnarray*}
\hat{\rho}(t)=\hat{\rho}^{\dag}(t)=\ket{\psi(t)}\bra{\psi(t)}=\nonumber \\ =\hat{U}(t,t_0)\hat{\rho}(t_0)\hat{U}(t,t_0)^{-1}= \nonumber \\
=e^{\frac{1}{i \hbar}\int_{t_0}^{t}\hat{H}(t_1)dt_1}(\ket{\psi(t_0)}\bra{\psi(t_0)})e^{-\frac{1}{i \hbar}\int_{t_0}^{t}\hat{H}(t_1)dt_1}= \nonumber \\
=e^{\frac{1}{i \hbar}\int_{t_0}^{t}\hat{H}(t_1)dt_1}
\bigg(
\begin{pmatrix}
\alpha(0) \\
\beta(0) \\
\end{pmatrix} 
\begin{pmatrix}
\alpha^{*}(0) & \beta^{*}(0) \\
\end{pmatrix}
\bigg)e^{-\frac{\int_{t_0}^{t}\hat{H}(t_1)dt_1}{i \hbar}} \nonumber \\
=\hat{U}(t,t_0)  
\begin{pmatrix}
|\alpha(0)|^2 & \alpha(0)\beta^{*}(0)  \\
\beta(0)\alpha(0)^{*} &  |\beta(0)|^2 \\
\end{pmatrix} \hat{U}(t,t_0)^{\dag}.\nonumber \\
\end{eqnarray*}

\paragraph{Pauli decomposition (rephrased).}
Any Hermitian $2\times 2$ matrix admits the expansion $A=\sum_{k=0}^{3} c_k\,\sigma_k$ with
$c_0=\tfrac{1}{2}\mathrm{Tr}\,A$ and $\vec c=\tfrac{1}{2}\mathrm{Tr}(A\,\vec\sigma)$.
For composite systems, $A\in\mathbb{C}^{2^N\times 2^N}$ can be written as
$A=\sum_{\mathbf{k}} c_{\mathbf{k}}\;\sigma_{k_1}\!\otimes\!\cdots\!\otimes\!\sigma_{k_N}$,
where $\mathbf{k}=(k_1,\ldots,k_N)$ and $k_i\in\{0,1,2,3\}$.

Using the above property for matrix of size 2$\times$2 we obtain $e^{\frac{1}{i\hbar}\int_{t_0}^{t}\hat{H}(t_1)dt_1}=\hat{U}(t,t_0), $ and assuming $E_{p1}(t)=E_{p2}(t)=E_{p}(t)$ and we  are given matrix $e^{\frac{1}{i\hbar}\int_{t_0}^{t}\hat{H}(t_1)dt_1}= 
 $ 
\begin{eqnarray*}
\begin{pmatrix}
e^{\frac{-i \int_{t_0}^{t}E_p(t')dt'}{\hbar}}ch\left(\frac{\sqrt{-\int_{t_0}^{t}(|t_{s}(t')|^2)dt'}}{\hbar}\right) & \frac{e^{\frac{-i \int_{t_0}^{t}E_{p}(t')dt'}{\hbar}} (\int_{t_0}^{t}(t_{s}^{*}(t'))dt')sh\left(\frac{\sqrt{-\int_{t_0}^{t}|t_{s}(t')|^2)}}{\hbar}\right)}{\sqrt{-\int_{t_0}^{t}((t_{si}(t')^2+t_{sr}(t'))^2)dt'}}  \\
\frac{e^{\frac{-i \int_{t_0}^{t}E_{p}(t')dt'}{\hbar}} (\int_{t_0}^{t}(-t_{s}(t'))dt')sh\left(\frac{\sqrt{-\int_{t_0}^{t}|t_{s}(t')|^2dt'}}{\hbar}\right)}{\sqrt{-\int_{t_0}^{t}((t_{si}(t')^2+t_{sr}(t'))^2)dt'}} & e^{\frac{-i\int_{t_0}^{t}E_p(t')dt'}{\hbar}}ch\left(\frac{\sqrt{-\int_{t_0}^{t}(|t_{s}(t')|^2)dt'}}{\hbar}\right)  \end{pmatrix}, \nonumber \\
\end{eqnarray*}
where $sh$(.) and $ch$(.) are $\sinh$ and $\cosh$ hyperbolic functions, where $|t_s(t)|^2=|t_{sr}(t)|^2+|t_{si}(t)|^2$. This matrix is unitary so $\hat{U}^{\dag}(t,t_0)=\hat{U}^{-1}(t,t_0)$.
At the very end we will also consider more general case when $E_{p1}(t) \neq E_{p2}(t)$. At first let us consider the case of two localized states in the left and right quantum well so there is no hopping which implies $t_s=0$.
In such case the evolution matrix $\hat{U}(t,t_0)$ is unitarian and has the following form 
\begin{eqnarray}
 \hat{U}(t,t_0)=\nonumber \\
e^{\frac{1}{i\hbar}\int_{t_0}^{t}\hat{H}(t_1)dt_1}= \nonumber \\
\begin{pmatrix}
e^{\frac{-i \int_{t_0}^{t}E_{p1}(t')dt'}{\hbar}} & 0 \\
0 & e^{\frac{-i \int_{t_0}^{t}E_{p2}(t')dt'}{\hbar}}
\end{pmatrix}= \nonumber \\
\frac{(e^{\frac{-i \int_{t_0}^{t}E_{p1}(t')dt'}{\hbar}}+e^{\frac{-i \int_{t_0}^{t}E_{p2}(t')dt'}{\hbar}})}{2} 
\begin{pmatrix}
1 & 0 \\
0 & 1 \\
\end{pmatrix}- \nonumber \\
\frac{(e^{\frac{-i \int_{t_0}^{t}E_{p1}(t')dt'}{\hbar}}-e^{\frac{-i \int_{t_0}^{t}E_{p2}(t')dt'}{\hbar}})}{2}
\begin{pmatrix}
1 & 0 \\
0 & -1 \\
\end{pmatrix}\nonumber \\ =
\frac{(e^{\frac{-i \int_{t_0}^{t}E_{p1}(t')dt'}{\hbar}}+e^{\frac{-i \int_{t_0}^{t}E_{p2}(t')dt'}{\hbar}})}{2} \sigma_0 + \nonumber \\
\frac{(e^{\frac{-i \int_{t_0}^{t}E_{p1}(t')dt'}{\hbar}}-e^{\frac{-i \int_{t_0}^{t}E_{p2}(t')dt'}{\hbar}})}{2} \sigma_3,
\end{eqnarray}
which implies that the left and right quantum dots are two disconnected systems, each undergoing its own time evolution. However since one electron is distributed between those physical systems the measurement conducted on the left quantum dot will have its immediate effect on the right quantum dot. Another extreme example is the situation when hopping energy is considerably bigger than localization energy. In such case we set $E_{p1}=E_{p2}=0$ and in case of non-zero hopping terms we obtain $\hat{U}(t,t_0)=$ given below 
\tiny 
\begin{eqnarray}
\hat{U}(t,t_0)=e^{\frac{1}{i\hbar}\int_{t_0}^{t}\hat{H}(t_1)dt_1}= \nonumber \\
\begin{pmatrix}
ch\left(\frac{\sqrt{-\int_{t_0}^{t}(|t_{s}(t')|^2)dt'}}{\hbar}\right) & \frac{ (\int_{t_0}^{t}(t_{s}^{*}(t'))dt')sh\left(\frac{\sqrt{-\int_{t_0}^{t}|t_{s}(t')|^2)}}{\hbar}\right)}{\sqrt{-\int_{t_0}^{t}((t_{si}(t')^2+t_{sr}(t'))^2)dt'}}\\
\frac{ (\int_{t_0}^{t}(-t_{s}(t'))dt')sh\left(\frac{\sqrt{-\int_{t_0}^{t}|t_{s}(t')|^2dt'}}{\hbar}\right)}{\sqrt{-\int_{t_0}^{t}((t_{si}(t')^2+t_{sr}(t'))^2)dt'}} & ch\left(\frac{\sqrt{-\int_{t_0}^{t}(|t_{s}(t')|^2)dt'}}{\hbar}\right)
\end{pmatrix}, \nonumber \\
\end{eqnarray}
\normalsize
$ $ \newline \newline \newline \newline $ $
Now it is time to move to most general situation of $E_{p1} \neq E_{p2}$, $t_{sr}, t_{si} \neq 0$. We have 4 elements of evolution matrix given as
\begin{eqnarray}
 \hat{U}(t,t_0)=e^{\frac{1}{i\hbar}\int_{t_0}^{t}\hat{H}(t_1)dt_1}= \nonumber \\
\begin{pmatrix}
 U(t,t_0)_{1,1} & U(t,t_0)_{1,2} \nonumber \\
 U(t,t_0)_{2,1}= U(t,t_0)_{1,2}^{*} & U(t,t_0)_{2,2}
\end{pmatrix}.
\end{eqnarray}
\begin{eqnarray}
 U(t,t_0)_{1,1}= \nonumber \\
\frac{\exp \left(-  \frac{\sqrt{-\hbar^2 \left(|\int_{t_0}^{t}dt'(E_{p1}(t')-E_{p2}(t'))|^2+4 \left(|\int_{t_0}^{t}dt't_{si}(t')|^2+|\int_{t_0}^{t}dt't_{sr}(t')|^2\right)\right)}+i \hbar \int_{t_0}^{t}dt'(E_{p1}(t')+E_{p2}(t'))}{2 \hbar^2}\right)}{2 \hbar
   \left( ( \int_{t_0}^{t}dt'(E_{p1}(t')-E_{p2}(t')))^2+4 \left(|\int_{t_0}^{t}dt't_{si}(t')|^2+|\int_{t_0}^{t}dt't_{sr}(t')|^2\right)\right)} \times \nonumber \\
\times  \Bigg[-i (\int_{t_0}^{t}dt'E_{p1}(t')) \sqrt{-\hbar^2 \left(|\int_{t_0}^{t}dt'(E_{p1}(t')-E_{p2}(t'))|^2+ 4 \left(|\int_{t_0}^{t}dt't_{si}(t')|^2+|\int_{t_0}^{t}dt't_{sr}(t')|^2\right)\right)}+ \nonumber \\  +\hbar \left(|\int_{t_0}^{t}dt'(E_{p1}(t')-E_{p2}(t'))|^2+4
   \left(|\int_{t_0}^{t}dt't_{si}(t')|^2+|\int_{t_0}^{t}dt't_{sr}(t')|^2\right)\right) \times \nonumber \\ e^{\frac{\sqrt{-\hbar^2 \left(|\int_{t_0}^{t}dt'(E_{p1}(t')-E_{p2}(t'))|^2+4 \left(|\int_{t_0}^{t}dt't_{si}(t')|^2+|\int_{t_0}^{t}dt't_{sr}(t')|^2\right)\right)}}{\hbar^2}}+ \nonumber\\  + \left(
   \left(( \int_{t_0}^{t}dt'(E_{p1}(t')-E_{p2}(t')))^2+4 \left(|\int_{t_0}^{t}dt't_{si}(t')|^2+|\int_{t_0}^{t}dt't_{sr}(t')|^2\right)\right) \right) + \nonumber \\
+i (\int_{t_0}^{t}dt'E_{p1}(t')) e^{\frac{\sqrt{-h^2 \left(|\int_{t_0}^{t}dt'(E_{p1}(t')-E_{p2}(t'))|^2+4 \left(|\int_{t_0}^{t}dt't_{si}(t')|^2+|\int_{t_0}^{t}dt't_{sr}(t')|^2\right)\right)}}{\hbar^2}} \times \nonumber \\ \sqrt{-\hbar^2 \left(|\int_{t_0}^{t}dt'(E_{p1}(t')-E_{p2}(t'))|^2+4
   \left(|\int_{t_0}^{t}dt't_{si}(t')|^2+|\int_{t_0}^{t}dt't_{sr}(t')|^2\right)\right)} \nonumber \\ -i (\int_{t_0}^{t}dt'E_{p2}(t')) e^{\frac{\sqrt{-\hbar^2 \left(|\int_{t_0}^{t}dt'(E_{p1}(t')-E_{p2}(t'))|^2+4 \left(|\int_{t_0}^{t}dt't_{si}(t')|^2+|\int_{t_0}^{t}dt't_{sr}(t')|^2\right)\right)}}{\hbar^2}}\times \nonumber \\
   \sqrt{-\hbar^2 \left(|\int_{t_0}^{t}dt'(E_{p1}(t')-E_{p2}(t'))|^2+4 \left(|\int_{t_0}^{t}dt't_{si}(t')|^2+|\int_{t_0}^{t}dt't_{sr}(t')|^2\right)\right)}+ \nonumber \\ + i (\int_{t_0}^{t}dt' E_{p2}(t')) \sqrt{-\hbar^2 \left(|\int_{t_0}^{t}dt'(E_{p1}(t')-E_{p2}(t'))|^2+4
   \left(|\int_{t_0}^{t}dt't_{si}(t')|^2+|\int_{t_0}^{t}dt't_{sr}(t')|^2\right) \right)} \Bigg].
\end{eqnarray}

\begin{eqnarray*}
U(t,t_0)_{1,2}=\nonumber \\
2 \hbar (\int_{t_0}^{t}dt'(t_{si}(t')-i t_{sr}(t'))) e^{-\frac{i \int_{t_0}^{t}dt'(E_{p1}(t')+E_{p2}(t'))}{2 \hbar}} \times \\
 \sinh \left(\frac{\sqrt{-\hbar^2 \left(|\int_{t_0}^{t}dt'(E_{p1}(t')-E_{p2}(t'))|^2+4
   \left(|\int_{t_0}^{t}dt't_{si}(t')|^2+|\int_{t_0}^{t}dt't_{sr}(t')|^2\right)\right)}}{2 h^2}\right) \times \\
\frac{1}{\sqrt{-\hbar^2 \left(|\int_{t_0}^{t}dt'(E_{p1}(t')-E_{p2}(t'))|^2+4 \left(|\int_{t_0}^{t}dt't_{si}(t')|^2+|\int_{t_0}^{t}dt't_{sr}(t')|^2\right)\right)}}= \nonumber \\
   =U(t,t_0)_{2,1}^{*}. \nonumber \\
\end{eqnarray*}

\section{Description of 2 qubit interaction in general static case}
We consider most minimalist model of electrostatically interacting two position-based qubits that are double quantum dots A (with nodes 1 and 2  and named as U-upper qubit) and B (with nodes 1' and 2' and named as L-lower qubit) with local confinement potentials as given in Fig.5 and Fig7.
By introducing notation $\ket{1,0}_x=\ket{1},\ket{0,1}_x=\ket{2},\ket{1',0'}_x=\ket{1'},\ket{0',1'}_x=\ket{1'}$ the minimalistic Hamiltonian of the system of electrostatically interacting position based qubits can be written as
\begin{eqnarray}
\hat{H}=(t_{s21}(t)\ket{2}\bra{1}+t_{s12}(t)\ket{1}\bra{2})\hat{I}_{b})+(\hat{I}_a(t_{s2'1'}(t)\ket{2'}\bra{1'}+t_{s1'2'}(t)\ket{2'}\bra{1'})+ \nonumber \\
+(E_{p1}(t)\ket{1}\bra{1}+E_{p2}(t)\ket{2}\bra{2})\hat{I}_b+ \hat{I}_a(E_{p1'}(t)\ket{1'}\bra{1'}+E_{p2'}(t)\ket{2'}\bra{2'})+ \nonumber \\ 
+\frac{q^2}{d_{11'}}\ket{1,1'}\bra{1,1'}+\frac{q^2}{d_{22'}}\ket{2,2'}\bra{2,2'}+
\frac{q^2}{d_{12'}}\ket{1,2'}\bra{1,2'}+\frac{q^2}{d_{21'}}\ket{2,1'}\bra{2,1'}= \nonumber \\
H_{kinetic1}+H_{pot1}+H_{kinetic2}+H_{pot2}+H_{A-B}
\end{eqnarray} described by parameters
 $E_{p1}(t)$,$E_{p2}(t)$,$E_{p1'}(t)$,$E_{p2'}(t)$, $t_{s12}(t)$, $t_{s1'2'}(t)$ and distances between nodes k and l': $d_{11'}$,$d_{22'}$,$d_{21'}$,$d_{12'}$.
In such case q-state of the system is given as
\begin{eqnarray}
\ket{\psi}_t=\gamma_1(t)\ket{1,0}_U\ket{1,0}_L+\gamma_2(t)\ket{1,0}_U\ket{0,1}_L+\gamma_3(t)\ket{0,1}_U\ket{1,0}_L+\gamma_4(t)\ket{0,1}_U\ket{0,1}_L, \nonumber \\
\end{eqnarray}
where normalization condition gives $|\gamma_1(t)|^2+..|\gamma_4(t)|^2$. Probability of finding electron in upper system at node 1 is by action of projector $\hat{P}_{1U}=\bra{1,0}_U\bra{1,0}_L+\bra{1,0}_U\bra{0,1}_L$ on q-state $\hat{P}_{1U} \ket{\psi}$ so
it gives probability amplitude $|\gamma_1(t)+\gamma_3(t)|^2$ . On the other hand probability of finding electron from qubit A (U) at node 2 and electron from qubit B(L) at node 1 is obtained by projection $\hat{P}_{2U,1L}=\bra{0,1}_U\bra{1,0}_L$ acting on q-state
giving $(\bra{0,1}_U\bra{1,0}_L)\ket{\psi}$ that gives probability amplitude $|\gamma_3(t)|^2$.
Referring to picture from Fig.5
we set distances between nodes as $d_{11'}=d_{22'}=d_1$,$d_{12'}=d_{21'}=\sqrt{(a+b)^2+d_1^2}$ and assume Coulomb electrostatic energy to
 be of the form $E_c(k,l)=\frac{q^2}{d_{kl'}}$ and hence we obtain the matrix  Hamiltonian given as $ \hat{H}(t)= $ \tiny
\begin{eqnarray*}
\label{2bodies}
\begin{pmatrix}
E_{p1}(t)+E_{p1'}(t) + \frac{q^2}{d_1} & t_{s1'2'}(t) & t_{s12}(t) & 0 \\
t_{s1'2'}(t)^{*} & E_{p1}(t)+E_{p2'}(t)+\frac{q^2}{\sqrt{(d1)^2+(b+a)^2}} & 0 & t_{s12}(t) \\
t_{s12}^{*}(t) & 0 & E_{p2}(t)+E_{p1'}(t)+ \frac{q^2}{\sqrt{(d1)^2+(b+a)^2}} & t_{s1'2'}(t) \\
0 & t_{s12}^{*}(t) & t_{s1'2'}(t)^{*} & E_{p2}(t)+E_{p2'}(t)+ \frac{q^2}{d1} \\
\end{pmatrix} \nonumber \\
\end{eqnarray*}
\normalsize
We can introduce notation $E_{c1}=\frac{q^2}{d_1}$ and $E_{c2}=\frac{q^2}{\sqrt{d_1^2+(b+a)^2}}$. In most general case of 2 qubit electrostatic interaction one of which has 4 different Coulomb terms on matrix diagonal $E_{c1}=\frac{q^2}{d_{11'}}$, $E_{c2}\frac{q^2}{d_{12'}}$, $E_{c3}=\frac{q^2}{d_{21'}}$, $E_{c4}=\frac{q^2}{d_{22'}}$ and $\ket{\psi,t}=\hat{U}(t,t_0)\ket{\psi,t_0}$.
We introduce $q_1=E_{p1}(t)+E_{p1'}(t)+E_{c11'}$,$q_2=E_{p1}(t)+E_{p2'}(t)+E_{c12'}$, $q_3=E_{p2}(t)+E_{p1'}(t)+E_{c21'}$,$q_4=E_{p2}(t)+E_{p2'}(t)+E_{c22'}$ and in such case by using formula 8 one can decompose 2 particle Hamiltonian \ref{2bodies} as
\begin{eqnarray}
\hat{H}=\Big[\frac{(q_1+q_2+q_3+q_4)}{4}\sigma_0 \times \sigma_0 +
\frac{(q_1-q_2+q_3-q_4)}{4}\sigma_0 \times \sigma_3 + \nonumber \\
\frac{(q_1+q_2-q_3-q_4)}{4}\sigma_3 \times \sigma_0 +
\frac{(q_1-q_2-q_3+q_4)}{4}\sigma_3 \times \sigma_3 + \nonumber \\
+t_{sr1}(t)\sigma_0 \times \sigma_1 -t_{si1}(t) \sigma_0 \times \sigma_2+t_{sr2}(t)\sigma_1 \times \sigma_0 \nonumber \\
 - t_{si2}(t) \sigma_2 \times \sigma_0 \Big.
\end{eqnarray}
A very similar procedure is for the case of 3 or N interacting particles so one deals with tensor product of 3 or N Pauli matrices.
In order to simplify representation of unitary matrix describing physical system of 2 particles evolution with time it is helpful to define  $Q_1(t)=\int_{t_0}^{t}(E_{p1}(t')+E_{p1'}(t')+E_{c11'})dt'$,$Q_2(t)=\int_{t_0}^{t}(E_{p1}(t')+E_{p2'}(t')+E_{c12'})dt'$, $Q_3(t)=\int_{t_0}^{t}(E_{p2}(t')+E_{p1'}(t')+E_{c21'})dt'$,$Q_4(t)=\int_{t_0}^{t}(E_{p2}(t')+E_{p2'}(t')+E_{c22'})dt'$ and $TR1(t)=\int_{t_0}^{t}dt't_{s1r}(t')$ , $TI1(t)=\int_{t_0}^{t}dt't_{s1i}(t')$. We consider the situation when there is no hopping between q-wells $t_{s2}=0$ so, the second particle is localized among two quantum wells and first particle can move freely among 2 q-wells.
We obtain the following unitary matrix evolution with time with following $\hat{U}(t,t_0)_{1,2}=\hat{U}(t,t_0)_{1,4}=0=\hat{U}(t,t_0)_{2,3}=\hat{U}_{3,4}$ and

\tiny
\begin{eqnarray}
\hat{U}(t,t_0)_{1,1}=   \frac{1}{2 \sqrt{
(Q_1(t)-Q_3(t))^2+4 \left(TR_1(t)^2+TI_1(t)^2\right)}} \times
 \Bigg[Q_1(t)
\left(-e^{i \hbar \sqrt{
(Q_1(t)-Q_3(t))^2+4 \left(TR_1(t)^2+TI_1(t)^2\right)
}}\right) \nonumber \\ +
\left(\sqrt{ |Q_1(t)-Q_3(t)|^2+4(TR_1(t)^2+TI_1(t)^2)}
+Q_3(t) \right) \times  \left(-e^{i \hbar \sqrt{
(Q_1(t)-Q_3(t))^2+4 \left(TR_1(t)^2+TI_1(t)^2\right)
}}\right)+ \nonumber \\ \sqrt{(Q_1(t)-Q_3(t))^2+4 \left(TR_1(t)^2+TI_1(t)^2\right)} +(Q_1(t)-Q_3(t)))
e^{-\frac{1}{2} i \hbar \left(\sqrt{|\int_{t_0}^{t}dt'(q_{1}(t')-q_{3}(t'))|^2+4
 \left(t_{s1r}^2+t_{si1}^2\right)}+(Q_1(t)+Q_3(t))\right)} \Bigg] \nonumber \\
\end{eqnarray}
\normalsize
\begin{eqnarray}
\hat{U}(t,t_0)_{1,3}= \frac{2 (TI_1(t)-iTR_1(t)) e^{-\frac{1}{2}(Q_1(t)+Q_3(t)) i \hbar } \sin \left(\frac{1}{2} \hbar
   \sqrt{|Q_1(t)-Q_3(t)|^2+4(TR_1(t)^2+TI_1(t)^2)}\right)}{\sqrt{|Q_1(t)-Q_3(t)|^2+4(TR_1(t)^2+TI_1(t)^2)}}, \nonumber \\
\end{eqnarray}
\\,
\begin{eqnarray}
\hat{U}(t,t_0)_{2,2}=\Bigg[e^{(\frac{1}{2} i \hbar \left(\sqrt{  (Q_2(t)-Q_4(t))^2 +4(TR_1(t)^2+TI_1(t)^2)}-(Q_2(t)+Q_4(t))\right))} \times \nonumber \\ \times
 \frac{\left(\sqrt{(Q_2(t)-Q_4(t))^2 +4(TR_1(t)^2+TI_1(t)^2)}-Q_2(t)+Q_4(t)\right) }{2
   \sqrt{(Q_2-Q_4)^2+4(TR_1(t)^2+TI_1(t)^2)}} \nonumber \\
   -e^{\left(\frac{1}{2} i \hbar
   \left(-\sqrt{(Q_2(t)-Q_4(t))^2+ 4(TR_1(t)^2+TI_1(t)^2)}-(Q_2(t)+Q_4(t))\right)\right)} \times \nonumber \\ \times
\frac{\left(-\sqrt{ (Q_2(t)-Q_4(t))^2+ 4(TR_1(t)^2+TI_1(t)^2) }-Q_2(t)+Q_4(t)\right) }{2\sqrt{(Q_2-Q_4)^2+4(TR_1(t)^2+TI_1(t)^2)}}
\end{eqnarray}
$ $
\\
\\
$ $
\begin{eqnarray}
\hat{U}(t,t_0)_{3,3}=\frac{\exp \left(-\frac{1}{2} i \hbar \left(\sqrt{(Q_1(t)^2-Q_3(t))^2+4
   \left(TR_1(t)^2+TI_1(t)^2\right)}+Q_1(t)+Q_3(t)\right)
   \right)}{2 \sqrt{(Q_1(t)^2-Q_3(t))^2+4
   \left( TR_1(t)^2+TI_1(t)^2 \right)}} \times \nonumber \\
\Bigg[Q_1(t) \left(-1+e^{i \hbar \sqrt{(Q_1(t)-Q_3(t))^2+4
   \left(TR_1(t)^2+TI_1(t)^2 \right)}}\right)+ \nonumber \\
   \left(\sqrt{(Q_1(t)-Q_3(t))^2+4
   \left(TR_1(t)^2+TI_1(t)^2\right)}-\text{q3}\right) e^{i \hbar
   \sqrt{(Q_1(t)-Q_3(t))^2+4
   \left(TR_1(t)^2+TI_1(t)^2\right)}}+\nonumber \\+\sqrt{(Q_1(t)-Q_3(t))^2+4 \left(TR_1(t)^2+TI_1(t)^2\right)}+Q_3(t)\Bigg]
\end{eqnarray}

\begin{eqnarray*}
 \hat{U}(t,t_0)_{4,4}=\frac{\exp \left(-\frac{1}{2} i \hbar \left(\sqrt{(Q_2(t)-Q_4(t))^2+4
   \left(TR_1(t)^2+TI_1(t)^2\right)}+Q_2(t)+Q_4(t)\right)
   \right)}{2 \sqrt{(Q_2(t)-Q_4(t))^2+4
   \left(TR_1(t)^2+TI_1(t)^2\right)}} \times \nonumber \\
\times \Bigg[Q_2(t) \left(-1+e^{i \hbar \sqrt{(Q_2(t)-Q_4(t))^2+4
   \left(TR_1(t)^2+TI_1(t)^2\right)}}\right)+\nonumber \\ +\left(\sqrt{(Q_2(t)-Q_4(t)^2)^2 +4
   \left( TR_1(t)^2+TI_1(t)^2\right)}-Q_4(t)\right) e^{i \hbar
   \sqrt{(Q_2(t)-Q_4(t))^2+ 4
   \left(TR_1(t)^2+TI_1(t)^2\right)}}+ \nonumber \\
   \sqrt{(Q_2(t)-Q_4(t))^2 +4 \left(TR_1(t)^2+TI_1(t)^2\right)}+Q_4(t)\Bigg]
\end{eqnarray*}
\begin{eqnarray}
\hat{U}(t,t_0)_{2,4}=\frac{2 (TI_1(t)-i TR_1(t)) e^{-\frac{1}{2} i \hbar
   (Q_2(t)+Q_4(t))} \sin \left(\frac{1}{2} \hbar
   \sqrt{(Q_2(t)-Q_4(t))^2+4
   \left(TR_1(t)^2+TI_1(t)^2\right)}\right)}{\sqrt{(Q_2(t)-Q_4(t))^2+4 \left(TR_1(t)^2+TI_1(t)^2\right)}} \nonumber \\
\end{eqnarray}

\begin{figure}
\centering
\includegraphics[scale=0.4]{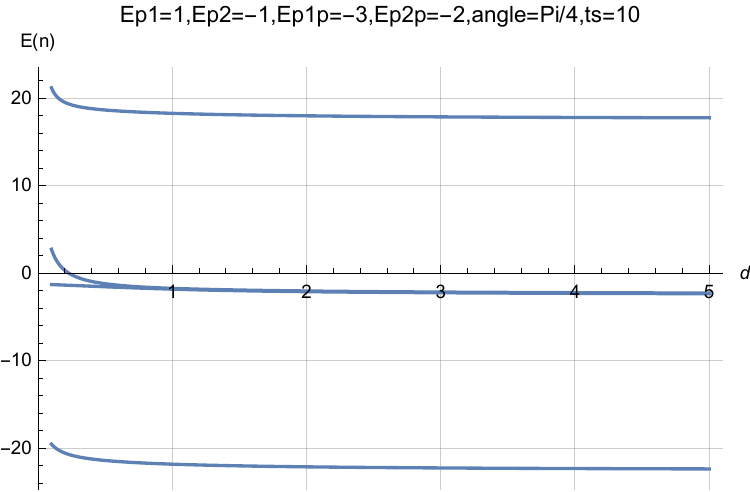}
\includegraphics[scale=0.4]{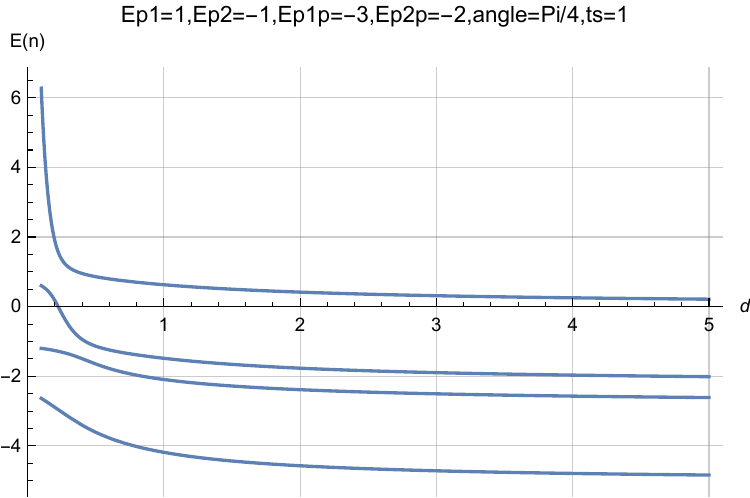}
\includegraphics[scale=0.4]{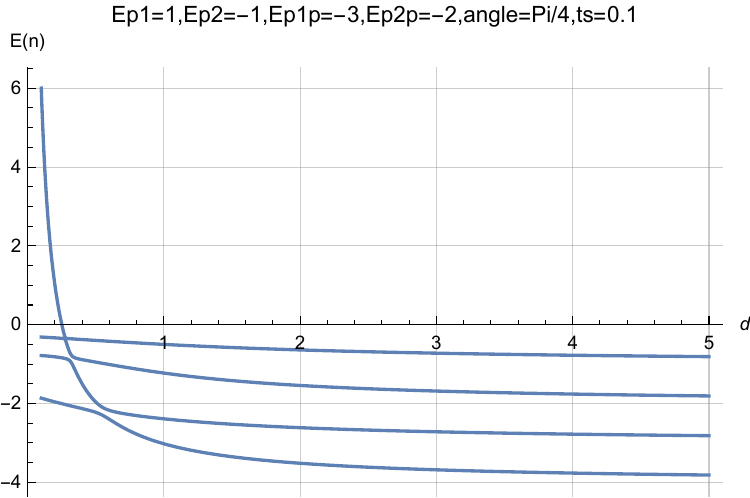}
\includegraphics[scale=0.4]{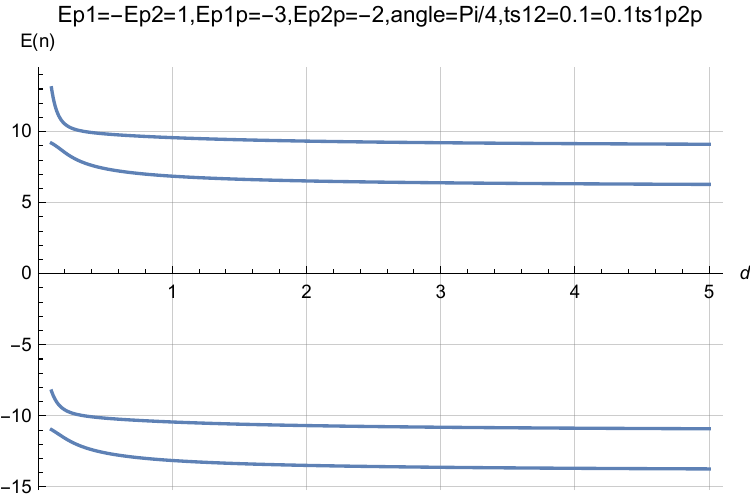}
\includegraphics[scale=0.4]{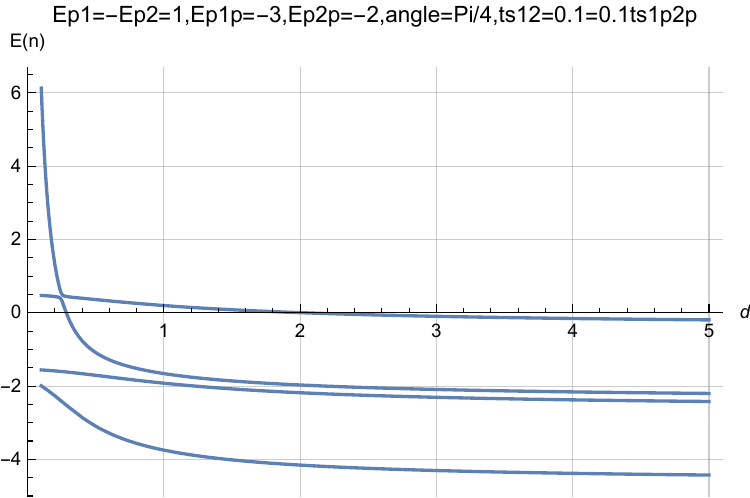}
\includegraphics[scale=0.4]{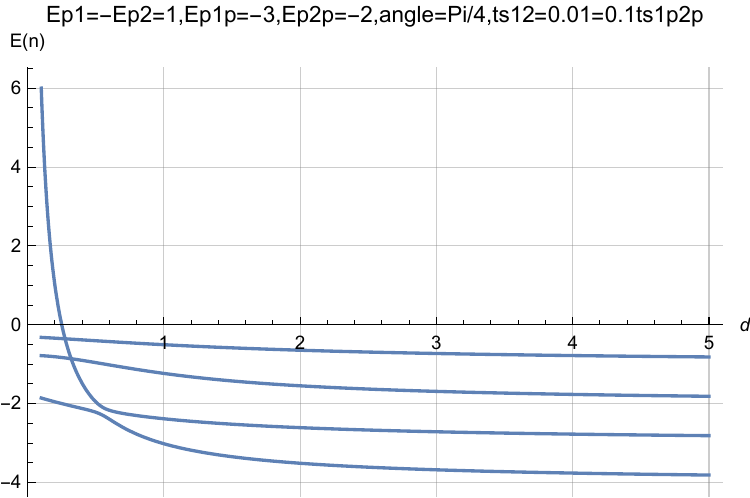}
\includegraphics[scale=0.4]{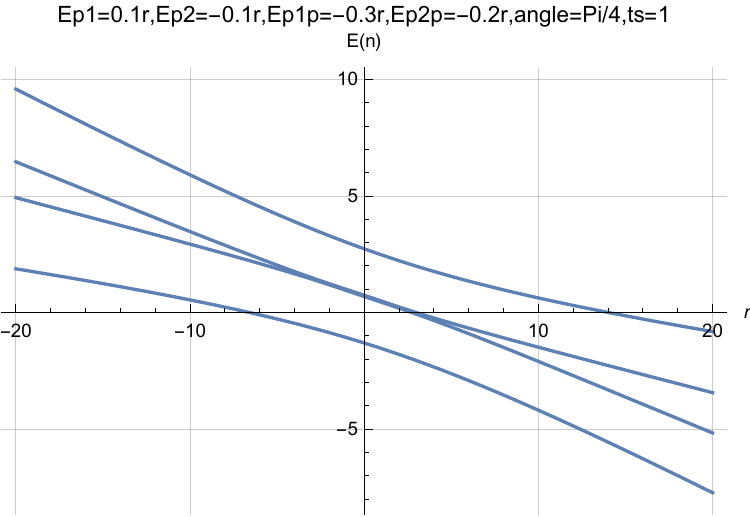}
\includegraphics[scale=0.4]{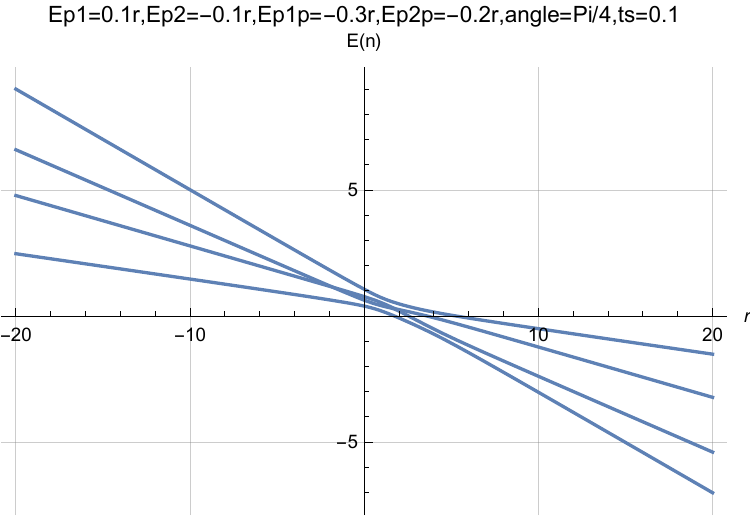}
\includegraphics[scale=0.4]{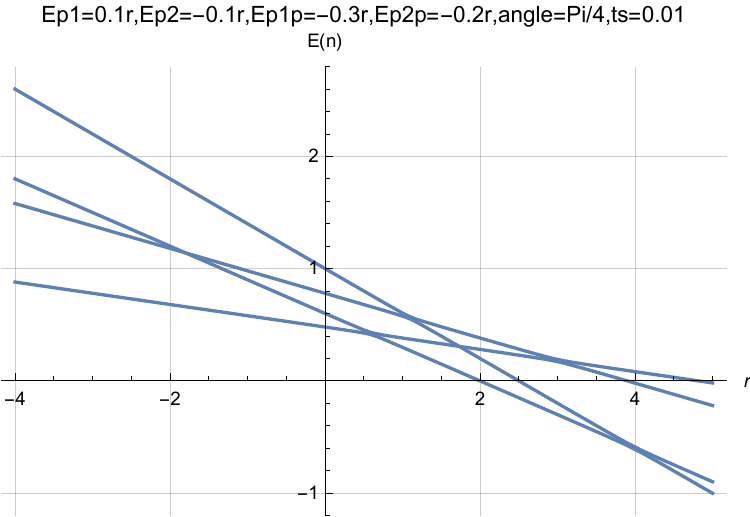}
\includegraphics[scale=0.4]{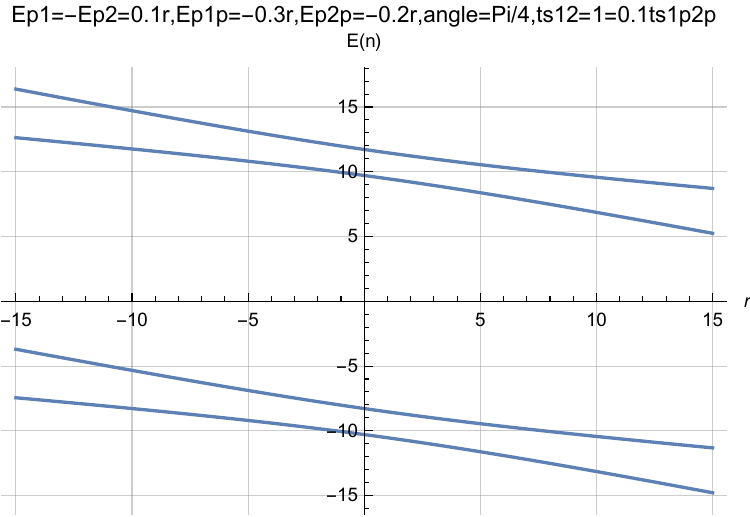}
\includegraphics[scale=0.4]{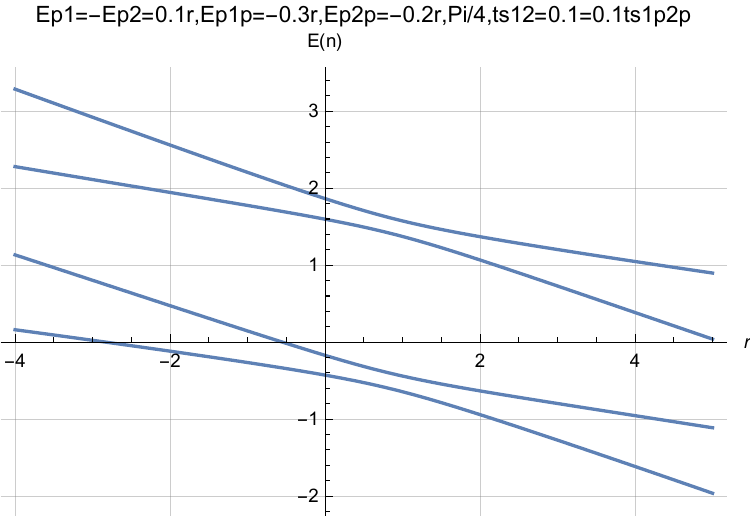}
\includegraphics[scale=0.4]{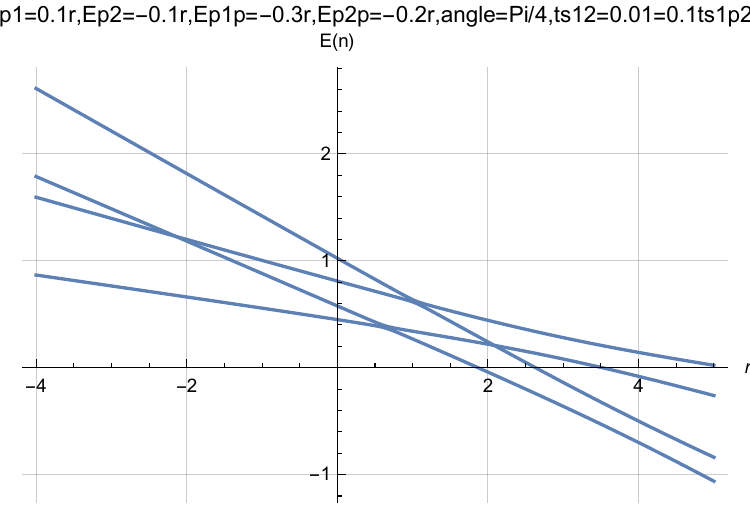}
\caption{Case of metallic-insulator transition seen from energy spectra of quantum swap gate in dependence on the distance $d$ (case of 2 electrostatically interacting qubits from Fig.\,\ref{GQSwapGate}). No level crossings are observed, even after careful magnification of the plots.}
\label{fig:spectra}
\end{figure}

The example of function dependence of eigenenergy spectra of 2 electrostatically interacting qubits on distance is given by Fig.\ref{fig:spectra}.

\begin{figure}
\centering
\includegraphics[scale=0.3]{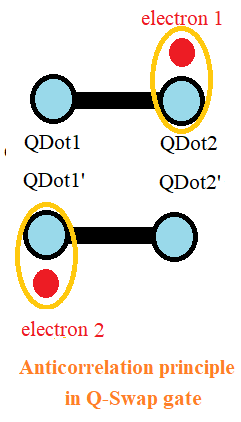}\includegraphics[scale=0.3]{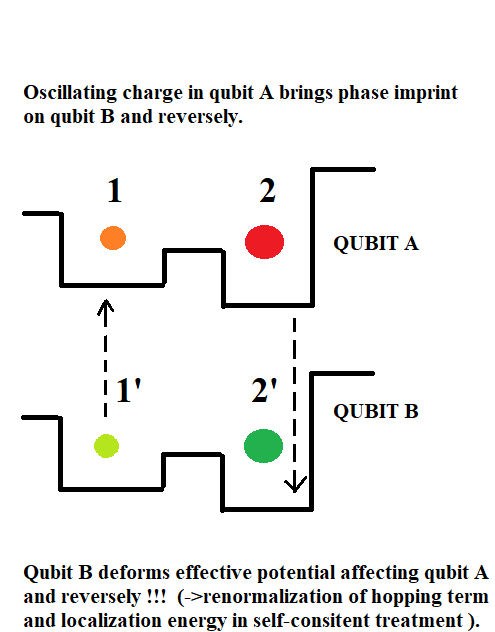}\includegraphics[scale=0.3]{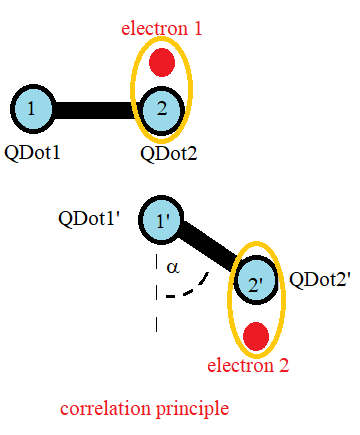}\includegraphics[scale=0.3]{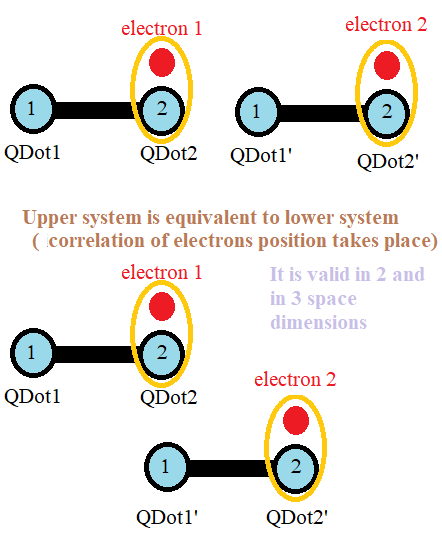}\includegraphics[scale=0.3]{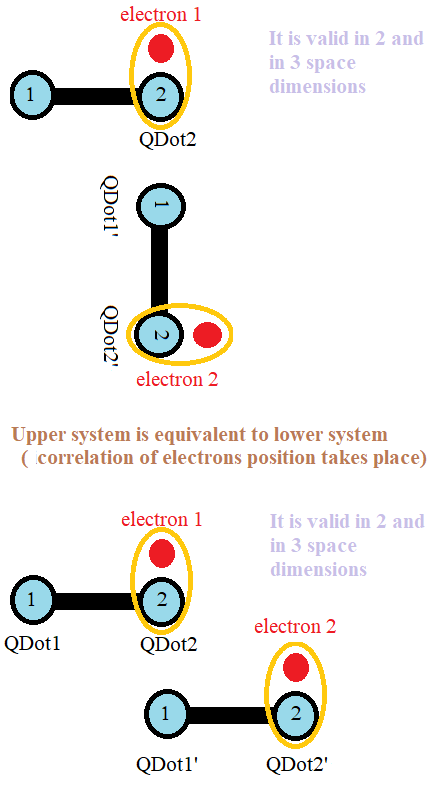}
\caption{(Very Left):Anticorrelation principle in Q-Swap gate, (Left): Scheme of renormalization procedure in the system of coupled qubits accounted by procedure given in work entitled"Analytic view on N body interaction in electrostatic
quantum gates and decoherence effects in
tight-binding model" [ \cite{b5} ], (Right):Illustration of anticorrelation principle in classical or quantum swap gate, (Right): Correlation principle in chain of coupled dots on one line (parallel lines), (Very Right):Correlation principle for the system of qubits in perpendicular alignment. This system is also equivalent to generalized electrostatic quantum swap gate from Fig.5.}
\label{renormalization}
\end{figure}
An important observation is that any element of matrix $\hat{H}(t')$ for $t' \in (t_0,t)$ denoted as $H_{k,l}(t')$ is transferred to element $\hat{U}_{k,l}(t,t_0)=e^{\frac{1}{\hbar i}\int_{t_0}^{t}dt'(H_{k,l}(t'))}$ of matrix $\hat{U}(t,t_0)$. 
We can easily generalize the presented reasoning for the system of N electrostatically coupled electrons confined by some local potentials. However we need to know the position dependent Hamiltonian eigenstate at the initial time $t_0$. In case $N>2$ finding such eigenstate is the numerical problem since
analytical solutions for roots of polynomials of one variable for higher order than 4 does not exist. Using numerical eigenstate at time instance $t_0$ we can compute the system quantum dynamics in analytical way.
This give us a strong and relatively simple mathematical tool giving full determination of quantum dynamical state at the any instance of time.
The act of measurement on position based qubit is represented by the operator $P_{Left}=\ket{1,0}_{E_1,E_2}\bra{1,0}_{E_1,E_2}$ and $P_{Right}=\ket{0,1}_{E_1,E_2}\bra{0,1}_{E_1,E_2}$.

\subsection{Simplified picture of symmetric Q-Swap gate}
Now we need to find a system 4 eigenvalues and eigenstates \newline
 (4 orthogonal 4-dimensional vectors)  so we are dealing with a matrix eigenvalue problem), a standard topic in linear algebra. Let us assume that 2 double quantum dot systems are symmetric and biased by the same voltages generating potential bottoms $V_s$ so we have $E_{p1}=E_{p2}=E_{p1'}=E_{p2'}=E_p=V_s$ and that $t_{s12}=t_{s1'2'}=t_s$. Denoting $E_c(1,1')=E_c(2,2')=E_{c1}$ and $E_c(1,2')=E_c(2,1')=E_{c2}$  we are obtaining 4 orthogonal Hamiltonian eigenvectors
\begin{eqnarray}
\ket{E_1}=
\begin{pmatrix}
-1 \\
0 \\
0 \\
+1
\end{pmatrix}= \nonumber \\
=-\ket{1,0}_U\ket{1,0}_L+\ket{0,1}_U\ket{0,1}_L \nonumber \\
\neq (a_1\ket{1,0}_U+a_2\ket{0,1}_U)(a_3\ket{1,0}_U+a_4\ket{0,1}_U),
\end{eqnarray}
\begin{eqnarray}
\ket{E_2}=
\begin{pmatrix}
1 \\
0 \\
0 \\
-1
\end{pmatrix}= \nonumber \\
=\ket{1,0}_U\ket{0,1}_L-\ket{0,1}_U\ket{1,0}_L \neq (a_1\ket{1,0}_U+a_2\ket{0,1}_U)(a_3\ket{1,0}_U+a_4\ket{0,1}_U)\nonumber \\.
\end{eqnarray}
We observe that two first energetic states are degenerated so the same quantum state corresponds to 2 different eigenenergies $E_1$ and $E_2$.
 This degeneracy is non-present if we come back to Schroedinger picture and  observe that localized energy and hopping terms for one particle are depending on another particle presence that will bring renormalization of wavevectors.
Situation is depicted in Fig.\ref{renormalization}. Degeneracy of eigenstates is lifted if we set $E_{p1}(|\psi(1')|^2,|\psi(2')|^2), E_{p2}(|\psi(1')|^2,|\psi(1')|^2)$, $E_{p1'}(|\psi(1)|^2,|\psi(2)|^2), E_{p2'}(|\psi(1)|^2,|\psi(1)|^2)$ and $t_{1 \rightarrow 2}(|\psi(1')|^2,|\psi(2')|^2)$, \newline
 $t_{1 \rightarrow 2}(|\psi(1)|^2,|\psi(2)|^2)$.

The same argument is for another
wavevectors as given below.
\begin{eqnarray}
\ket{E_{3(4)}}=
\begin{pmatrix}
1 \\
\mp\frac{4 t_s}{\pm(-E_{c1} + E_{c2}) + \sqrt{(E_{c1} - E_{c2})^2 + 16 t_s^2}} \\
\mp\frac{4 t_s}{\pm(-E_{c1} + E_{c2}) + \sqrt{(E_{c1} - E_{c2})^2 + 16 t_s^2}} \\
1
\end{pmatrix}= \nonumber \\
=\ket{1,0}_U\ket{1,0}_L+\ket{0,1}_U\ket{0,1}_L+c (\ket{1,0}_U\ket{0,1}_L +\ket{0,1}_U\ket{1,0}_L )= \nonumber \\
=(\ket{1,0}_U+\ket{0,1}_U)(\ket{1,0}_L+\ket{0,1}_L)+(c-1)(\ket{1,0}_U\ket{0,1}_L +\ket{0,1}_U\ket{1,0}_L) \nonumber \\
\neq (a_1\ket{1,0}_U+a_2\ket{0,1}_U)(a_3\ket{1,0}_U+a_4\ket{0,1}_U), \nonumber \\
\end{eqnarray}
where c=$\mp\frac{4 t_s}{\pm(-E_{c1} + E_{c2}) + \sqrt{(E_{c1} - E_{c2})^2 + 16 t_s^2}}$.
First two $\ket{E_1}$ and $\ket{E_2}$ energy eigenstates are always entangled, while  $\ket{E_3}$ and $\ket{E_4}$ eigenenergies are only partially entangled if $\mp\frac{4 t_s}{\pm(-E_{c1} + E_{c2}) + \sqrt{(E_{c1} - E_{c2})^2 + 16 t_s^2}} \neq 1$. If
$c=1=\mp\frac{4 t_s}{\pm(-E_{c1} + E_{c2}) + \sqrt{(E_{c1} - E_{c2})^2 + 16 t_s^2}}$ last two energy eigenstates are not entangled. The situation of c=1 takes place when $E_{c1}=E_{c2}$ so when two qubits are infinitely far away so when they are electrostatically decoupled. Situation of c=0 is interesting because it means that $\ket{E_3}$ and $\ket{E_4}$ are maximally entangled and it occurs when $t_s=0$ so when two electrons are maximally localized in each of the qubit so there is no hopping between left and right well.

The obtained eigenenergy states correspond to 4 eigenenergies
\begin{eqnarray}
E_1=E_{c1} + 2 V_s, E_2=E_{c2} + 2 V_s, E_1 > E_2 \nonumber \\
E_3= \frac{1}{2} ( (E_{c1} + E_{c2})  - \sqrt{(E_{c1} -E_{c2})^2 + 16 t_s^2} + 4 V_s)=  \nonumber \\
=\frac{1}{2} ( (q^2(\frac{1}{d_1}+\frac{1}{\sqrt{d_1^2+(a+b)^2}} ))  - \sqrt{(q^2(\frac{1}{d_1}-\frac{1}{\sqrt{d_1^2+(a+b)^2}} ))^2 + 16 t_s^2} + 4 V_s), \nonumber \\
E_4 =\frac{1}{2} ( (E_{c1} + E_{c2})  + \sqrt{(E_{c1} -E_{c2})^2 + 16 t_s^2} + 4 V_s)=  \nonumber \\
 \frac{1}{2} ( (q^2(\frac{1}{d_1}+\frac{1}{\sqrt{d_1^2+(a+b)^2}})+ \sqrt{(q^2(\frac{1}{d_1}-\frac{1}{\sqrt{d_1^2+(a+b)^2}}))^2 + 16 t_s^2} + 4 V_s), E_4 > E_3 \nonumber \\ .
\end{eqnarray}

We also notice that the eigenenergy states $\ket{E_1}$, $\ket{E_2}$ ,$\ket{E_3}$, $\ket{E_4}$ do not have its classical counterpart since upper electron exists at both positions 1 and 2 and lower electron exists at both positions at the same time. We observe that when distance between two systems of double quantum dots goes into infinity the energy difference between quantum state corresponding to $\ket{E_3}$ and $\ket{E_4}$ goes to zero. This makes those two entangled states degenerated.




Normalized 4 eigenvectors of 2 interacting qubits in SWAP Q-Gate configuration are of the following form
\begin{eqnarray}
 \ket{E_1}_n=
\frac{1}{\sqrt{\left(8\left(\frac{t_{sr1}-t_{sr2}}{\sqrt{(E_{c1}-E_{c2})^2+4
   (t_{sr1}-t_{sr2})^2}-E_{c1}+E_{c2}}\right)\right)^2+2
   }}
\begin{pmatrix}
-1,\\
-\frac{2(t_{sr1}-t_{sr2})}{\sqrt{(E_{c1}-E_{c2})^2+4 (t_{sr1}-t_{sr2})^2}-E_{c1}+E_{c2}}, \\
\frac{2(t_{sr1}-t_{sr2})}{\sqrt{(E_{c1}-E_{c2})^2+4(t_{sr1}-t_{sr2})^2}-E_{c1}+E_{c2}} \\
1
\end{pmatrix}=\nonumber \\
\frac{1}{\sqrt{\left(8\left(\frac{t_{sr1}-t_{sr2}}{\sqrt{(E_{c1}-E_{c2})^2+4
   (t_{sr1}-t_{sr2})^2}-E_{c1}+E_{c2}}\right)\right)^2+2
   }} \ket{E_1}
\end{eqnarray}

\begin{eqnarray}
 \ket{E_2}_{n}= -\frac{1}{\sqrt{\left(8\left(\frac{t_{sr1}-t_{sr2}}{\sqrt{(E_{c1}-E_{c2})^2+4
   (t_{sr1}-t_{sr2})^2}+E_{c1}-E_{c2}}\right)\right)^2+2
   }}
\begin{pmatrix}
-1 \\
\frac{2
   (t_{sr1}-t_{sr2})}{\sqrt{(E_{c1}-E_{c2})^2+4
   (t_{sr1}-t_{sr2})^2}+E_{c1}-E_{c2}} \\
-\frac{2(t_{sr1}-t_{sr2})}{\sqrt{(E_{c1}-E_{c2})^2+4
   (t_{sr1}-t_{sr2})^2}+E_{c1}-E_{c2}} \\,1
\end{pmatrix}=\nonumber \\ =-
\frac{1}{\sqrt{\left(8\left(\frac{t_{sr1}-t_{sr2}}{\sqrt{(E_{c1}-E_{c2})^2+4
   (t_{sr1}-t_{sr2})^2}+E_{c1}-E_{c2}}\right)\right)^2+2
   }} \ket{E_2}
\end{eqnarray}

\begin{eqnarray}
\ket{E_3}_{n}= \frac{1}{\sqrt{\left(8\left(\frac{\text{tsr1}+\text{tsr2}}{\sqrt{(\text{Ec1}-\text{Ec2})^2+4
   (\text{tsr1}+\text{tsr2})^2}-\text{Ec1}+\text{Ec2}}\right)\right)^2+2
   }}
\begin{pmatrix}
1, \\
-\frac{2(t_{sr1}+t_{sr2})}{\sqrt{(E_{c1}-E_{c2})^2+4
   (t_{sr1}+t_{sr2})^2}-E_{c1}+E_{c2}}, \\
-\frac{2(t_{sr1}+t_{sr2})}{\sqrt{(E_{c1}-E_{c2})^2+4
   (t_{sr1}+t_{sr2})^2}-E_{c1}+E_{c2}}, \\
1
\end{pmatrix}= \nonumber \\ =
\frac{1}{\sqrt{\left(8\left(\frac{\text{tsr1}+\text{tsr2}}{\sqrt{(\text{Ec1}-\text{Ec2})^2+4
   (\text{tsr1}+\text{tsr2})^2}-\text{Ec1}+\text{Ec2}}\right)\right)^2+2
   }} \ket{E_3}
\end{eqnarray}

\begin{eqnarray}
\ket{E_4}_{n}=  \frac{1}{\sqrt{\left(8\left(\frac{t_{sr1}+t_{sr2}}{\sqrt{(E_{c1}-E_{c2})^2+4
   (t_{sr1}+t_{sr2})^2}-E_{c2}+E_{c1}}\right)\right)^2+2
   }}
\begin{pmatrix}
1, \\
\frac{2(t_{sr1}+t_{sr2})}{\sqrt{(E_{c1}-E_{c2})^2+4(t_{sr1}+t_{sr2})^2}+E_{c1}-E_{c2}}, \\
\frac{2(t_{sr1}+t_{sr2})}{\sqrt{(E_{c1}-E_{c2})^2+4(t_{sr1}+t_{sr2})^2}+E_{c1}-E_{c2}},  \\
1
\end{pmatrix}= \nonumber \\
\frac{1}{\sqrt{\left(8\left(\frac{t_{sr1}+t_{sr2}}{\sqrt{(E_{c1}-E_{c2})^2+4
   (t_{sr1}+t_{sr2})^2}-E_{c2}+E_{c1}}\right)\right)^2+2
   }} \ket{E_4}.
   \end{eqnarray}

We are obtaining simplifications after assuming $t_{sr1}(t)=t_{sr2}(t)$ so we obtain
\begin{equation}
\ket{E_1}_{n}=\frac{1}{\sqrt{2}}
\begin{pmatrix}
-1 \\
0 \\
0 \\
1
\end{pmatrix},
\ket{E_2}_{n}=\frac{1}{\sqrt{2}}
\begin{pmatrix}
1 \\
0 \\
0 \\
-1
\end{pmatrix},
\end{equation}
\begin{eqnarray}
\ket{E_3}_{n}=\sqrt{\frac{4t_s}{(E_{c2}-E_{c1}) + 8 t_s - \sqrt{ (E_{c1} - E_{c2})^2 + 16 t_s^2}}}
\begin{pmatrix}
1 \\
-\frac{4 t_s}{(-E_{c1} + E_{c2}) + \sqrt{(E_{c1} - E_{c2})^2 + 16 t_s^2}} \\
-\frac{4 t_s}{(-E_{c1} + E_{c2}) + \sqrt{(E_{c1} - E_{c2})^2 + 16 t_s^2}} \\
1
\end{pmatrix}, \nonumber \\
\end{eqnarray}

\begin{eqnarray}
\ket{E_4}_{n}=\sqrt{\frac{4t_s}{(E_{c1}-E_{c2}) + 8 t_s - \sqrt{ (E_{c1} - E_{c2})^2 + 16 t_s^2}}}
\begin{pmatrix}
1 \\
\frac{4 t_s}{(E_{c1} - E_{c2}) + \sqrt{(E_{c1} - E_{c2})^2 + 16 t_s^2}} \\
\frac{4 t_s}{(E_{c1} - E_{c2}) + \sqrt{(E_{c1} - E_{c2})^2 + 16 t_s^2}} \\
1
\end{pmatrix}, \nonumber \\.
\end{eqnarray}

It is worth mentioning that if we want to bring two electrostatic qubits to the entangled state we need to cool down (or heat-up)  the system of interacting qubits to the energy $E_1$ (or to energy $E_2$).
Otherwise we might also wish to disentangle two electrostatically interacting qubits. In such way one of the scenario is to bring the quantum system either to energy $E_3$ or $E_4$ so only partial entanglement will be achieved.
Other scenario would be by bringing the occupancy of different energetic levels so net entanglement is reduced. One can use the entanglement witness in quantifying the existence of entanglement. One of the simplest q-state entanglement measurement is von Neumann entanglement entropy as it is expressed by formula \ref{entropyS} that requires the knowledge of q-system density matrix with time. Such matrices can be obtained analytically for the case of 2 electrostatically interacting qubits.

It is interesting to spot the dependence of eigenergies on distance between interacting qubits in the general case as it is depicted in Fig.6. Now we are moving towards description the procedure of cooling down or heating up in Q-Swap gate.
The procedure was discussed previously in the case of single qubit. Now it is exercised in the case of 2-qubit electrostatic interaction. For the sake of simplicity we will change the occupancy of the energy level $E_1$ and energy level level $E_2$ and keep
the occupancy of other energy levels unchanged.  We can write the $\ket{E_2}\bra{E_1} $ as
\begin{eqnarray}
\ket{E_2}_n\bra{E_1}_n=\frac{1}{2}
\begin{pmatrix}
1 \\ 0 \\ 0 \\ -1
\end{pmatrix}
\begin{pmatrix}
-1 & 0 & 0 & 1
\end{pmatrix}=
\begin{pmatrix}
-1 & 0 & 0 & +1 \\
0 & 0 & 0 & 0 \\
0 & 0 & 0 & 0 \\
+1 & 0 & 0 & -1 \\
\end{pmatrix}, \nonumber \\
\ket{E_1}_n\bra{E_2}_n=\frac{1}{2}
\begin{pmatrix}
-1 \\ 0 \\ 0 \\ 1
\end{pmatrix}
\begin{pmatrix}
1 & 0 & 0 & -1
\end{pmatrix}=
\begin{pmatrix}
-1 & 0 & 0 & +1 \\
0 & 0 & 0 &  0 \\
0 & 0 & 0 & 0 \\
+1 & 0 & 0 & -1 \\
\end{pmatrix}. \nonumber \\
\end{eqnarray}
We are introducing $f_1$ and $f_2$ real valued functions of small magnitude $f(t)=f_1(t)=f_2(t), (|f_1|,|f_2|<<(E_1,E_2))$  and we are considering the following Hamiltonian having $H_0$ that is time-independent and other part dependent part as
\begin{eqnarray}
\hat{H}=\hat{H}_0+f_1(t)\ket{E_2}_n\bra{E_1}_n+f_2(t)\ket{E_1}_n\bra{E_2}_n= \nonumber \\
=E_1\ket{E_1}\bra{E_1}+E_2\ket{E_2}\bra{E_2}+f_1(t)\ket{E_2}_n\bra{E_1}_n+f_2(t)\ket{E_1}_n\bra{E_2}_n= \nonumber
\end{eqnarray}
\begin{eqnarray}
=
\begin{pmatrix}
2E_{p}+ \frac{q^2}{d_1} & t_{s} & t_{s} & 0 \\
t_{s}^{*} & 2E_{p}+\frac{q^2}{\sqrt{(d1)^2+(b+a)^2}} & 0 & t_{s} \\
t_{s}^{*} & 0 & 2E_{p}+ \frac{q^2}{\sqrt{(d1)^2+(b+a)^2}} & t_{s} \\
0 & t_{s}^{*} & t_{s}^{*} & 2 E_{p}+ \frac{q^2}{d1} \\
\end{pmatrix}   \nonumber \\
+\frac{1}{2}\left(f_1
\begin{pmatrix}
-1 & 0 & 0 & 1 \\
0 & 0 & 0 & 0 \\
0 & 0 & 0 & 0 \\
1 & 0 & 0 & -1 \\
\end{pmatrix}
+f_2
\begin{pmatrix}
-1 & 0 & 0 & 1 \\
0 & 0 & 0 &  0 \\
0 & 0 & 0 & 0 \\
1 & 0 & 0 & -1 \\
\end{pmatrix}\right)=
\nonumber \\
=
\begin{pmatrix}
2E_{p}+ \frac{q^2}{d_1}-f(t) & t_{s} & t_{s} & f(t) \\
t_{s}^{*} & 2E_{p}+\frac{q^2}{\sqrt{(d1)^2+(b+a)^2}} & 0 & t_{s} \\
t_{s}^{*} & 0 & 2E_{p}+ \frac{q^2}{\sqrt{(d1)^2+(b+a)^2}} & t_{s} \\
f(t) & t_{s}^{*} & t_{s}^{*} & 2 E_{p}+ \frac{q^2}{d1}-f(t) \\
\end{pmatrix} = \nonumber \\
=\hat{H}(t)_{E_1<->E_2,Q-Swap}.\nonumber \\
\end{eqnarray}

\normalsize
Initially we have established the following parameters of tight-binding model as $t_{s12}=t_{s1'2'}$. Changing $t_{s12}$ into $t_{s12}-\frac{f(t)}{2}$ and $t_{s1'2'}$ into $t_{s1'2'}+\frac{f(t)}{2}$ while keeping other parameters of tight-binding model unchanged will result
in the heating up (cooling down) of q-state of SWAP gate so population of energy level $E_1$ and $E_2$ are time-depenent, while populations of energy levels $E_3$ and $E_4$ are unchanged. Practically our results mean that we need to keep all our confiment potential bottoms constant, while changing barrier height between neighbouring q-dots in each of position based qubits.
In such way we have established the procedure of perturbative cooling (heating up) of q-state. Non-perturbative approach is absolutely possible but it requires full knowledge of time dependent eigenstates and eigenenergies (solutions of eigenenergies of 4th order polynomial are very lengthy in general case) and therefore corresponding expression are very lengthy. In similar fashion we can heat up or cool down two coupled Single Electron Lines \cite{SEL} as in Fig.5 and Fig.7 or any other q-system having N interacting q-bodies that can be represented by the system of N-interacting position based qubits.

\subsection{Case of density matrix in case of 2 interacting particles in symmetric case}
We consider the simplifying matrix and highly symmetric matrix of the form
\tiny
\begin{eqnarray}
\hat{H}(t)= \nonumber \\
\begin{pmatrix}
2E_{p}(t)+ \frac{q^2}{d_1}=q_{11}+q_{22}  & t_{sr2}(t) & t_{sr1}(t) & 0 \\
t_{sr2}(t)  & 2E_{p}(t)+\frac{q^2}{\sqrt{(d_1)^2+(b+a)^2}}=q_{11}-q_{22} & 0 & t_{sr1}(t) \\
t_{sr1}(t) & 0 & 2E_{p}(t)+ \frac{q^2}{\sqrt{(d_1)^2+(b+a)^2}}=q_{11}-q_{22} & t_{sr2}(t) \\
0 & t_{sr1}(t) & t_{sr2}(t) & 2 E_{p}(t)+ \frac{q^2}{d1}=q_{11}+q_{22} \\
\end{pmatrix} =\nonumber \\
=\hat{\sigma}_0 \times \hat{\sigma}_0 q_{11} + \hat{\sigma}_{3} \times \hat{\sigma}_{3} q_{22} +t_{sr2}(t)  \hat{\sigma}_0 \times \hat{\sigma}_3 + t_{sr1}(t) \hat{\sigma}_3  \times \hat{\sigma}_0 \nonumber \\
\end{eqnarray}
\normalsize
that has only real value components $H_{k,l}$ with $q_{11}=E_p(t)+\frac{E_{c1}+E_{c2}}{2}=E_p(t)+\frac{1}{2}(\frac{q^2}{d_1}+\frac{q^2}{\sqrt{(d_1)^2+(b+a)^2}})$, $q_{22}=\frac{E_{c1}-E_{c2}}{2}=\frac{1}{2}(\frac{q^2}{d_1}-\frac{q^2}{\sqrt{(d_1)^2+(b+a)^2}})$ and $Q_{11}(t)=\int_{t_0}^{t}dt'q_{11}(t')$, $Q_{22}(t)=\int_{t_0}^{t}dt'q_{22}(t')$, $TR1(t)=\int_{t_0}^{t}dt't_{sr1}(t')$, $TR2(t)=\int_{t_0}^{t}dt't_{sr2}(t')$.  We obtain the density matrix

\begin{eqnarray}
\hat{U}(t)=
\begin{pmatrix}
U_{1,1}(t) & U_{1,2}(t) & U_{1,3}(t) & U_{1,4}(t) \\
U_{2,1}(t) & U_{2,2}(t) & U_{2,3}(t) & U_{2,4}(t) \\
U_{3,1}(t) & U_{3,2}(t) & U_{3,3}(t) & U_{3,4}(t) \\
U_{4,1}(t) & U_{4,2}(t) & U_{4,3}(t) & U_{4,4}(t) \\
\end{pmatrix},
\hat{\rho}(t)=\hat{U}(t,t_0)
\begin{pmatrix}
\rho_{1,1}(t_0) & \rho_{1,2}(t_0) & \rho_{1,3}(t_0) & \rho_{1,4}(t_0) \\
\rho_{2,1}(t_0) & \rho_{2,2}(t_0) & \rho_{2,3}(t_0) & \rho_{2,4}(t_0) \\
\rho_{3,1}(t_0) & \rho_{3,2}(t_0) & \rho_{3,3}(t_0) & \rho_{3,4}(t_0) \\
\rho_{4,1}(t_0) & \rho_{4,2}(t_0) & \rho_{4,3}(t_0) & \rho_{4,4}(t_0) \\
\end{pmatrix}\hat{U}^{-1}(t,t_0)\nonumber \\
\end{eqnarray}
with the following components of unitary matrix
\begin{eqnarray}
U_{1,1}(t)=\frac{e^{-i \hbar Q_{11}(t)}}{2}\Bigg[
-iQ_{22}(t)\times \nonumber \\ \times \left(\frac{\sin \left(\hbar
   \sqrt{|Q_{22}(t)|^2+(TR1(t)-TR2(t))^2}\right)}{\sqrt{|Q_{22}(t)|^2+(TR1(t)-TR2(t))^2}}+\frac{\sin \left(\hbar
   \sqrt{|Q_{22}(t)|^2+(TR1(t)+TR2(t))^2}\right)}{\sqrt{|Q_{22}(t)|^2+(TR1(t)+TR2(t))^2}}\right)+ \nonumber \\ +\cos \left(\hbar
   \sqrt{|Q_{22}(t)|^2+(TR1(t)-TR2(t))^2}\right)+\cos \left(\hbar \sqrt{|Q_{22}(t)|^2+(TR1(t)+TR2(t))^2}\right) \Bigg].
\end{eqnarray}
\begin{eqnarray}
U_{1,2}(t)=\frac{i e^{-i \hbar Q_{11}(t)} \Bigg(  (TR1(t)-TR2(t))  \sin \left(\hbar
   \sqrt{  |Q_{22}(t)|^2  +(TR1(t)-TR2(t))^2}\right) }{2 \sqrt{|Q_{22}(t)|^2+(TR1(t)-TR2(t))^2}  }
\nonumber \\
-\frac{(TR1(t)+TR2(t))  \sin \left(\hbar
   \sqrt{|Q_{22}(t)|^2+(TR1(t)+TR2(t))^2}\right) \Bigg)  }{2  \sqrt{|Q_{22}(t)|^2+(TR1(t)+TR2(t))^2}},
\end{eqnarray}
\begin{eqnarray}
U_{1,3}(t)=-i e^{-i \hbar Q_{11}(t)}\frac{ \Bigg[ (TR1(t)-TR2(t))  \sin \left(\hbar
   \sqrt{|Q_{22}(t)|^2+(TR1(t)-TR2(t))^2}\right)}{2 \sqrt{|Q_{22}(t)|^2+(TR1(t)-TR2(t))^2} } \nonumber \\
+  \frac{(TR1(t)+TR2(t))  \sin \left(\hbar
   \sqrt{|Q_{22}(t)|^2+|TR1(t)+TR2(t)|^2}\right)\Bigg]  }{2  \sqrt{|Q_{22}(t)|^2+(TR1(t)+TR2(t))^2}}. 
\end{eqnarray}

\begin{eqnarray}
U_{1,4}(t)=
\frac{1}{2} e^{-i \hbar Q_{11}(t)} \Bigg[i Q_{22}(t) \Bigg[\frac{\sin \left(\hbar
   \sqrt{ |Q_{22}(t)|^2+(TR1(t)-TR2(t))^2}\right)}{\sqrt{|Q_{22}(t)|^2+(TR1(t)-TR2(t))^2}}  \nonumber \\
   -\frac{\sin \left(\hbar
   \sqrt{|Q_{22}(t)|^2+(TR1(t)+TR2(t))^2}\right)}{\sqrt{|Q_{22}(t)|^2+(TR1(t)+TR2(t))^2}}\Bigg] \nonumber \\ -\cos \left(\hbar
   \sqrt{|Q_{22}(t)|^2+(TR1(t)-TR2(t))^2}\right)+\cos \left(\hbar \sqrt{|Q_{22}(t)|^2+(TR1(t)+TR2(t))^2}\right)\Bigg]
\end{eqnarray}

\begin{eqnarray}
U_{2,1}(t)=-\frac{i}{2}e^{-i \hbar Q_{11}(t)}
\frac{\Bigg[ (TR1(t)-TR2(t)) \sin \left(\hbar
   \sqrt{|Q_{22}(t)|^2+(TR1(t)-TR2(t))^2}\right)}{\sqrt{|Q_{22}(t)|^2+(TR1(t)-TR2(t))^2}} \nonumber \\
-\frac{ (TR1(t)+TR2(t)) \sin \left(\hbar
   \sqrt{|Q_{22}(t)|^2+(TR1(t)+TR2(t))^2}\right)\Bigg] }{\sqrt{|Q_{22}(t)|^2+(TR1(t)+TR2(t))^2}}
\end{eqnarray}

\begin{eqnarray}
U_{2,2}(t)=
\frac{1}{2} e^{-i \hbar Q_{11}(t)} \Bigg[ i Q_{22}(t) \Bigg[\frac{\sin \left(\hbar
   \sqrt{|Q_{22}(t)|^2+(TR1(t) -TR2(t))^2}\right)}{\sqrt{|Q_{22}(t)|^2+(TR1(t)-TR2(t))^2}}+\nonumber \\ \frac{\sin \left(\hbar
   \sqrt{|Q_{22}(t)|^2+(TR1(t)+TR2(t))^2}\right)}{\sqrt{|Q_{22}(t)|^2+(TR1(t)+TR2(t))^2}}\Bigg]+  \nonumber \\ +\cos \left(\hbar
   \sqrt{|Q_{22}(t)|^2+(TR1(t)-TR2(t))^2}\right)+\cos \left(\hbar \sqrt{|Q_{22}(t)|^2+(TR1(t)+TR2(t))^2}\right)\Bigg]
\end{eqnarray}

\begin{eqnarray}
U_{2,3}(t)=e^{-i \hbar Q_{11}(t)} 
\Bigg[\frac{  
 -Q_{22}(t)  \sin \left(\hbar
   \sqrt{|Q_{22}(t)|^2+(TR1(t)-TR2(t))^2}\right) 
}{2 \sqrt{|Q_{22}(t)|^2+(TR1(t)-TR2(t))^2} } + \nonumber \\ +
\frac{Q_{22}(t)  \sin \left(\hbar
   \sqrt{|Q_{22}(t)|^2+(TR1(t)+TR2(t))^2}\right)}{2  \sqrt{|Q_{22}(t)|^2+(TR1(t)+TR2(t))^2}}+\nonumber \\ +\frac{ i   \cos \left(\hbar \sqrt{|Q_{22}(t)|^2+(TR1(t)-TR2(t))^2}
   \right)-i   \cos \left(\hbar
   \sqrt{|Q_{22}(t)|^2+(TR1(t)+TR2(t))^2}\right)}{2  }\Bigg]
\end{eqnarray}

\begin{eqnarray}
U_{2,4}(t)=-\frac{i e^{-i \hbar Q_{11}(t) } \Bigg[(TR1(t)-TR2(t)) \sin \left(\hbar
   \sqrt{|Q_{22}(t)|^2+(TR1(t)-TR2(t))^2}\right)}{2 \sqrt{|Q_{22}(t)|^2+(TR1(t)-TR2(t))^2} }+\nonumber \\+\frac{(TR1(t)+TR2(t))  \sin \left(\hbar
   \sqrt{|Q_{22}(t)|^2+(TR1(t)-TR2(t))^2}\right)\Bigg]}{2  \sqrt{|Q_{22}(t)|^2+(TR1(t)+TR2(t))^2}}
\end{eqnarray}

\begin{eqnarray}
U_{3,1}(t)=-i e^{-i \hbar Q_{11}(t) }\frac{ \Bigg[(TR1(t)-TR2(t))  \sin \left(\hbar
   \sqrt{|Q_{22}(t)|^2+(TR1(t)-TR2(t))^2}\right)}{2 \sqrt{|Q_{22}(t)|^2+(TR1(t)-TR2(t))^2} }+\nonumber\\+\frac{(TR1(t)+TR2(t))  \sin \left(\hbar
   \sqrt{|Q_{22}(t)|^2+(TR1(t)+TR2(t))^2}\right)\Bigg]}{2  \sqrt{|Q_{22}(t)|^2+(TR1(t)+TR2(t))^2}}
\end{eqnarray}

\begin{eqnarray}
U_{3,2}(t)= e^{-i \hbar Q_{11}(t) } 
\Bigg[ \frac{ -Q_{22}(t)  \sin \left(\hbar
   \sqrt{|Q_{22}(t)|^2+(TR1(t)-TR2(t))^2}\right) }{2 \sqrt{|Q_{22}(t)|^2+(TR1(t)-TR2(t))^2}  }+ \nonumber \\ \frac{Q_{22}(t)  \sin \left(\hbar
   \sqrt{|Q_{22}(t)|^2+(TR1(t)+TR2(t))^2}\right)}{2  \sqrt{|Q_{22}(t)|^2+(TR1(t)+TR2(t))^2}}+\nonumber \\ +\frac{i   \cos \left(\hbar
   \sqrt{|Q_{22}(t)|^2+(TR1(t)-TR2(t))^2}\right)-i   \cos \left(\hbar
   \sqrt{|Q_{22}(t)|^2+(TR1(t)+TR2(t))^2}\right)\Bigg]}{2  }
\end{eqnarray}

\begin{eqnarray}
U_{3,3}(t)=\frac{1}{2} e^{-i \hbar Q_{11}(t)} \Bigg[ i Q_{22}(t) \Bigg[\frac{\sin \left(\hbar
   \sqrt{|Q_{22}(t)|^2+(TR1(t)-TR2(t))^2}\right)}{\sqrt{|Q_{22}(t)|^2+(TR1(t)-TR2(t))^2}}\nonumber \\ +\frac{\sin \left(\hbar
   \sqrt{|Q_{22}(t)|^2+(TR1(t)+TR2(t))^2}\right)}{\sqrt{|Q_{22}(t)|^2+(TR1(t)+TR2(t))^2}}\Bigg]+\nonumber \\ +\cos \left(\hbar
   \sqrt{|Q_{22}(t)|^2+(TR1(t)-TR2(t))^2}\right)+\cos \left(\hbar \sqrt{|Q_{22}(t)|^2+(TR1(t)+TR2(t))^2}\right)\Bigg]
\end{eqnarray}

\begin{eqnarray}
U_{3,4}(t)=(\sin (\hbar Q_{11}(t))+i \cos (\hbar Q_{11}(t) ))\frac{ \Bigg[(TR1(t)-TR2(t))  \sin \left(\hbar
   \sqrt{|Q_{22}(t)|^2+(TR1(t)-TR2(t))^2}\right)}{2 \sqrt{|Q_{22}(t)|^2 +(TR1(t)-TR2(t))^2} }+\nonumber \\   -\frac{(TR1(t)+TR2(t))  \sin \left(\hbar
   \sqrt{|Q_{22}(t)|^2+(TR1(t)+TR2(t))^2}\right)\Bigg]}{2  \sqrt{|Q_{22}(t)|^2+(TR1(t)+TR2(t))^2}}
\end{eqnarray}

\begin{eqnarray}
U_{4,1}(t)=\frac{1}{2} e^{-i \hbar Q_{11}(t)} \Bigg[i Q_{22}(t) \Bigg[\frac{\sin \left(\hbar
   \sqrt{|Q_{22}(t)|^2+(TR1(t)-TR2(t))^2}\right)}{\sqrt{|Q_{22}(t)|^2+(TR1(t)-TR2(t))^2}}\nonumber \\ -\frac{\sin \left(\hbar
   \sqrt{|Q_{22}(t)|^2+(TR1(t)+TR2(t))^2}\right)}{\sqrt{|Q_{22}(t)|^2+(TR1(t)+TR2(t))^2}}\Bigg]+\nonumber \\  -\cos \left(\hbar
   \sqrt{|Q_{22}(t)|^2+(TR1(t)-TR2(t))^2}\right)+\cos \left(\hbar \sqrt{|Q_{22}(t)|^2+(TR1(t)+TR2(t))^2}\right)\Bigg]
\end{eqnarray}

\begin{eqnarray}
U_{4,2}(t)=-\frac{i e^{-i \hbar Q_{11}(t) } \Bigg[(TR1(t)-TR2(t))  \sin \left(\hbar
   \sqrt{|Q_{22}(t)|^2+(TR1(t)-TR2(t))^2}\right)}{2 \sqrt{|Q_{22}(t)|^2+(TR1(t)-TR2(t))^2} }+\nonumber \\+\frac{(TR1(t)+TR2(t))  \sin \left(\hbar
   \sqrt{|Q_{22}(t)|^2+(TR1(t)+TR2(t))^2}\right)\Bigg]}{2  \sqrt{|Q_{22}(t)|^2+(TR1(t)+TR2(t))^2}}
\end{eqnarray}

\begin{eqnarray}
U_{4,3}(t)=i e^{-i \hbar Q_{11}(t)}\Bigg[ \frac{  (TR1(t)-TR2(t))  \sin \left(\hbar
   \sqrt{|Q_{22}(t)|^2+(TR1(t)-TR2(t))^2}\right)}{2 \sqrt{|Q_{22}(t)|^2+(TR1(t)-TR2(t))^2} }+\nonumber \\ -\frac{(TR1(t)+TR2(t))  \sin \left(\hbar
   \sqrt{|Q_{22}(t)|^2+(TR1(t)+TR2(t))^2}\right)}{2  \sqrt{|Q_{22}(t)|^2+(TR1(t)+TR2(t))^2}}\Bigg]
\end{eqnarray}

\begin{eqnarray}
U_{4,4}(t)=\frac{1}{2} e^{-i \hbar Q_{11}(t)} \Bigg[-i Q_{22}(t) \Bigg[\frac{\sin \left(\hbar
   \sqrt{|Q_{22}(t)|^2+(TR1(t)-TR2(t))^2}\right)}{\sqrt{|Q_{22}(t)|^2+(TR1(t)-TR2(t))^2}}+ \nonumber \\
   \frac{\sin \left(\hbar
   \sqrt{|Q_{22}(t)|^2+(TR1(t)+TR2(t))^2}\right)}{\sqrt{|Q_{22}(t)|^2+(TR1(t)+TR2(t))^2}}\Bigg]+ \nonumber \\ + \cos \left(\hbar
   \sqrt{|Q_{22}(t)|^2+(TR1(t)-TR2(t))^2}\right)+\cos \left(\hbar \sqrt{|Q_{22}(t)|^2+(TR1(t)+TR2(t))^2}\right)\Bigg] \nonumber \\
\end{eqnarray}
We set the quantum state to be $\ket{\psi,t_0}=\ket{E_1}$ at time $t_0$ so it is maximally entangled and its density matrix is $\rho(t_0)=\ket{\psi,t_0}\bra{\psi,t_0}=\ket{E_1}\bra{E_1}=\frac{1}{2}
\begin{pmatrix}
+1 & 0 & 0 & -1 \\
0 & 0 & 0 & 0  \\
0 & 0 & 0 & 0  \\
-1 & 0 & 0 & 1 \\
\end{pmatrix}
$.
 Finally we obtain the following density matrix
\begin{eqnarray}
\rho_{1,1}(t)=\frac{(TR1(t)-TR2(t))^2 \cos \left(2 \hbar \sqrt{|Q_{22}(t)|^2+(TR1(t)-TR2(t))^2}\right)+2 |Q_{22}(t)|^2+(TR1(t)-TR2(t))^2}{4
   \left(|Q_{22}(t)|^2+(TR1(t)-TR2(t))^2\right)}\nonumber \\
\end{eqnarray}

\begin{eqnarray}
\rho_{1,2}(t)=\frac{(TR1(t)-TR2(t)) \Bigg[-i \sqrt{|Q_{22}(t)|^2+(TR1(t)-TR2(t))^2} \sin \left(2 \hbar \sqrt{|Q_{22}(t)|^2+(TR1(t)-TR2(t))^2}\right)}{4 \left(|Q_{22}(t)|^2+(TR1(t)-TR2(t))^2\right)}+ \nonumber \\ +\frac{Q_{22}(t) \cos
   \left(2 \hbar \sqrt{|Q_{22}(t)|^2+(TR1(t)-TR2(t))^2}\right)-Q_{22}(t)\Bigg]}{4 \left(|Q_{22}(t)|^2+(TR1(t)-TR2(t))^2\right)}\nonumber \\
\end{eqnarray}

\begin{eqnarray}
\rho_{1,3}(t)=-(TR1(t)-TR2(t))\frac{ \Bigg[-i \sqrt{|Q_{22}(t)|^2+(TR1(t)-TR2(t))^2} \sin \left(2 \hbar \sqrt{|Q_{22}(t)|^2+(TR1(t)-TR2(t))^2}\right)}{4 \left(|Q_{22}(t)|^2+(TR1(t)-TR2(t))^2\right)}+ \nonumber \\ +\frac{Q_{22}(t) \cos
   \left(2 \hbar \sqrt{|Q_{22}(t)|^2+(TR1(t)-TR2(t))^2}\right)-Q_{22}(t)\Bigg]}{4 \left(|Q_{22}(t)|^2+(TR1(t)-TR2(t))^2\right)}\nonumber \\
\end{eqnarray}

\begin{eqnarray}
\rho_{1,4}(t)=-\frac{(TR1(t)-TR2(t))^2 \cos \left(2 \hbar \sqrt{|Q_{22}(t)|^2+(TR1(t)-TR2(t))^2}\right)+2 |Q_{22}(t)|^2+(TR1(t)-TR2(t))^2}{4
   \left(|Q_{22}(t)|^2+(TR1(t)-TR2(t))^2\right)}\nonumber \\
\end{eqnarray}

\begin{eqnarray}
\rho_{2,1}(t)=
\frac{(TR1(t)-TR2(t)) \Bigg(i \sqrt{|Q_{22}(t)|^2+(TR1(t)-TR2(t))^2} \sin \left(2 \hbar \sqrt{|Q_{22}(t)|^2+(TR1(t)-TR2(t))^2}\right)}{4 \left(|Q_{22}(t)|^2+(TR1(t)-TR2(t))^2\right)}+\nonumber \\+\frac{Q_{22}(t) \cos
   \left(2 \hbar \sqrt{|Q_{22}(t)|^2+(TR1(t)-TR2(t))^2}\right)-Q_{22}(t)\Bigg]}{4 \left(|Q_{22}(t)|^2+(TR1(t)-TR2(t))^2\right)}\nonumber \\
\end{eqnarray}

\begin{eqnarray}
\rho_{2,2}(t)=\frac{(TR1(t)-TR2(t))^2 \sin ^2\left(\hbar \sqrt{|Q_{22}(t)|^2+(TR1(t)-TR2(t))^2}\right)}{2 \left(|Q_{22}(t)|^2+(TR1(t)-TR2(t))^2\right)}
\end{eqnarray}

\begin{eqnarray}
\rho_{2,3}(t)=-\frac{(TR1(t)-TR2(t))^2 \sin ^2\left(\hbar \sqrt{|Q_{22}(t)|^2+(TR1(t)-TR2(t))^2}\right)}{2 \left(|Q_{22}(t)|^2+(TR1(t)-TR2(t))^2\right)}
\end{eqnarray}

\begin{eqnarray}
\rho_{2,4}(t)=-\frac{(TR1(t)-TR2(t)) \Bigg[ i \sqrt{|Q_{22}(t)|^2+(TR1(t)-TR2(t))^2} \sin \left(2 \hbar \sqrt{|Q_{22}(t)|^2+(TR1(t)-TR2(t))^2}\right)}{4 \left(|Q_{22}(t)|^2+(TR1(t)-TR2(t))^2\right)}+\nonumber \\+\frac{Q_{22}(t)\cos
   \left(2 \hbar \sqrt{|Q_{22}(t)|^2+(TR1(t)-TR2(t))^2}\right)-Q_{22}(t)\Bigg]}{4 \left(|Q_{22}(t)|^2+(TR1(t)-TR2(t))^2\right)}\nonumber \\
\end{eqnarray}

\begin{eqnarray}
\rho_{3,1}(t)=-(TR1(t)-TR2(t))\frac{ \Bigg[i \sqrt{|Q_{22}(t)|^2+(TR1(t)-TR2(t))^2} \sin \left(2 \hbar \sqrt{|Q_{22}(t)|^2+(TR1(t)-TR2(t))^2}\right)}{4 \left(|Q_{22}(t)|^2+(TR1(t)-TR2(t))^2\right)}+ \nonumber \\ +\frac{Q_{22}(t) \cos
   \left(2 \hbar \sqrt{|Q_{22}(t)|^2+(TR1(t)-TR2(t))^2}\right)-Q_{22}(t)\Bigg]}{4 \left(|Q_{22}(t)|^2+(TR1(t)-TR2(t))^2\right)}\nonumber \\
\end{eqnarray}

\begin{eqnarray}
\rho_{3,2}(t)=-\frac{(TR1(t)-TR2(t))^2 \sin ^2\left(\hbar \sqrt{|Q_{22}(t)|^2+(TR1(t)-TR2(t))^2}\right)}{2 \left(|Q_{22}(t)|^2+(TR1(t)-TR2(t))^2\right)}
\end{eqnarray}

\begin{eqnarray}
\rho_{3,3}(t)=\frac{(TR1(t)-TR2(t))^2 \sin ^2\left(\hbar \sqrt{|Q_{22}(t)|^2+(TR1(t)-TR2(t))^2}\right)}{2 \left(|Q_{22}(t)|^2+(TR1(t)-TR2(t))^2\right)}
\end{eqnarray}

\begin{eqnarray}
\rho_{3,4}(t)=\frac{(TR1(t)-TR2(t)) \Bigg(i \sqrt{|Q_{22}(t)|^2+(TR1(t)-TR2(t))^2} \sin \left(2 \hbar \sqrt{|Q_{22}(t)|^2+(TR1(t)-TR2(t))^2}\right)}{4 \left(|Q_{22}(t)|^2+(TR1(t)-TR2(t))^2\right)}+ \nonumber \\ \frac{Q_{22}(t) \cos
   \left(2 \hbar \sqrt{|Q_{22}(t)|^2+(TR1(t)-TR2(t))^2}\right)-Q_{22}(t)\Bigg]}{4 \left(|Q_{22}(t)|^2+(TR1(t)-TR2(t))^2\right)}\nonumber \\
\end{eqnarray}

\begin{eqnarray}
\rho_{4,1}(t)=-\frac{(TR1(t)-TR2(t))^2 \cos \left(2 \hbar \sqrt{|Q_{22}(t)|^2+(TR1(t)-TR2(t))^2}\right)+2|Q_{22}(t)|^2+(TR1(t)-TR2(t))^2}{4
   \left(|Q_{22}(t)|^2+(TR1(t)-TR2(t))^2\right)}\nonumber \\
\end{eqnarray}

\begin{eqnarray}
\rho_{4,2}(t)=-(TR1(t)-TR2(t))\frac{ \Bigg[ -i \sqrt{|Q_{22}(t)|^2+(TR1(t)-TR2(t))^2} \sin \left(2 \hbar \sqrt{|Q_{22}(t)|^2+(TR1(t)-TR2(t))^2}\right)}{4 \left(|Q_{22}(t)|^2+(TR1(t)-TR2(t))^2\right)}+ \nonumber \\ + \frac{Q_{22}(t) \cos
   \left(2 \hbar \sqrt{|Q_{22}(t)|^2+(TR1(t)-TR2(t))^2}\right)-Q_{22}(t)\Bigg]}{4 \left(|Q_{22}(t)|^2+(TR1(t)-TR2(t))^2\right)}\nonumber \\
\end{eqnarray}

\begin{eqnarray}
\rho_{4,3}(t)=(TR1(t)-TR2(t))\frac{ \Bigg[-i \sqrt{|Q_{22}(t)|^2+(TR1(t)-TR2(t))^2} \sin \left(2 \hbar \sqrt{|Q_{22}(t)|^2+(TR1(t)-TR2(t))^2}\right)}{4 \left(|Q_{22}(t)|^2+(TR1(t)-TR2(t))^2\right)}+\nonumber \\ \frac{Q_{22}(t) \cos
   \left(2 \hbar \sqrt{|Q_{22}(t)|^2+(TR1(t)-TR2(t))^2}\right)-Q_{22}(t)\Bigg]}{4 \left(|Q_{22}(t)|^2+(TR1(t)-TR2(t))^2\right)}\nonumber \\
\end{eqnarray}

\begin{eqnarray}
\rho_{4,4}(t)=\frac{(TR1(t)-TR2(t))^2 \cos \left(2 \hbar \sqrt{|Q_{22}(t)|^2+(TR1(t)-TR2(t))^2}\right)+2|Q_{22}(t)|^2+(TR1(t)-TR2(t))^2}{4
   \left(|Q_{22}(t)|^2+(TR1(t)-TR2(t))^2\right)}\nonumber \\
\end{eqnarray}

It turns out that $\rho^n(t)=\rho(t)$ so one deals with a pure quantum state.
Now we are obtaining reduced matrices describing the state of particle B  from 2 particle density matrix.

\begin{eqnarray}
\rho_{B}(t)=
\begin{pmatrix}
\rho_{11}(t)+\rho_{22}(t) & \rho_{13}(t)+\rho_{24}(t) \\
\rho_{31}(t)+\rho_{42}(t) & \rho_{33}(t)+\rho_{44}(t) \\
\end{pmatrix}= \nonumber \\
\begin{pmatrix}
\frac{1}{2} & \frac{Q_{22}(t) (TR1(t)-TR2(t)) \sin ^2\left(\hbar \sqrt{|Q_{22}(t)|^2+(TR1(t)-TR2(t))^2}\right)}{|Q_{22}(t)|^2+(TR1(t)-TR2(t))^2} \\
\frac{ Q_{22}(t) (TR1(t)-TR2(t)) \sin ^2\left(\hbar \sqrt{|Q_{22}(t)|^2+(TR1(t)-TR2(t))^2}\right)}{|Q_{22}(t)|^2+(TR1(t)-TR2(t))^2} & \frac{1}{2} \\
\end{pmatrix}. \nonumber \\
\end{eqnarray}
Consequently we can compute entanglement entropy.
At first we evaluate
\begin{eqnarray}
Log(\rho_{B}(t))=
\begin{pmatrix}
a & b \\
c & d \\
\end{pmatrix},
\end{eqnarray}
\small
\begin{eqnarray*}
a=\frac{1}{2} \Bigg[\log \Bigg[ Q_{22}(t) (TR1(t)-TR2(t)) \cos \left(2 \hbar \sqrt{|Q_{22}(t)|^2+(TR1(t)-TR2(t))^2}\right)+\nonumber \\ +|Q_{22}(t)|^2+Q_{22}(t)
   (TR2(t)-TR1(t))+(TR1(t)-TR2(t))^2 \Bigg] \nonumber \\
   -2\log \Bigg[
   |Q_{22}(t)|^2+(TR1(t)-TR2(t))^2
   \Bigg] \nonumber \\
   +\log \Bigg[\Bigg[Q_{22}(t) (TR2(t)-TR1(t)) \cos \left(2 \hbar
   \sqrt{|Q_{22}(t)|^2+(TR1(t)-TR2(t))^2}\right)+ \nonumber \\
   |Q_{22}(t)|^2+Q_{22}(t)
   (TR1(t)-TR2(t))+(TR1(t)-TR2(t))^2\Bigg] 
   -\log (4)\Bigg]
\end{eqnarray*}
\normalsize
$b=-\tanh ^{-1}\left(\frac{Q_{22}(t)(TR1(t)-TR2(t)) \left(\cos \left(2 \hbar
   \sqrt{|Q_{22}(t)|^2+(TR1(t)-TR2(t))^2}\right)-1\right)}{|Q_{22}(t)|^2+(TR1(t)-TR2(t))^2}\right)=c$

\small
\begin{eqnarray}
d=\frac{1}{2}  \Bigg[\log\Bigg[ Q_{22}(t) (TR1(t)-TR2(t)) \cos \left(2 \hbar
\sqrt{| Q_{22}(t)|^2+(TR1(t)-TR2(t))^2}\right)+\nonumber \\ | Q_{22}(t)|^2+Q_{22}(t)
   (TR2(t)-TR1(t))+(TR1(t)-TR2(t))^2\Bigg] \nonumber \\  -2\log\Bigg[|Q_{22}(t)|^2+(TR1(t)-TR2(t))^2\Bigg]\Bigg] \nonumber \\
+\log \Bigg[] Q_{22}(t)(TR2(t)-TR1(t)) \cos \left(2 \hbar
   \sqrt{|Q_{22}(t)|^2+(TR1(t)-TR2(t))^2}\right)+ \nonumber \\ |Q_{22}(t)|^2+Q_{22}(t)
   (TR1(t)-TR2(t))+(TR1(t)-TR2(t))^2\Bigg] 
-\log (4)\Bigg]
\end{eqnarray}
\normalsize
and we obtain the formula when we start from $TR1(t_0)=TR2(t_0)$ as
\small
\begin{eqnarray}
S_{B}(t)=Tr[\rho_{B}(t) Log[\rho_{B}(t)]]= \nonumber \\
=Tr \Bigg[
\begin{pmatrix}
\frac{1}{2} & \frac{Q_{22}(t) (TR1(t)-TR2(t)) \sin ^2\left(\hbar \sqrt{|Q_{22}(t)|^2+(TR1(t)-TR2(t))^2}\right)}{|Q_{22}(t)|^2+(TR1(t)-TR2(t))^2} \\
\frac{ Q_{22}(t) (TR1(t)-TR2(t)) \sin ^2\left(\hbar \sqrt{|Q_{22}(t)|^2+(TR1(t)-TR2(t))^2}\right)}{|Q_{22}(t)|^2+(TR1(t)-TR2(t))^2} & \frac{1}{2} \\
\end{pmatrix} \times \nonumber \\
Log \Bigg[
\begin{pmatrix}
\frac{1}{2} & \frac{Q_{22}(t) (TR1(t)-TR2(t)) \sin ^2\left(\hbar \sqrt{|Q_{22}(t)|^2+(TR1(t)-TR2(t))^2}\right)}{|Q_{22}(t)|^2+(TR1(t)-TR2(t))^2} \\
\frac{ Q_{22}(t) (TR1(t)-TR2(t)) \sin ^2\left(\hbar \sqrt{|Q_{22}(t)|^2+(TR1(t)-TR2(t))^2}\right)}{|Q_{22}(t)|^2+(TR1(t)-TR2(t))^2} & \frac{1}{2} \\
\end{pmatrix}
\Bigg] \Bigg]=\nonumber \\
=-\log (4)\frac{1}{2}+ 
\frac{1}{2} \Bigg[\log \Bigg[Q_{22}(t) (TR1(t)-TR2(t)) \cos \left(2 \hbar \sqrt{|Q_{22}(t)|^2+(TR1(t)-TR2(t))^2}\right)+\nonumber \\ +|Q_{22}(t)|^2+Q_{22}(t)
   (TR2(t)-TR1(t))+(TR1(t)-TR2(t))^2\Bigg]+
   \nonumber \\ +\log \Bigg[Q_{22}(t)(TR2(t)-TR1(t)) \cos \left(2 \hbar
   \sqrt{|Q_{22}(t)|^2+(TR1(t)-TR2(t))^2}\right) \nonumber \\ +|Q_{22}(t)|^2+Q_{22}(t)
   (TR1(t)-TR2(t))+(TR1(t)-TR2(t))^2\Bigg]\nonumber \\ -2\log\Bigg[
   |Q_{22}(t)|^2+(TR1(t)-TR2(t))^2
   \Bigg] \nonumber \\ +\frac{4 Q_{22}(t) (TR2(t)-TR1(t)) \sin ^2\left(\hbar
   \sqrt{|Q_{22}(t)|^2+(TR1(t)-TR2(t))^2}\right) }{|Q_{22}(t)|^2+(TR1(t)-TR2(t))^2} \times \nonumber \\
\times   \tanh ^{-1}\left(\frac{Q_{22}(t) (TR1(t)-TR2(t)) \left(\cos \left(2 \hbar
   \sqrt{|Q_{22}(t)|^2+(TR1(t)-TR2(t))^2}\right)-1\right)}{|Q_{22}(t)|^2+(TR1(t)-TR2(t))^2}\right)\Bigg]
   \nonumber \\
\end{eqnarray}
\normalsize

\subsection{Analytical treatment of 2 interacting particles in asymmetric case}
We consider the situation as depicted in Fig.5. We obtain the following Hamiltonian
\begin{eqnarray}
H=
\begin{pmatrix}
E_c(1,1')+E_{p}(1)+E_{p}(1') & t_{s1'2p'} & t_{s12} & 0 \\
t_{s1'2p'}^{*} & E_c(1,2')+E_{p}(1)+E_{p}(2') & 0 &  t_{s12} \\
t_{s12}^{*} & 0 & E_c(2,1')+E_{p}(2)+E_{p}(1') & t_{s1'2p'} \\
0 & t_{s12}^{*} & t_{s1'2p'}{*} & E_c(2,2')+E_{p}(2)+E_{p}(2')
\end{pmatrix}= \nonumber \\
\begin{pmatrix}
\frac{q^2}{\sqrt{d^2+(a+b)^2}} & t_{s1'2p'} & t_{s12} & 0 \\
t_{s1'2p'}^{*} & \frac{q^2}{\sqrt{(d+Cos(\alpha)(a+b))^2+(1+Sin(\alpha))^2(a+b)^2}} & 0 &  t_{s12} \\
t_{s12}^{*} & 0 & \frac{q^2}{d} & t_{s1'2p'} \\
0 & t_{s12}^{*} & t_{s1'2p'}{*} & \frac{q^2}{\sqrt{(d+Cos(\alpha)(a+b))^2+(Sin(\alpha))^2(a+b)^2}}
\end{pmatrix}+\nonumber \\
\begin{pmatrix}
E_{p}(1)+E_{p}(1') & 0 & 0 & 0 \\
0 & E_{p}(1)+E_{p}(2') & 0 &  0 \\
0 & 0 & E_{p}(2)+E_{p}(1') & 0 \\
0 & 0 & 0 &E_{p}(2)+E_{p}(2')
\end{pmatrix}=\nonumber \\
\begin{pmatrix}
E_f(1,1') & t_{s1'2p'} & t_{s12} & 0 \\
t_{s1'2p'}^{*} & E_f(1,2') & 0 &  t_{s12} \\
t_{s12}^{*} & 0 & E_f(2,1') & t_{s1'2p'} \\
0 & t_{s12}^{*} & t_{s1'2p'}{*} & E_f(2,2')
\end{pmatrix}. \nonumber \\
\end{eqnarray}
In general situation all eigenvalues and eigenstates of this matrix can be determined in analytical way since roots of polynomials of 4th are given by explicit formulas. However those formulas are very complicated and thus no so practical. We will make radical simplification leading to simple formulas for quantum eigenstates and eigenenergies. First main assumption is setting $t_{s1'2p'}=t_{s12}=1$. We obtain simplified Hamiltonian of the form
\begin{eqnarray}H=
\begin{pmatrix}
E_f(1,1') & 1 & 1 & 0 \\
1 & E_f(1,2') & 0 & 1 \\
1 & 0 & E_f(2,1') & 1 \\
0 & 1 & 1 & E_f(2,2')
\end{pmatrix}. \nonumber \\
\end{eqnarray}
Now we will consider certain cases.
\subsection{Qubit-Qubit interaction with 2 symmetric conditions}
\subsubsection{Case I: $E_f(1,1')=E_f(1,2')=U$, $E_f(2,1')=E_f(2,2')=U_1$ }
We postulate
\begin{eqnarray}H=
\begin{pmatrix}
U & 1 & 1 & 0 \\
1 & U & 0 & 1 \\
1 & 0 & U_1 & 1 \\
0 & 1 & 1 & U_1
\end{pmatrix}. \nonumber \\
\end{eqnarray}
We obtain
\begin{eqnarray}
U=E_{p1}+E_{p1'}+\frac{q^2}{\sqrt{d^2+(a+b)^2}}, \nonumber \\
U=E_{p1}+E_{p2'}+\frac{q^2}{\sqrt{(d+Cos(\alpha)(a+b))^2+(1+Sin(\alpha))^2(a+b)^2}},
\end{eqnarray}
and it implies
\begin{eqnarray}
E_{p2'}-E_{p1'}=+\frac{q^2}{\sqrt{d^2+(a+b)^2}}-\frac{q^2}{\sqrt{(d+Cos(\alpha)(a+b))^2+(1+Sin(\alpha))^2(a+b)^2}}, \nonumber \\
\end{eqnarray}
We also have
\begin{eqnarray}
U_1=E_{p2}+E_{p1'}+\frac{q^2}{d}, \nonumber \\
U_1=E_{p2}+E_{p2'}+\frac{q^2}{\sqrt{(d+Cos(\alpha)(a+b))^2+(Sin(\alpha))^2(a+b)^2}}.
\end{eqnarray}
and we obtain
\begin{eqnarray}
E_{p2'}-E_{p1'}=+\frac{q^2}{d}-\frac{q^2}{\sqrt{(d+Cos(\alpha)(a+b))^2+(Sin(\alpha))^2(a+b)^2}}.
\end{eqnarray}
that implies condition
\begin{eqnarray}
\label{extra}
+\frac{q^2}{d}-\frac{q^2}{\sqrt{(d+Cos(\alpha)(a+b))^2+(Sin(\alpha))^2(a+b)^2}}= \nonumber \\
\frac{q^2}{\sqrt{d^2+(a+b)^2}}-\frac{q^2}{\sqrt{(d+Cos(\alpha)(a+b))^2+(1+Sin(\alpha))^2(a+b)^2}}.
\end{eqnarray}
that is fulfilled for two angles $\alpha$ as in accordance to numerical solutions. Usually $a+b << d$ since in most cases qubit size is small in comparison to the distance between qubits. Let us solve the problem with certain approximation by using Taylor expansion

\begin{eqnarray}
+\frac{q^2}{d}-\frac{q^2}{d}+\frac{q^2}{d^2}(\sqrt{d^2+2 d Cos(\alpha)(a+b)+(Cos(\alpha))^2(a+b)^2+(Sin(\alpha))^2(a+b)^2}-d) =\nonumber \\ 
\frac{q^2}{\sqrt{d^2+(a+b)^2}}-\frac{q^2}{d}+ \nonumber \\
+\frac{q^2}{d^2}(\sqrt{(Cos(\alpha))^2(a+b)^2+2(a+b)d Cos(\alpha)+(Sin(\alpha))^2(a+b)^2+(a+b)^2+d^2+2Sin(\alpha)(a+b)^2}-d). 
\end{eqnarray}
It implies
\begin{eqnarray}
+\frac{1}{d}+\frac{1}{d^2}(\sqrt{d^2+2 d Cos(\alpha)(a+b)+(a+b)^2}) =\nonumber \\ 
\frac{1}{\sqrt{d^2+(a+b)^2}}
+\frac{1}{d^2}(\sqrt{2(a+b)d Cos(\alpha)+2(a+b)^2+d^2+2Sin(\alpha)(a+b)^2}). 
\end{eqnarray}
and hence we have
\begin{eqnarray}
+1+(\sqrt{1+2 Cos(\alpha)\frac{(a+b)}{d}+(\frac{a+b}{d})^2}) =\nonumber \\ 
\frac{d}{\sqrt{d^2+(a+b)^2}}
+(\sqrt{2\frac{(a+b)}{d}Cos(\alpha)+2(\frac{a+b}{d})^2+1+2Sin(\alpha)(\frac{a+b}{d})^2}). 
\end{eqnarray}
Using Taylor expansion for square root function we obtain relation
\begin{eqnarray}
+1+1+\frac{1}{2}(2 Cos(\alpha)\frac{(a+b)}{d}+(\frac{a+b}{d})^2) =\nonumber \\ 
\frac{d}{\sqrt{d^2+(a+b)^2}}+1
+\frac{1}{2}(2\frac{(a+b)}{d}Cos(\alpha)+2(\frac{a+b}{d})^2+2Sin(\alpha)(\frac{a+b}{d})^2). 
\end{eqnarray}
that can be simplified into simple relation for sinusoidal function of the form
\begin{eqnarray}
\frac{d^2}{(a+b)^2}[+1-\frac{1}{2}(\frac{a+b}{d})^2-\frac{d}{\sqrt{d^2+(a+b)^2}}] 
=Sin(\alpha). 
\end{eqnarray}
Two angles fulfills such relation.
\newline \newline

We can also solve equation \ref{extra} in rigorous way (without approximation) by equivalent to the following relation
\begin{eqnarray}
+\frac{1}{d}-\frac{1}{d_1}= 
\frac{1}{d_2}-\frac{1}{d_1+x} \text{ or } \frac{1}{d}-\frac{1}{d_2}=\frac{1}{d_1}-\frac{1}{d_1+x}=\frac{x}{d_1 (d_1+x)}.
\end{eqnarray}
and this implies
\begin{eqnarray}
d_1^2\frac{d_2-d}{d_2 d}+xd_1\frac{d_2-d}{d_2 d}=x \text{ or }+(d_1\frac{d_2-d}{d_2 d}-1)x=-d_1^2\frac{d_2-d}{d_2 d}.
\end{eqnarray}
and finally we have
\begin{eqnarray}
(\frac{d_1d_2-d d_1 - d_2 d}{d_2 d})x=-d_1^2\frac{d_2-d}{d_2 d}, x=-d_1^2\frac{d_2-d}{d_1d_2-d d_1 - d_2 d}=\frac{-d_1^2 d_2+d d_1^2}{d_1d_2-d d_1 - d_2 d},.
\end{eqnarray}
We obtain
\begin{equation}
\frac{x+d_1}{d_1}=\frac{-d_1^2 d_2+d d_1^2+ d_1(d_1d_2-d d_1 - d_2 d)}{ (d_1d_2-d d_1 - d_2 d) d_1}=\frac{-d d_1 d_2 }{(d_1d_2-d d_1 - d_2 d) d_1}=\frac{-d d_2 }{(d_1d_2-d d_1 - d_2 d)}=\frac{d d_2 }{d_1(-d_2+d) + d_2 d}
\end{equation}
This implies
\begin{eqnarray}
(\frac{x+d_1}{d_1})^2=\frac{ (d d_2)^2 }{ [d_1(-d_2+d) + d_2 d ]^2}=\frac{(d+Cos(\alpha)(a+b))^2+(1+Sin(\alpha))^2(a+b)^2}{(d+Cos(\alpha)(a+b))^2+(Sin(\alpha))^2(a+b)^2}= \nonumber \\
=\frac{ (d \sqrt{ d^2 + (a+b)^2})^2 }{ [\sqrt{(d+Cos(\alpha)(a+b))^2+(Sin(\alpha))^2(a+b)^2}(-\sqrt{ d^2 + (a+b)^2}+d) + (d \sqrt{ d^2 + (a+b)^2}) ]^2}.
\end{eqnarray}

We recognize that we have 3 free parameters $E_{p1}\in R, E_{p2}\in R, E_{p1'}\in R$ that predetermines $E_{p2'}$ given as
\begin{eqnarray}
E_{p2'}=E_{p1'}+\frac{q^2}{d}-\frac{q^2}{\sqrt{(d+Cos(\alpha)(a+b))^2+(Sin(\alpha))^2(a+b)^2}}.
\end{eqnarray}
Using 3 free parameters $E_{p1}\in R, E_{p2}\in R, E_{p1'}\in R$ we obtain
\begin{eqnarray}
U_1=2E_{p1'}+\frac{2q^2}{d}-\frac{q^2}{\sqrt{(d+Cos(\alpha)(a+b))^2+(Sin(\alpha))^2(a+b)^2}}, 
\end{eqnarray}
\begin{eqnarray}
U=E_{p1}+E_{p1'}+\frac{q^2}{\sqrt{d^2+(a+b)^2}}.
\end{eqnarray}
We have 4 energy eigenvalues
\begin{equation}
E_1=\frac{1}{2} \left(-\sqrt{(U-U_1)^2+4}+U+U_1)-2 \right),
\end{equation}
\begin{equation}
E_2=\frac{1}{2} \left(-\sqrt{(U-U_1)^2+4}+U+U_1)+2 \right),
\end{equation}
\begin{equation}
E_3=\frac{1}{2} \left(+\sqrt{(U-U_1)^2+4}+U+U_1)-2 \right),
\end{equation}
\begin{equation}
E_4=\frac{1}{2} \left(+\sqrt{(U-U_1)^2+4}+U+U_1)+2 \right).
\end{equation}
and we obtain the following eigenstates
\begin{equation}
\ket{E_1}=
\begin{pmatrix}
\frac{1}{2} \left(\sqrt{U^2-2 U U_1+U_1^2+4}-U+U_1\right) \\
\frac{1}{2} \left(-\sqrt{U^2-2 U U_1+U_1^2+4}+U-U_1\right) \\
-1 \\
+1
\end{pmatrix},
\end{equation}
\begin{equation}
\ket{E_2}=
\begin{pmatrix}
\frac{1}{2} \left(-\sqrt{U^2-2 U U_1+U_1^2+4}+(U-U_1)\right) \\
\frac{1}{2} \left(-\sqrt{U^2-2 U U_1+U_1^2+4}+(U-U_1)\right) \\
1 \\
1 \\
\end{pmatrix},
\end{equation}
\begin{equation}
\ket{E_3}=
\begin{pmatrix}
-\frac{1}{2} \left(\sqrt{U^2-2 U U_1+U_1^2+4}+(U-U_1)\right) \\
+\frac{1}{2} \left(\sqrt{U^2-2 U U_1+U_1^2+4}+(U-U_1)\right) \\
-1 \\
+1 \\
\end{pmatrix},
\end{equation}
\begin{equation}
\ket{E_4}=
\begin{pmatrix}
\frac{1}{2} \left(\sqrt{U^2-2 U U_1+U_1^2+4}+U-U_1\right) \\
\frac{1}{2} \left(\sqrt{U^2-2 U U_1+U_1^2+4}+U-U_1\right) \\
+1 \\
+1 \\
\end{pmatrix}.
\end{equation}
\subsubsection{Case II: $E_f(1,1')=E_f(2,2')=U$, $E_f(1,2')=E_f(2,1')=U_1$ }
We postulate
\begin{eqnarray}H=
\begin{pmatrix}
U & 1 & 1 & 0 \\
1 & U_1 & 0 & 1 \\
1 & 0 & U_1 & 1 \\
0 & 1 & 1 & U
\end{pmatrix}, \nonumber \\
\end{eqnarray}
that implies
\begin{eqnarray}
U=E_{p1}+E_{p1'}+\frac{q^2}{\sqrt{d^2+(a+b)^2}}, \nonumber \\
U=E_{p2}+E_{p2'}+\frac{q^2}{\sqrt{(d+Cos(\alpha)(a+b))^2+(Sin(\alpha))^2(a+b)^2}}.
\end{eqnarray}
and consequently we obtain
\begin{eqnarray}
(E_{p1}-E_{p2})+(E_{p1'}-E_{p2'})=\frac{q^2}{\sqrt{(d+Cos(\alpha)(a+b))^2+(Sin(\alpha))^2(a+b)^2}} 
-\frac{q^2}{\sqrt{d^2+(a+b)^2}}.
\end{eqnarray}

We also have
\begin{eqnarray}
U_1=E_{p1}+E_{p2'}+\frac{q^2}{\sqrt{(d+Cos(\alpha)(a+b))^2+(1+Sin(\alpha))^2(a+b)^2}}, \nonumber \\
U_1=E_{p2}+E_{p1'}+\frac{q^2}{d},
\end{eqnarray}
and it implies
\begin{eqnarray}
-(E_{p1}-E_{p2})+(E_{p1'}-E_{p2'})=\frac{q^2}{\sqrt{(d+Cos(\alpha)(a+b))^2+(1+Sin(\alpha))^2(a+b)^2}}-\frac{q^2}{d}.
\end{eqnarray}
This results in
\begin{eqnarray}
(E_{p1'}-E_{p2'})=\frac{1}{2}[\frac{q^2}{\sqrt{(d+Cos(\alpha)(a+b))^2+(1+Sin(\alpha))^2(a+b)^2}}-\frac{q^2}{d} \nonumber \\
+\frac{q^2}{\sqrt{(d+Cos(\alpha)(a+b))^2+(Sin(\alpha))^2(a+b)^2}}-\frac{q^2}{\sqrt{d^2+(a+b)^2}}].
\end{eqnarray}
and
\begin{eqnarray}
(E_{p1}-E_{p2})=\frac{1}{2}[\frac{q^2}{\sqrt{(d+Cos(\alpha)(a+b))^2+(1+Sin(\alpha))^2(a+b)^2}}-\frac{q^2}{d} \nonumber \\
-\frac{q^2}{\sqrt{(d+Cos(\alpha)(a+b))^2+(Sin(\alpha))^2(a+b)^2}}+\frac{q^2}{\sqrt{d^2+(a+b)^2}}].
\end{eqnarray}
Therefore we have 3 open parameters $E_{p1} \in R$, $E_{p1'} \in R$ and for any $\alpha \in (0, 2\pi)$ we have predetermined conditions for $E_{p2}$ and $E_{p2'}$ given as
\begin{eqnarray}
E_{p2}=E_{p1}-\frac{1}{2}[\frac{q^2}{\sqrt{(d+Cos(\alpha)(a+b))^2+(1+Sin(\alpha))^2(a+b)^2}}-\frac{q^2}{d} \nonumber \\
-\frac{q^2}{\sqrt{(d+Cos(\alpha)(a+b))^2+(Sin(\alpha))^2(a+b)^2}}+\frac{q^2}{\sqrt{d^2+(a+b)^2}}].
\end{eqnarray}
and
\begin{eqnarray}
E_{p2'}=E_{p1'}-\frac{1}{2}[\frac{q^2}{\sqrt{(d+Cos(\alpha)(a+b))^2+(1+Sin(\alpha))^2(a+b)^2}}-\frac{q^2}{d} \nonumber \\
+\frac{q^2}{\sqrt{(d+Cos(\alpha)(a+b))^2+(Sin(\alpha))^2(a+b)^2}}-\frac{q^2}{\sqrt{d^2+(a+b)^2}}].
\end{eqnarray}
and implies $U \in R$
\begin{eqnarray}
U=E_{p1}+E_{p1'}+\frac{q^2}{\sqrt{d^2+(a+b)^2}},
\end{eqnarray}
as well as
\begin{eqnarray}
U_1=E_{p2}+E_{p1'}+\frac{q^2}{d}= \nonumber
2E_{p1'}+\frac{q^2}{d} 
-\frac{1}{2}[\frac{q^2}{\sqrt{(d+Cos(\alpha)(a+b))^2+(1+Sin(\alpha))^2(a+b)^2}}-\frac{q^2}{d} \nonumber \\
+\frac{q^2}{\sqrt{(d+Cos(\alpha)(a+b))^2+(Sin(\alpha))^2(a+b)^2}}-\frac{q^2}{\sqrt{d^2+(a+b)^2}}].
\end{eqnarray}

We have the following eigenvalues
\begin{eqnarray}
E_1=U, E_2=U_1, E_3=\frac{(U+U_1-\sqrt{(4)^2+(U-U_1)^2})}{2}, E_4=\frac{(U+U_1+\sqrt{(4)^2+(U-U_1)^2})}{2}.
\end{eqnarray}
and eigenstates
\begin{eqnarray} \ket{E_1}=
\begin{pmatrix}
-1 \\ 0 \\ 0 \\ 1 \\
\end{pmatrix}, 
\end{eqnarray}
\begin{eqnarray} \ket{E_2}=
\begin{pmatrix}
0 \\ -1 \\ +1 \\ 0 \\
\end{pmatrix}, 
\end{eqnarray}
\begin{eqnarray} \ket{E_3}=
\begin{pmatrix}
1 \\ \frac{4}{-U+U_1+\sqrt{(4)^2+(U-U_1)^2}} \\ \frac{4}{-U+U_1+\sqrt{(4)^2+(U-U_1)^2}} \\ 1 \\
\end{pmatrix}, 
\end{eqnarray}
\begin{eqnarray} \ket{E_4}=
\begin{pmatrix}
1 \\ \frac{4}{U-U_1+\sqrt{(4)^2+(U-U_1)^2}} \\ \frac{4}{U-U_1+\sqrt{(4)^2+(U-U_1)^2}} \\ 1 \\
\end{pmatrix}, 
\end{eqnarray}
\subsubsection{Case III: $E_f(1,1')=E_f(2,1')=U$, $E_f(1,2')=E_f(2,2')=U_1$ }
We postulate
\begin{eqnarray}H=
\begin{pmatrix}
U & 1 & 1 & 0 \\
1 & U_1 & 0 & 1 \\
1 & 0 & U & 1 \\
0 & 1 & 1 & U_1
\end{pmatrix}. \nonumber \\
\end{eqnarray}
that implies $U=\frac{q^2}{\sqrt{d^2+(a+b)^2}}+E_{p}(1)+E_{p}(1')$, $U=\frac{q^2}{d}+E_{p2}+E_{p}(1')$
and we obtain
\begin{eqnarray}
E_{p2}=E_{p1}+\frac{q^2}{\sqrt{d^2+(a+b)^2}}-\frac{q^2}{d},
\end{eqnarray}
.
We also have
\begin{eqnarray}
U_1=\frac{q^2}{\sqrt{(d+Cos(\alpha)(a+b))^2+(1+Sin(\alpha))^2(a+b)^2}}+E_{p1}+E_{p2'}, \nonumber \\
U_1=\frac{q^2}{\sqrt{(d+Cos(\alpha)(a+b))^2+(Sin(\alpha))^2(a+b)^2}}+E_{p2}+E_{p2'},
\end{eqnarray}
what implies
\begin{eqnarray}
E_{p2}=E_{p1}+\frac{q^2}{\sqrt{(d+Cos(\alpha)(a+b))^2+(1+Sin(\alpha))^2(a+b)^2}}-\frac{q^2}{\sqrt{(d+Cos(\alpha)(a+b))^2+(Sin(\alpha))^2(a+b)^2}}.
\end{eqnarray}
and consequently we obtain the condition
\begin{eqnarray}
\frac{q^2}{\sqrt{(d+Cos(\alpha)(a+b))^2+(1+Sin(\alpha))^2(a+b)^2}}-\frac{q^2}{\sqrt{(d+Cos(\alpha)(a+b))^2+(Sin(\alpha))^2(a+b)^2}}= \nonumber \\
\frac{q^2}{\sqrt{d^2+(a+b)^2}}-\frac{q^2}{d}
\end{eqnarray}
that is fulfilled for one angle $\alpha$.
It gives us 3 controlling parameters $E_{p1}$, $E_{p1'}$, $E_{p2'}$ under given $(a+b),d$ that determine
\begin{eqnarray}
U_1=\frac{q^2}{\sqrt{(d+Cos(\alpha)(a+b))^2+(1+Sin(\alpha))^2(a+b)^2}}+E_{p1}+E_{p2'}, \nonumber
U=\frac{q^2}{d}+E_{p2}+E_{p}(1').
\end{eqnarray}
We obtain the following Hamiltonian eigenvalues and eigenstates
\begin{eqnarray}
E_1=\frac{-2+U+U_1-\sqrt{(4)^2+(U-U_1)^2}}{2},  \nonumber \\
E_2=\frac{+2+U+U_1-\sqrt{(4)^2+(U-U_1)^2}}{2},  \nonumber \\
E_3=\frac{-2+U+U_1+\sqrt{(4)^2+(U-U_1)^2}}{2},   \nonumber \\
E_4=\frac{+2+U+U_1+\sqrt{(4)^2+(U-U_1)^2}}{2}.
\end{eqnarray}
and
\begin{eqnarray}
\ket{E_1}=
\begin{pmatrix}
\frac{1}{2}(-U + \sqrt{4 + (U - U_1)^2} + U_1) \\
-1 \\
\frac{1}{2}(U - \sqrt{4 + (U - U_1)^2} - U_1) \\
1 \\
\end{pmatrix},
\end{eqnarray}
\begin{eqnarray}
\ket{E_2}=
\begin{pmatrix}
\frac{1}{2}(U - \sqrt{4 + (U - U_1)^2} - U_1)\\
1 \\
\frac{1}{2}(U - \sqrt{4 + (U - U_1)^2} - U_1)\\
1
\end{pmatrix},
\end{eqnarray}
\begin{eqnarray}
\ket{E_3}=
\begin{pmatrix}
\frac{1}{2}(-U - \sqrt{4 + (U - U_1)^2} + U_1) \\
-1 \\
\frac{1}{2}(U + \sqrt{4 + (U - U_1)^2} - U_1) \\
1 \\
\end{pmatrix},
\end{eqnarray}
\begin{eqnarray}
\ket{E_4}=
\begin{pmatrix}
\frac{1}{2}(-U - \sqrt{4 + (U - U_1)^2} + U_1) \\
1 \\
\frac{1}{2}(U + \sqrt{4 + (U - U_1)^2} - U_1) \\
1 \\
\end{pmatrix},
\end{eqnarray}

\section{Analytical method for prediction of correlation-anticorrelation in generalized Swap Gate}
We can predict correlation or anticorrelation by monitoring the sign of the correlation function $f(C)$ with correlation function operator C
\begin{eqnarray}
  C=\frac{N_{1,1'}+N_{2,2'}-N_{1,2'}-N_{2,1'}}{N_{1,1'}+N_{2,2'}+N_{1,2'}+N_{2,1'}}= 
  \begin{pmatrix}
  1 & 0 & 0 & 0 \\
  0 & -1 & 0 & 0 \\
  0 & 0 & -1 & 0 \\
  0 & 0 & 0 & 1
  \end{pmatrix}, 1=N_{1,1'}+N_{2,2'}+N_{1,2'}+N_{2,1'},
\end{eqnarray}
and under the circumstances of time-independent Hamiltonian we have
\begin{eqnarray}
f(C)=\bra{\psi(t)}C\ket{\psi(t)}=(\bra{E_1}c_{E1}^{*}e^{-i\phi_{E1}}e^{-\frac{E_1 t}{i \hbar}}+\bra{E_2}c_{E2}^{*}e^{-i\phi_{E2}}e^{-\frac{E_2 t}{i\hbar}}+\bra{E_3}c_{E3}^{*}e^{-i\phi_{E3}}e^{-\frac{E_3 t}{i\hbar}}+\bra{E_4}c_{E4}^{*}e^{-i\phi_{E4}}e^{-\frac{E_4 t}{\hbar}}) \nonumber \\
C(\ket{E_1}c_{E1}e^{i\phi_{E1}}e^{\frac{E_1 t}{i\hbar}}+\ket{E_2}c_{E2}e^{i\phi_{E2}}e^{\frac{E_2 t}{i\hbar}}+\ket{E_3}c_{E3}e^{i\phi_{E3}}e^{\frac{E_3 t}{i\hbar}}+\ket{E_4}c_{E4}e^{i\phi_{E4}}e^{\frac{E_4 t}{i\hbar}}). 
\end{eqnarray}

When $f(C)$ has negative values we are dealing with Swap Gate, while ANTISWAP gate requires positive value.
\subsubsection{Correlation function for case I:$E_f(1,1')=E_f(1,2')=U$, $E_f(2,1')=E_f(2,2')=U_1$}
For qubit-qubit Hamiltonian of structure
\begin{eqnarray}H=
\begin{pmatrix}
U & 1 & 1 & 0 \\
1 & U & 0 & 1 \\
1 & 0 & U_1 & 1 \\
0 & 1 & 1 & U_1
\end{pmatrix}. \nonumber \\
\end{eqnarray}
we obtain the following correlation function
\begin{eqnarray}
C= \frac{1}{((4 + (U - U_1)^2)^{(\frac{3}{2})})}2 e^{-i(\phi_{E1} + \phi_{E2} + \phi_{E3} +
    \phi_{E4} + (E_{1} + E_{2} + E_{3} +
       E_{4}) t)} (c_{E1} c_{E2} (4 + (U - U_1)^2) (U - U_1) \times \nonumber \\ \times
         Cos(
     \phi_{E1} - \phi_{E2} - E_{1}t + E_{2}t) + \nonumber \\
   c_{E2} c_{E3} \sqrt{4 + (U - U_1)^2} \sqrt{
    4 + (U - U_1) (U + \sqrt{4 + (U - U_1)^2} - U_1)} \sqrt{
    4 - (U - U_1) (-U +\sqrt{4 + (U - U_1)^2} + U_1)} \times \nonumber \\ \times
     Cos[\phi_{E2} - \phi_{E3} - E_{2} t + E_{3} t] + \nonumber \\
   c_{E1} c_{E4} \sqrt{4 + (U - U_1)^2} \sqrt{
    4 + (U - U1) (U + \sqrt{4 + (U - U_1)^2} - U_1)} \sqrt{
    4 - (U - U1) (-U + Sqrt[4 + (U - U_1)^2] + U_1)} \times \nonumber \\ \times
     Cos(\phi_{E1} - \phi_{E_4} - E_1 t  + E_4 t)  \nonumber \\
   -c_{E3} c_{E4} (4 + (U - U_1)^2) (U - U_1) Cos(
     \phi_{E3} - \phi_{E4} - E_3t + E_4 t)) (Cos[
    \phi_{E1} + \phi_{E2} + \phi_{E3} +
     \phi_{E4} + (E_{1} + E_{2} + E_{3} + E_{4})t] + \nonumber \\
   i Sin(\phi_{E1} + \phi_{E2} + \phi_{E3} +
      \phi_{E4} + (E_1 + E_2 + E_3 + E_4) t)),\nonumber \\
\end{eqnarray}
where $|c_{E1}|^2$,$|c_{E2}|^2$, $|c_{E3}|^2$, $|c_{E4}|^2$ are probabilities for the system of coupled qubits to occupy energy level $E_1$, $E_2$, $E_3$, $E_4$ and normalization condition is fulfilled $1=|c_{E1}|^2+|c_{E2}|^2+|c_{E3}|^2+|c_{E4}|^2$. We encounter mixture of time-dependent cos or sin oscillations of different frequencies as $(E_{1} + E_{2} + E_{3} + E_{4})$, $- E_{1} + E_{2}$,  $- E_{2} + E_{3}$, $- E_1 + E_4$, $- E_3+ E_4$, where quantum state is given as
\begin{eqnarray}
\ket{\psi}_t=c_{E1}e^{\frac{E_1 t}{i}} e^{i \phi_{E1}}\ket{E_1}+c_{E2}e^{\frac{E_2 t}{i}} e^{i \phi_{E2}}\ket{E_2}+c_{E3}e^{\frac{E_3 t}{i}} e^{i \phi_{E3}}\ket{E_3}+c_{E4}e^{\frac{E_4 t}{i}} e^{i \phi_{E4}}\ket{E_4}.
 \end{eqnarray}
 Here we have tunning coefficients $U, U_1$, $\phi_{E1}$,$\phi_{E2}$, $\phi_{E3}$, $\phi_{E4}$.
The probability of occupancy of $1,1'$ is of the following form
\small
\begin{eqnarray}
\rho_{1,1}=p(1,1')=\frac{1}{ [4(4 + (U - U_1)^2)]} \times \nonumber \\
\times \Bigg[ ((c_{E3}^2 + c_{E4}^2) (4 + (U - U_1) (U + \sqrt{4 + (U - U_1)^2} - U_1)) +
    c_{E1}^2 (4 - (U - U_1) (-U + \sqrt{4 + (U - U_1)^2} + U_1)) + \nonumber \\
    +c_{E2}^2 (4 - (U - U_1) (-U + \sqrt{4 + (U - U_1)^2} +
          U_1)) \Bigg] \nonumber \\
           - \frac{(c_{E1} c_{E2} (-U + \sqrt{
      4 + (U - U_1)^2} + U_1) Cos[
     \phi_{E1} - \phi_{E2} + (-E_1 + E_2) t])}{(2 \sqrt{
    4 + (U -
       U_1)^2})} \nonumber \\
        - \frac{(2 c_{E_1} (c_{E_3} Cos[
        \phi_{E1} - \phi_{E3} - E_{1} t + E_{3} t] -
      c_{E4} Cos[\phi_{E1} - \phi_{E4} - E_1 t + E_4 t]))}{(\sqrt{
    4 + (U - U1) (U + \sqrt{4 + (U - U_1)^2} - U_1)} \sqrt{
    4 - (U - U1) (-U + \sqrt{4 + (U - U_1)^2} + U_1)})} + \nonumber \\
    +\frac{(2 c_{E2} (c_{E3} Cos[
        \phi_{E2} - \phi_{E3} - E_2 t + E_3 t] -
      c_{E4} Cos[\phi_{E2} - \phi_{E4} - E_2 t + E_4 t]))}{(\sqrt{
    4 + (U - U_1) (U + \sqrt{4 + (U - U_1)^2} - U_1)} \sqrt{
    2 + 1/2 (-U + \sqrt{4 + (U - U_1)^2} + U_1)^2})} \nonumber \\
    - \frac{(c_{E3} c_{E4} (U + \sqrt{
      4 + (U - U_1)^2} - U_1) Cos[
     \phi_{E3} - \phi_{E4} + (-E_3 + E_4) t])}{(2 \sqrt{4 + (U - U_1)^2})},
\end{eqnarray}

\begin{eqnarray}
\rho_{2,2}=p(1,2')=\frac{1}{(4 (4 + (U - U_1)^2))} ((c_{E3}^2 +
       c_{E4}^2) (4 + (U - U_1) (U + \sqrt{4 + (U - U_1)^2} - U_1)) + \nonumber \\
+  c_{E1}^2 (4 - (U - U_1) (-U + \sqrt{4 + (U - U_1)^2} + U_1)) +
    c_{E2}^2 (4 - (U - U_1) (-U + \sqrt{4 + (U - U_1)^2} +
          U_1)))+   \nonumber \\
 + \frac{(c_{E1} c_{E2} (-U + \sqrt{4 + (U - U_1)^2} + U_1) Cos[
     \phi_{E1} - \phi_{E2} + (-E_1 + E_2) t])}{(2 \sqrt{[
    4 + (U -
       U_1)^2]})}    \nonumber \\
 - \frac{(2 c_{E1} (c_{E3} Cos[
        \phi_{E1} - \phi_{E3} - E_1 t + E_3 t] +
      c_{E4} Cos[\phi_{E1} - \phi_{E4} - E_1 t + E_4 t]))}{(\sqrt{[
    4 + (U - U_1) (U + \sqrt{4 + (U - U_1)^2} - U_1)]} \sqrt{[
    4 - (U - U_1) (-U + \sqrt{4 + (U - U_1)^2} + U_1)]})}   \nonumber \\
 - \frac{(2 c_{E2} (c_{E3} Cos[
        \phi_{E2} - \phi_{E3} - E_2 t + E_3 t] +
      c_{E4} Cos[\phi_{E2} - \phi_{E4} - E_2 t + E_4 t]))}{(\sqrt{
    4 + (U - U_1) (U + \sqrt{4 + (U - U_1)^2} - U_1)} \sqrt{
    2 + \frac{1}{2} (-U + \sqrt{[4 + (U - U_1)^2]} + U_1)^2})} \nonumber \\
 + \frac{(c_{E3} c_{E4} (U + \sqrt{
      4 + (U - U_1)^2} - U_1) Cos[
     \phi_{E3} -\phi_ {E4} + (-E_3 + E_4) t])}{(2 \sqrt{4 + (U - U_1)^2})},
\end{eqnarray}

and 
\begin{eqnarray}
\rho_{3,3}=p(2,1')=\frac{(c_{E3}^2 + c_{E4}^2)}{(4 + (U - U_1) (U + \sqrt{[4 + (U - U_1)^2]} - U_1))} + \nonumber \\
 + \frac{(
 c_{E1}^2 + c_{E2}^2)}{(4 - (U - U_1) (-U + \sqrt{[4 + (U - U_1)^2]} + U_1))} + \nonumber \\
 -\frac{ (
 2 c_{E1} c_{E2} Cos[ \phi_{E1} -\phi_{E2} + (-E_1 + E_2) t])}{(
 4 - (U - U_1) (-U + \sqrt{[4 + (U - U_1)^2]} +
     U_1))}+  \nonumber \\
 + \frac{(2 c_{E1} (c_{E3} Cos[ \phi_{E1} -\phi_{E3} - E_1 t + E_3 t] -
      c_{E4} Cos[\phi_{E1} - \phi_{E4} - E_1 t + E_4 t]))}{(\sqrt{[
    4 + (U - U_1) (U + \sqrt{[4 + (U - U_1)^2]} - U_1)]} \sqrt{[
    4 - (U - U_1) (-U + \sqrt{[4 + (U - U_1)^2]} + U_1)]})} \nonumber \\
-\frac{ (2 c_{E2} (c_{E3} Cos[
        \phi_{E2} - \phi_{E3} - E_2 t + E_3 t]
 -  c_{E4} Cos[ \phi_{E2} - \phi_{E4} - E_2 t + E_4 t]))}{(\sqrt{[
    4 + (U - U_1) (U + \sqrt{[4 + (U - U_1)^2]} - U_1)]} \sqrt{[
    2 + \frac{1}{2} (-U + \sqrt[4 + (U - U_1)^2] + U_1)^2]})} \nonumber \\
 - \frac{(
 2 c_{E3} c_{E4} Cos[\phi_{E3} - \phi_{E4} + (-E_3 + E_4) t])}{(
 4 + (U - U1) (U + \sqrt{4 + (U - U_1)^2} - U1))}
\end{eqnarray}

\begin{eqnarray}
\rho_{4,4}(t)=p(2,2')=\frac{(c_{E3}^2 + c_{E4}^2)}{(4 + (U - U_1) (U + \sqrt{4 + (U - U_1)^2} - U_1))} + \frac{(
 c_{E1}^2 + c_{E2}^2)}{(4 - (U - U_1) (-U + \sqrt{4 + (U - U_1)^2} + U_1))} \nonumber \\
 - \frac{(
 2 c_{E1} c_{E2} Cos[
   \phi_{E1} -
    \phi_{E2} + (-E_1 + E_2) t])}{(-4 + (U - U_1) (-U + \sqrt{
     4 + (U - U_1)^2} +
     U_1))} + \nonumber \\
+\frac{(2 c_{E1} (c_{E3} Cos[\phi_{E1} - \phi_{E3} - E_1 t + E_3 t] +
      c_{E4} Cos[\phi_{E1} - \phi_{E4} - E_1 t + E_4 t]))}{(\sqrt{
    4 + (U - U_1) (U + \sqrt{[4 + (U - U_1)^2]} - U_1)} \sqrt{[
    4 - (U - U_1) (-U + \sqrt{[4 + (U - U_1)^2]} + U_1)]})}+ \nonumber \\ + \frac{(2 c_{E2} (c_{E3} Cos[
        \phi_{E2} - \phi_{E3} - E_2 t + E_3 t] +
      c_{E4} Cos[\phi_{E2} -\phi_ {E4} - E_2 t + E_4 t]))}{(\sqrt{
    4 + (U - U_1) (U + \sqrt{4 + (U - U_1)^2} - U_1)} \sqrt{
    2 + \frac{1}{2} (-U + \sqrt{4 + (U - U_1)^2} + U_1)^2})} + \nonumber \\ +\frac{(
 2 c_{E3} c_{E4} Cos[\phi_{E3} - \phi_{E4} + (-E_3 + E_4) t])}{(
 4 + (U - U_1) (U + \sqrt{4 + (U - U_1)^2} - U_1))}
\end{eqnarray}

Density matrix of 2 interacting parciles allow us to obtain density matrices of particles A and B. The elements of density matrix A are given as
\begin{eqnarray}
\rho_{A: (1,1)}=\frac{((c_{E3}^2 + c_{E4}^2) (U + \sqrt{4 + (U - U_1)^2} - U_1) +
    c_{E1}^2 (-U + \sqrt{[4 + (U - U_1)^2]} + U_1) +
    c_{E2}^2 (-U + \sqrt{[4 + (U - U_1)^2]} + U_1))}{(2 \sqrt{[
    4 + (U -
       U_1)^2]})} \nonumber \\  - \frac{(4 (c_{E1} c_{E3} Cos[
        \phi_{E1} - \phi_{E3} - E_1 t + E_3 t] +
      c_{E2} c_{E4} Cos[\phi_{E2} - \phi_{E4} - E_2 t + E_4 t]))}{(\sqrt{[
    4 + (U - U_1) (U + \sqrt{[4 + (U - U_1)^2]} - U_1)]} \sqrt{[
    4 - (U - U_1) (-U + \sqrt{[4 + (U - U_1)^2]} + U_1)]})}=p(1,t),
\end{eqnarray}

\begin{eqnarray}
\rho_{A: (2,2)}= (   
\frac{  ((c_{E1}^2+c_{E2}^2) (U + \sqrt{[4 + (U - U_1)^2]} - U_1) +
 (c_{E3}^2 + c_{E4}^2) (-U +
         \sqrt{[4 + (U - U_1)^2]} +
         U_1))}{2 \sqrt{[
     4 + (U - U_1)^2]}} + \nonumber \\
+\frac{(4 (c_{E1} c_{E3} Cos[\phi_{E1} -\phi_{E3} - E_1 t + E_3 t] +
         c_{E2} c_{E4} Cos[\phi_{E2} - \phi_{E4} - E_2 t + E_4 t]))}{(\sqrt{[
      4 + (U - U_1) (U + \sqrt{[4 + (U - U_1)^2]} - U_1)]} \sqrt{[
      4 - (U - U_1) (-U + \sqrt{[4 + (U - U_1)^2]} + U_1)]})})=p(2,t).
\end{eqnarray}

\begin{eqnarray}
\rho_{A: (1,2)}=\frac{1}{(\sqrt{[
   4 + (U - U_1)^2]} \sqrt{[4 + (U - U_1) (U + \sqrt{[4 + (U - U_1)^2]} - U_1)]}
    \sqrt{[2 + \frac{1}{2} (-U + \sqrt{[4 + (U - U_1)^2]} + U_1)^2]})} \times \nonumber \\
\times \Bigg[(-c_{E1}^2 - c_{E2}^2 + c_{E3}^2 + c_{E4}^2) \sqrt{[
    4 + (U - U_1) (U + \sqrt{[4 + (U - U_1)^2]} - U_1)]} \sqrt{[
    4 - (U - U_1) (-U + \sqrt{[4 + (U - U_1)^2]} + U_1)]} + \nonumber \\
   +2 (c_{E1} c_{E3} \sqrt{[
       4 + (U - U_1)^2]} (U - U_1) Cos[
        \phi_{E1} - \phi_{E3} - E_1 t + E_3 t] + \nonumber \\
      +c_{E2} c_{E4} \sqrt{[
       4 + (U - U_1)^2]} (U - U_1) Cos[
        \phi_{E2} - \phi_{E4} - E_2 t + E_4 t] + \nonumber \\
     + i (4 + (U - U_1)^2) (c_{E1} c_{E3} Sin[
           \phi_{E1} - \phi_{E3} - E_1 t + E_3 t] +
         c_{E2} c_{E4} Sin[\phi_{E2} - \phi_{E4} - E_2 t + E_4 t]))\Bigg]
\end{eqnarray}


\begin{eqnarray}
\rho_{A: (2,1)}= \frac{1}{(\sqrt{[
   4 + (U - U_1)^2]} \sqrt{[4 + (U - U_1) (U + \sqrt{[4 + (U - U_1)^2]} - U_1)]}
    \sqrt{[2 + 1/2 (-U + \sqrt{[4 + (U - U_1)^2]} + U_1)^2]})}
\times \nonumber \\ \times
\Bigg[(-c_{E1}^2 - c_{E2}^2 + c_{E3}^2 + c_{E4}^2) \sqrt{[
    4 + (U - U_1) (U + \sqrt{[4 + (U - U_1)^2]} - U_1)]} \sqrt{[
    4 - (U - U_1) (-U + \sqrt{[4 + (U - U_1)^2]} + U_1)]} + \nonumber \\
   +2 (c_{E1} c_{E3} \sqrt{[
       4 + (U - U_1)^2]} (U - U_1) Cos[
        \phi_{E1} - \phi_{E3} - E_1 t + E_3 t] + \nonumber \\
     + c_{E2} c_{E4} \sqrt{[
       4 + (U - U_1)^2]} (U - U_1) Cos[
        \phi_{E2} - \phi_{E4} - E_2 t + E_4 t] \nonumber \\
-  i (4 + (U - U_1)^2) (c_{E1} c_{E3} Sin[
           \phi_{E1} - \phi_{E3} - E_1 t + E_3 t] +
         c_{E2} c_{E4} Sin[\phi_{E2} - \phi_{E4} - E_2 t + E_4 t]))\Bigg]
\end{eqnarray}
We have also obtained the density matrix of particle B (electron at nodes 1' and 2') in the form as
\begin{eqnarray}
\rho_{B: (1,1)}=\frac{1}{2}(1 - 2 c_{E1} c_{E2} Cos[\phi_{E1} -\phi_{E2} - E_1 t + E_2 t]  \nonumber \\ -
   2 c_{E3} \sqrt{[1 - c_{E1}^2 - c_{E2}^2 - c_{E3}^2]}
     Cos[\phi_{E3} - \phi_{E4} - E_3 t + E_4 t])
\end{eqnarray}

\begin{eqnarray}
\rho_{B: (2,2)}=\frac{1}{2}(1 + 2 c_{E1} c_{E2} Cos[\phi_{E1} - \phi_{E2} - E_1 t + E_2 t]  \nonumber \\ +
   2 c_{E3} \sqrt{[1 - c_{E1}^2 - c_{E2}^2 - c_{E3}^2]}
     Cos[\phi_{E3} - \phi_{E4} - E_3 t + E_4 t])
\end{eqnarray}

\begin{eqnarray}
\rho_{B: (1,2)}=\frac{1}{2} (1 - 2 c_{E1}^2 - 2 c_{E3}^2 +
   2 i c_{E1} c_{E2} Sin[\phi_{E1} - \phi_{E2} - E_1 t + E_2 t] + \nonumber \\
   2 i c_{E3} \sqrt{[1 - c_{E1}^2 - c_{E2}^2 - c_{E3}^2]}
     Sin[\phi_{E3} - \phi_{E4} - E_3 t + E_4 t])
\end{eqnarray}

\begin{eqnarray}
\rho_{B: (2,1)}=\frac{1}{2}(1 - 2 c_{E1}^2 - 2 c_{E3}^2 -
   2 i c_{E1} c_{E2} Sin[\phi_{E1} - \phi_{E2} - E_1 t + E_2 t] \nonumber \\
 -    2 i c_{E3} \sqrt{[1 - c_{E1}^2 - c_{E2}^2 - c_{E3}^2]}
     Sin[\phi_{E3} - \phi_{E4} - E_3 t + E_4 t])
\end{eqnarray}
It  remarkable to notice that hopping constant that was constant for position dependent qubit was modified and it has two parts that depends on the frequency $E_2-E_1=\frac{1}{2} \left(-\sqrt{(U-U_1)^2+4}+U+U_1)+2 \right)-\frac{1}{2} \left(-\sqrt{(U-U_1)^2+4}+U+U_1)-2 \right)$ and $E_4-E_3=\frac{1}{2} \left(+\sqrt{(U-U_1)^2+4}+U+U_1)+2 \right)-\frac{1}{2} \left(+\sqrt{(U-U_1)^2+4}+U+U_1)-2 \right)$
For given density matrix
\begin{equation}
\rho=
\begin{pmatrix}
\rho_{1,1}=1-\rho_{2,2} & \rho_{1,2} \\
\rho_{2,1}=\rho_{1,2}^{*} & \rho_{2,2} \\
\end{pmatrix},
\end{equation}
where $\rho_{1,1}$ , $\rho_{2,2}$, $\rho(1,2)_r$, $\rho(1,2)_i$ in R with $\rho_{1,2}=\rho(1,2)_r + i\rho(1,2)_i $
we identify von Neumann entanglement $S$ entropy expressed as
\begin{eqnarray}
\label{entropyS}
-S(\rho_{2,2},|\rho_{1,2}|)= \frac{1}{(2 \sqrt{
  4 (|\rho(1,2)_r|^2+|\rho(1,2)_i|^2) + (1 -
     2 \rho_{2,2})^2})} \times [((-1 - 4 (|\rho(1,2)_r|^2+|\rho(1,2)_i|^2)+ \nonumber \\ + \sqrt{4 (|\rho(1,2)_r|^2+|\rho(1,2)_i|^2) + (1 - 2 \rho_{2,2})^2} -
      4 (-1 + \rho_{2,2}) \rho_{2,2}) Log[
     \frac{1}{2} (1 - \sqrt{4 (|\rho(1,2)_r|^2+|\rho(1,2)_i|^2) + (1 - 2 \rho_{2,2})^2})] + \nonumber \\ +(4 (|\rho(1,2)_r|^2+|\rho(1,2)_i|^2) + \sqrt{
      4 (|\rho(1,2)_r|^2+|\rho(1,2)_i|^2) + (1 - 2 \rho_{2,2})^2} + \nonumber \\ + (1 - 2 \rho_{2,2})^2) Log[
     \frac{1}{2} (1 + \sqrt{4 (|\rho(1,2)_r|^2+|\rho(1,2)_i|^2) + (1 - 2 \rho_{2,2})^2})])] .
     \end{eqnarray}
\begin{equation}
E_1=\frac{1}{2} \left(-\sqrt{(U-U_1)^2+4}+U+U_1)-2 \right),
E_2=\frac{1}{2} \left(-\sqrt{(U-U_1)^2+4}+U+U_1)+2 \right),
\end{equation}
\begin{equation}
E_3=\frac{1}{2} \left(+\sqrt{(U-U_1)^2+4}+U+U_1)-2 \right),
E_4=\frac{1}{2} \left(+\sqrt{(U-U_1)^2+4}+U+U_1)+2 \right).
\end{equation}
\normalsize

\subsubsection{Correlation function for case II: $E_f(1,1')=E_f(2,1')=U$, $E_f(1,2')=E_f(2,2')=U_1$ }

We postulate
\begin{eqnarray}H=
\begin{pmatrix}
U & 1 & 1 & 0 \\
1 & U_1 & 0 & 1 \\
1 & 0 & U_1 & 1 \\
0 & 1 & 1 & U
\end{pmatrix}, \nonumber \\
\end{eqnarray}
that implies
\begin{eqnarray}
U=E_{p1}+E_{p1'}+\frac{q^2}{\sqrt{d^2+(a+b)^2}}, \nonumber \\
U=E_{p2}+E_{p2'}+\frac{q^2}{\sqrt{(d+Cos(\alpha)(a+b))^2+(Sin(\alpha))^2(a+b)^2}}.
\end{eqnarray}
and consequently we obtain
\begin{eqnarray}
(E_{p1}-E_{p2})+(E_{p1'}-E_{p2'})=\frac{q^2}{\sqrt{(d+Cos(\alpha)(a+b))^2+(Sin(\alpha))^2(a+b)^2}} 
-\frac{q^2}{\sqrt{d^2+(a+b)^2}}.
\end{eqnarray}

We also have
\begin{eqnarray}
U_1=E_{p1}+E_{p2'}+\frac{q^2}{\sqrt{(d+Cos(\alpha)(a+b))^2+(1+Sin(\alpha))^2(a+b)^2}}, \nonumber \\
U_1=E_{p2}+E_{p1'}+\frac{q^2}{d},
\end{eqnarray}
and it implies
\begin{eqnarray}
-(E_{p1}-E_{p2})+(E_{p1'}-E_{p2'})=\frac{q^2}{\sqrt{(d+Cos(\alpha)(a+b))^2+(1+Sin(\alpha))^2(a+b)^2}}-\frac{q^2}{d}.
\end{eqnarray}
This results in
\begin{eqnarray}
(E_{p1'}-E_{p2'})=\frac{1}{2}[\frac{q^2}{\sqrt{(d+Cos(\alpha)(a+b))^2+(1+Sin(\alpha))^2(a+b)^2}}-\frac{q^2}{d} \nonumber \\
+\frac{q^2}{\sqrt{(d+Cos(\alpha)(a+b))^2+(Sin(\alpha))^2(a+b)^2}}-\frac{q^2}{\sqrt{d^2+(a+b)^2}}].
\end{eqnarray}
and
\begin{eqnarray}
(E_{p1}-E_{p2})=\frac{1}{2}[\frac{q^2}{\sqrt{(d+Cos(\alpha)(a+b))^2+(1+Sin(\alpha))^2(a+b)^2}}-\frac{q^2}{d} \nonumber \\
-\frac{q^2}{\sqrt{(d+Cos(\alpha)(a+b))^2+(Sin(\alpha))^2(a+b)^2}}+\frac{q^2}{\sqrt{d^2+(a+b)^2}}].
\end{eqnarray}
Therefore we have 3 open parameters $E_{p1} \in R$, $E_{p1'} \in R$ and for any $\alpha \in (0, 2\pi)$ we have predetermined conditions for $E_{p2}$ and $E_{p2'}$ given as
\begin{eqnarray}
E_{p2}=E_{p1}-\frac{1}{2}[\frac{q^2}{\sqrt{(d+Cos(\alpha)(a+b))^2+(1+Sin(\alpha))^2(a+b)^2}}-\frac{q^2}{d} \nonumber \\
-\frac{q^2}{\sqrt{(d+Cos(\alpha)(a+b))^2+(Sin(\alpha))^2(a+b)^2}}+\frac{q^2}{\sqrt{d^2+(a+b)^2}}].
\end{eqnarray}
and
\begin{eqnarray}
E_{p2'}=E_{p1'}-\frac{1}{2}[\frac{q^2}{\sqrt{(d+Cos(\alpha)(a+b))^2+(1+Sin(\alpha))^2(a+b)^2}}-\frac{q^2}{d} \nonumber \\
+\frac{q^2}{\sqrt{(d+Cos(\alpha)(a+b))^2+(Sin(\alpha))^2(a+b)^2}}-\frac{q^2}{\sqrt{d^2+(a+b)^2}}].
\end{eqnarray}
and implies $U \in R$
\begin{eqnarray}
U=E_{p1}+E_{p1'}+\frac{q^2}{\sqrt{d^2+(a+b)^2}},
\end{eqnarray}
as well as
\begin{eqnarray}
U_1=E_{p2}+E_{p1'}+\frac{q^2}{d}= \nonumber
2E_{p1'}+\frac{q^2}{d} 
-\frac{1}{2}[\frac{q^2}{\sqrt{(d+Cos(\alpha)(a+b))^2+(1+Sin(\alpha))^2(a+b)^2}}-\frac{q^2}{d} \nonumber \\
+\frac{q^2}{\sqrt{(d+Cos(\alpha)(a+b))^2+(Sin(\alpha))^2(a+b)^2}}-\frac{q^2}{\sqrt{d^2+(a+b)^2}}].
\end{eqnarray}

We have the following eigenvalues
\begin{eqnarray}
E_1=U, E_2=U_1, E_3=\frac{(U+U_1-\sqrt{(4)^2+(U-U_1)^2})}{2}, E_4=\frac{(U+U_1+\sqrt{(4)^2+(U-U_1)^2})}{2}.
\end{eqnarray}
and eigenstates
\begin{eqnarray} \ket{E_1}=
\begin{pmatrix}
-1 \\ 0 \\ 0 \\ 1 \\
\end{pmatrix}, 
\end{eqnarray}
\begin{eqnarray} \ket{E_2}=
\begin{pmatrix}
0 \\ -1 \\ +1 \\ 0 \\
\end{pmatrix}, 
\end{eqnarray}
\begin{eqnarray} \ket{E_3}=
\begin{pmatrix}
1 \\ \frac{4}{-U+U_1+\sqrt{(4)^2+(U-U_1)^2}} \\ \frac{4}{-U+U_1+\sqrt{(4)^2+(U-U_1)^2}} \\ 1 \\
\end{pmatrix}, 
\end{eqnarray}
\begin{eqnarray} \ket{E_4}=
\begin{pmatrix}
1 \\ \frac{4}{U-U_1+\sqrt{(4)^2+(U-U_1)^2}} \\ \frac{4}{U-U_1+\sqrt{(4)^2+(U-U_1)^2}} \\ 1 \\
\end{pmatrix}, 
\end{eqnarray}

\subsubsection{Correlation function for case III: $E_f(1,1')=E_f(2,1')=U$}
\begin{eqnarray}
C(t,\phi_{E_1},\phi_{E_2},\phi_{E_3},\phi_{E_4},U,U_1,E_1(U,U_1),E_2(U,U_1),E_3(U,U_1),E_4(U,U_1))= \nonumber \\
(2 (c_{E1}c_{E2}\sqrt{(U-U_1)^2+4}(U-U_1) \cos (-E_{1n}t+E_{2n}t+\phi_{E1}-\phi_{E2})+ \nonumber \\
+c_{E4} (c_{E1} \sqrt{(U-U_1) \left(\sqrt{(U-U_1)^2+4}+U-U_1\right)+4} \sqrt{4-(U-U_1) \left(\sqrt{(U-U_1)^2+4}-U+U_1\right)}\times \nonumber \\ \times  \cos
   ((- E_{1}+E_4)t+\phi_{E1}-\phi_{E4})+  \nonumber \\
   +c_{E3} \sqrt{(U-U_1)^2+4} (U_1-U) \cos ( (- E_3+E_{4})t
   +\phi_{E3}-\phi_{E4}))+ \nonumber \\
   +c_{E2} c_{E3} \sqrt{(U-U_1) \left(\sqrt{(U-U_1)^2+4}+U-U_1\right)+4} \sqrt{4-(U-U_1)
   \left(\sqrt{(U-U_1)^2+4}-U+U_1\right)} \times \nonumber \\
 \times \cos ((-E_{2}+E_{3})t+\phi_{E2}-\phi_{E3}))) \frac{1}{(U-U_1)^2+4},
\end{eqnarray}
where
\begin{eqnarray}H=
\begin{pmatrix}
U & 1 & 1 & 0 \\
1 & U_1 & 0 & 1 \\
1 & 0 & U & 1 \\
0 & 1 & 1 & U_1
\end{pmatrix}. \nonumber \\
\end{eqnarray}
and
\begin{eqnarray}
U_1=\frac{q^2}{\sqrt{(d+Cos(\alpha)(a+b))^2+(1+Sin(\alpha))^2(a+b)^2}}+E_{p1}+E_{p2'}, \nonumber
U=\frac{q^2}{d}+E_{p2}+E_{p}(1'),
\end{eqnarray}
\begin{eqnarray}
E_1=\frac{-2+U+U_1-\sqrt{(4)^2+(U-U_1)^2}}{2},  \nonumber \\
E_2=\frac{+2+U+U_1-\sqrt{(4)^2+(U-U_1)^2}}{2},  \nonumber \\
E_3=\frac{-2+U+U_1+\sqrt{(4)^2+(U-U_1)^2}}{2},   \nonumber \\
E_4=\frac{+2+U+U_1+\sqrt{(4)^2+(U-U_1)^2}}{2}.
\end{eqnarray}

Now we generalize the Hamiltonian matrix of the form
\begin{eqnarray}H=
\begin{pmatrix}
U & t_{s2} & t_{s1} & 0 \\
t_{s2} & U_1 & 0 & t_{s1} \\
t_{s1} & 0 & U & t_{s2} \\
0 & t_{s1} & t_{s2} & U_1
\end{pmatrix}. \nonumber \\
\end{eqnarray}
and energy eigenvalues depending on $t_{s1}$, $t_{s2}$, $U$ and $U_1$ are given as
\begin{eqnarray}
E_1=\frac{1}{2}(-2t_{s1}+U+U_1-\sqrt{(U-U_1)^2+(2t_{s2})^2}), \nonumber \\
E_2=\frac{1}{2}(+2t_{s1}+U+U_1-\sqrt{(U-U_1)^2+(2t_{s2})^2}), \nonumber \\
E_3=\frac{1}{2}(-2t_{s1}+U+U_1+\sqrt{(U-U_1)^2+(2t_{s2})^2}), \nonumber \\
E_4=\frac{1}{2}(+2t_{s1}+U+U_1+\sqrt{(U-U_1)^2+(2t_{s2})^2}).
\end{eqnarray}
It is surprising to discover that the corresponding energy eigenstates are depending only on $t_{s2}$ hopping constant and are not depending on $t_{s1}$ hopping constant and are given as
\begin{eqnarray}
\ket{E_1}=
\begin{pmatrix}
\frac{\sqrt{[8 t_{s2}^2 + 2 (-U + \sqrt{[4 t_{s2}^2 + (U - U_1)^2]} + U_1)^2]}}{(
  4 \sqrt{[4 t_{s2}^2 + (U - U_1)^2]})}, \\
  -(\frac{(2 t_{s2})}{\sqrt{[
   8 t_{s2}^2 + 2 (-U + \sqrt{[4 t_{s2}^2 + (U - U_1)^2]} + U_1)^2]}}), \\
   -(\frac{\sqrt{[
   8 t_{s2}^2 + 2 (-U + \sqrt{[4 t_{s2}^2 + (U - U_1)^2]} + U_1)^2]}}{(
   4 \sqrt{[4 t_{s2}^2 + (U - U_1)^2]})} ), \\
   \frac{(2 t_{s2})}{\sqrt{[
  8 t_{s2}^2 + 2 (-U + \sqrt{[4 t_{s2}^2 + (U - U_1)^2]} + U_1)^2]}},
  \end{pmatrix}, \nonumber \\
\ket{E_2}=
\begin{pmatrix}
-\frac{\sqrt{[
   8 t_{s2}^2 + 2 (-U + \sqrt{[4 t_{s2}^2 + (U - U_1)^2]} + U_1)^2]}}{(
   4 \sqrt{[4t_{s2}^2 + (U - U_1)^2]})}, \\
+\frac{2 t_{s2} }{\sqrt{
  8 t_{s2}^2 + 2 (-U + \sqrt{[4 t_{s2}^2 + (U - U_1)^2]} + U_1)^2}}, \\
  -\frac{\sqrt{
   8 t_{s2}^2 + 2 (-U + \sqrt{[4 t_{s2}^2 + (U - U_1)^2]} + U_1)^2}}{(
   4 \sqrt{[4 t_{s2}^2 + (U - U_1)^2]})}, \\
 +  \frac{2t_{s2}}{\sqrt{[
  8 t_{s2}^2 + 2 (-U + \sqrt{[4 t_{s2}^2 + (U - U_1)^2]} + U_1)^2]}},
  \end{pmatrix}, \nonumber \\
\ket{E_3}=
\begin{pmatrix}
-\frac{\sqrt{[8 t_{s2}^2 + 2 (U + \sqrt{[4 t_{s2}^2 + (U - U_1)^2]} - U_1)^2]}}{(
  4 \sqrt{[4 t_{s2}^2 + (U - U_1)^2]})}, \\
   -\frac{((2 t_{s2})}{\sqrt{[
  8 t_{s2}^2 + 2 (U + \sqrt{[4t_{s2}^2 + (U - U_1)^2]} - U_1)^2]}}),   \\
  \frac{\sqrt{[
 8 t_{s2}^2 + 2 (U + \sqrt{[4 t_{s2}^2 + (U - U_1)^2]} - U_1)^2]}}{(
 4 \sqrt{[4 t_{s2}^2 + (U - U_1)^2]})},  \\
 \frac{2 t_{s2}}{\sqrt{[
 8 t_{s2}^2 + 2 (U + \sqrt{[4 t_{s2}^2 + (U - U_1)^2]} - U_1)^2]}}
\end{pmatrix}
,  \\
\ket{E_4}=
\begin{pmatrix}
\frac{\sqrt{[8 t_{s2}^2 + 2 t_{s2}^2 (U + \sqrt{[4 t_{s2}^2 + (U - U_1)^2]} - U_1)^2]}}{(
 4 \sqrt{[4 t_{s2}^2 + (U - U_1)^2]})}, \\
 \frac{2 t_{s2}}{\sqrt{[
 8 t_{s2}^2 + 2 t_{s2}^2 (U + \sqrt{[4 t_{s2}^2 + (U - U_1)^2]} - U_1)^2]}}, \\
 \frac{\sqrt{[
 8 t_{s2}^2 + 2 t_{s2}^2 (U + \sqrt{[4 t_{s2}^2 + (U - U_1)^2]} - U_1)^2]}}{(
 4 \sqrt{[4 t_{s2}^2 + (U - U_1)^2]})}, \\
 \frac{2 t_{s2}}{\sqrt{[
 8 t_{s2}^2 + 2 t_{s2}^2 (U + \sqrt{[4 t_{s2}^2 + (U - U_1)^2]} - U_1)^2]}}
 \end{pmatrix}.
\end{eqnarray}

The minimalist density matrix of two electrostatically interacting qubits A and B in tight binding model in case of time-independent Hamiltonian is given as
\begin{eqnarray}
\rho_{AB}(t)= |c_{E1}|^2 \ket{E_1}\bra{E_1}+|c_{E2}|^2 \ket{E_2}\bra{E_2}+|c_{E3}|^2 \ket{E_3}\bra{E_3}+|c_{E4}|^2 \ket{E_4}\bra{E_4}+ \nonumber \\
+c_{E1}c_{E2}(\ket{E_1}\bra{E_2}e^{i(\phi_{E1(t_0)}-\phi_{E2(t_0)})}e^{\frac{1}{i\hbar}(E_1-E_2)(t-t_0)}+ \ket{E_2}\bra{E_1})e^{i(\phi_{E2(t_0)}-\phi_{E1(t_0)})}e^{\frac{1}{i\hbar}(E_2-E_1)(t-t_0)} + \nonumber \\ +c_{E1}c_{E3}(e^{i(\phi_{E1(t_0)}-\phi_{E3(t_0)})}e^{\frac{1}{i\hbar}(E_1-E_3)(t-t_0)}\ket{E_1}\bra{E_3} + e^{i(\phi_{E3(t_0)}-\phi_{E1(t_0)})}e^{\frac{1}{i\hbar}(E_3-E_1)(t-t_0)}\ket{E_3}\bra{E_1}) + \nonumber \\
+c_{E1}c_{E4}(e^{i(\phi_{E1(t_0)}-\phi_{E4(t_0)})}e^{\frac{1}{i\hbar}(E_1-E_4)(t-t_0)}\ket{E_1}\bra{E_4} + e^{i(\phi_{E4(t_0)}-\phi_{E1(t_0)})}e^{\frac{1}{i\hbar}(E_4-E_1)(t-t_0)}\ket{E_4}\bra{E_1})+ \nonumber \\
+c_{E2}c_{E3}(e^{i(\phi_{E2(t_0)}-\phi_{E3(t_0)})}e^{\frac{1}{i\hbar}(E_2-E_3)(t-t_0)}\ket{E_2}\bra{E_3}  
+e^{i(\phi_{E3(t_0)}-\phi_{E2(t_0)})}e^{\frac{1}{i\hbar}(E_3-E_2)(t-t_0)}\ket{E_3}\bra{E_2})+ \nonumber \\
+c_{E2}c_{E4}(e^{i(\phi_{E2(t_0)}-\phi_{E4(t_0)})}e^{\frac{1}{i\hbar}(E_2-E_4)(t-t_0)}\ket{E_2}\bra{E_4}  
+e^{i(\phi_{E4(t_0)}-\phi_{E2(t_0)})}e^{\frac{1}{i\hbar}(E_4-E_2)(t-t_0)}\ket{E_4}\bra{E_2})+ \nonumber \\
c_{E3}c_{E4}(e^{i(\phi_{E3(t_0)}-\phi_{E4(t_0)})}e^{\frac{1}{i\hbar}(E_3-E_4)(t-t_0)}\ket{E_3}\bra{E_4}  
+e^{i(\phi_{E4(t_0)}-\phi_{E3(t_0)})}e^{\frac{1}{i\hbar}(E_4-E_3)(t-t_0)}\ket{E_4}\bra{E_3}),
\end{eqnarray}
where $|c_{E1}|^2$, $|c_{E2}|^2$, $|c_{E3}|^2$ and $|c_{E4}|^2$ are probabilities of occupancy of eigenenergies $E_1$, $E_2$, $E_3$ and $E_4$ and
we obtain two particle  density matrix diagonal elements in the detailed form given below. We have probability of finding electron A at node 1 and electron B at node 1' given by formula
\small
\begin{eqnarray*}
\rho(1,1)=p(1,1',t)=
\frac{1}{4}[c_{E2}^2 + c_{E3}^2 + \frac{(c_{E4}^2 t_{s2}^2 (2 + 2 t_{s2}^2 + (U - U_1)^2))}{(
    4 t_{s2}^2 + (U - U_1)^2)} ]\nonumber \\
+\frac{(-(c_{E2}^2 - c_{E3}^2 - c_{E4}^2 t_{s2}^2) (U -
       U_1) + c_{E1}^2 (-U + \sqrt{[4 t_{s2}^2 + (U - U_1)^2]} + U_1) )}{(4 \sqrt{[
    4 t_{s2}^2 + (U - U_1)^2]})}\nonumber \\
%
 - \frac{(c_{E1} c_{E2} (-U + \sqrt{[
      4 t_{s2}^2 + (U - U_1)^2]} + U_1) Cos[
     \phi_{E1} -  \phi_{E2} + (-E_1 + E_2) t] )}{( 2 \sqrt{[
    4 t_{s2}^2 + (U - U_1)^2]})}\nonumber \\
 -\frac{(c_{E1} c_{E3} \sqrt{[
    8 t_{s2}^2 + 2 (U  + \sqrt{[4 t_{s2}^2 + (U - U_1)^2]} - U_1)^2]} \sqrt{[
    8 t_{s2}^2 + 2 (-U + \sqrt{[4 t_{s2}^2 + (U - U_1)^2]} + U_1)^2]}
     Cos[\phi_{E1} -
      \phi_{E3} + (-E_1 + E_3) t] )}{( 8 (4 t_{s2}^2 + (U -
        U_1)^2))} \nonumber \\
+ \frac{(c_{E2} c_{E3} \sqrt{[
    8 t_{s2}^2 + 2 (U + \sqrt{[4 t_{s2}^2 + (U - U_1)^2]} - U_1)^2]} \sqrt{[
    8 t_{s2}^2 + 2 (-U + \sqrt{[4 t_{s2}^2 + (U - U_1)^2]} + U_1)^2]}
     Cos[ \phi_{E2} -
      \phi_{E3} + (-E_2 + E_3) t])}{(8 (4 t_{s2}^2 + (U -
        U_1)^2) )}\nonumber \\
 + \frac{(c_{E1} c_{E4}  \sqrt{[
    t_{s2}^2 (4 + (U + \sqrt{[4 t_{s2}^2 + (U - U_1)^2]} - U_1)^2)]} \sqrt{[
    8 t_{s2}^2 + 2 (-U + \sqrt{[4 t_{s2}^2 + (U - U_1)^2]} + U_1)^2]}
     Cos[\phi_{E1} - \phi_{E4} + (-E_1 + E_4) t])}{(4 \sqrt{2} (4 t_{s2}^2 + (U - U_1)^2))} \nonumber \\
- \frac{(c_{E2} c_{E4} \sqrt{[
    t_{s2}^2 (4 + (U + \sqrt{[4 t_{s2}^2 + (U - U_1)^2]} - U_1)^2)]} \sqrt{[
    8 t_{s2}^2 + 2 (-U + \sqrt{[4 t_{s2}^2 + (U - U_1)^2]} + U_1)^2]}
     Cos[\phi_{E2} - \phi_{E4} + (-E_2 + E_4) t] )}{( 4 \sqrt{
    2} (4 t_{s2}^2 + (U - U_1)^2) )} \nonumber \\  
- \frac{(c_{E3} c_{E4} \sqrt{[
    t_{s2}^2 (4 + (U + \sqrt{[4 t_{s2}^2 + (U - U_1)^2]} - U_1)^2)]} \sqrt{[
    8 t_{s2}^2 + 2 (U + \sqrt{[4 t_{s2}^2 + (U - U_1)^2]} - U_1)^2 ]}
     Cos[ \phi_{E3} - \phi_{E4} + (-E_3 + E_4) t] )}{( 4 \sqrt{
    2} (4 t_{s2}^2 + (U - U_1)^2) )}
\end{eqnarray*}
and probability of finding electron A at node 1 and electron B at node 2' given by formula
\normalsize
\begin{eqnarray}
\rho(2,2)=p(1,2',t)=
\frac{(2 c_{E4}^2)}{(4 + (U + \sqrt{[4 t_{s2}^2 + (U - U_1)^2]} - U_1)^2)}+ \nonumber \\
+ \frac{(4 c_{E3}^2 t_{s2}^2)}{(
 8 t_{s2}^2 +
  2 (U + \sqrt{[4 t_{s2}^2 + (U - U_1)^2]} - U_1)^2)} + \frac{((c_{E1}^2 +
    c_{E2}^2) t_{s2}^2)}{(
 4 t_{s2}^2 - (U - U_1) (-U + \sqrt{[4 t_{s2}^2 + (U - U_1)^2]} + U_1))} \nonumber \\
- \frac{(
 2 c_{E1} c_{E2} t_{s2}^2 Cos[\phi_{E1} - \phi_{E2} + (-E_1 + E_2) t])}{(
 4 t_{s2}^2 - (U - U_1) (-U + \sqrt{[4 t_{s2}^2 + (U - U_1)^2]} +
     U_1))} + \nonumber \\
+\frac{(2 c_{E1} c_{E3} t_{s2}^2 Cos[
     \phi_{E1} - \phi_{E3} + (-E_1 + E_3) t])}{(\sqrt{[
    4 t_{s2}^2 + (U - U_1) (U + \sqrt{[4 t_{s2}^2 + (U - U_1)^2]} - U_1)]} \sqrt{[
    4 t_{s2}^2 - (U - U_1) (-U + \sqrt{[4 t_{s2}^2 + (U - U_1)^2]} +
        U_1)]})} \nonumber \\
 - \frac{(2 c_{E2} c_{E3} t_{s2}^2 Cos[
     \phi_{E2} - \phi_{E3} + (-E_2 + E_3) t])}{(\sqrt{[
    4 t_{s2}^2 + (U - U_1) (U + \sqrt{[4 t_{s2}^2 + (U - U_1)^2]} - U_1)]} \sqrt{[
    4 t_{s2}^2 - (U - U_1) (-U + \sqrt{[4 t_{s2}^2 + (U - U_1)^2]} +
        U_1)]})} + \nonumber \\
 - \frac{(2 c_{E1} c_{E4} t_{s2}^2 Cos[
     \phi_{E1} - \phi_{E4} + (-E_1 + E_4) t])}{(\sqrt{[
    t_{s2}^2 (2 +
       2 t_{s2}^2 + (U - U_1) (U + \sqrt{[4 t_{s2}^2 + (U - U_1)^2]} - U_1))]}
     \sqrt{[4 t_{s2}^2 - (U - U_1) (-U + \sqrt{[4 t_{s2}^2 + (U - U_1)^2]} +
        U_1)]})} + \nonumber \\
+ \frac{(2 c_{E2} c_{E4} t_{s2}^2 Cos[
     \phi_{E2} - \phi_{E4} + (-E_2 + E_4) t])}{(\sqrt{[
    t_{s2}^2 (2 +
       2 t_{s2}^2 + (U - U_1) (U + \sqrt{[4 t_{s2}^2 + (U - U_1)^2]} - U_1))]}
     \sqrt{[4 t_{s2}^2 - (U - U_1) (-U + \sqrt{[4 t_{s2}^2 + (U - U_1)^2]} +
        U_1)]})} + \nonumber \\
- \frac{(2 c_{E3} c_{E4} t_{s2}^2 Cos[
     \phi_{E3} - \phi_{E4} + (-E_3 + E_4) t])}{(\sqrt{[
    t_{s2}^2 (2 +
       2 t_{s2}^2 + (U - U_1) (U + \sqrt{[4 t_{s2}^2 + (U - U_1)^2]} - U_1))]}
     \sqrt{[4 t_{s2}^2 + (U - U_1) (U + \sqrt{[4 t_{s2}^2 + (U - U_1)^2]} - U_1)]})}
\end{eqnarray}
\small
and probability of finding electron A at node 2 and electron B at node 1' given as
\begin{eqnarray*}
\rho(3,3)=p(2,1',t)=
\frac{1}{4} (c_{E2}^2 + c_{E3}^2 + \frac{(c_{E4}^2 t_{s2}^2 (2 + 2 t_{s2}^2 + (U - U_1)^2))}{(
    4 t_{s2}^2 + (U - U_1)^2)}) + \nonumber \\
 + \frac{(-(c_{E2}^2 - c_{E3}^2 - c_{E4}^2 t_{s2}^2) (U -
       U_1) + c_{E1}^2 (-U + \sqrt{[4 ts2^2 + (U - U_1)^2]} + U_1))}{(4 \sqrt{[
    4 t_{s2}^2 + (U - U_1)^2]} )}+   \nonumber \\
 + \frac{(c_{E1}c_{E2} (-U + \sqrt{[
      4 t_{s2}^2 + (U - U_1)^2]} + U_1) Cos[
     \phi_{E1} - \phi_{E2} + (-E_1 + E_2) t])}{(2 \sqrt{[
    4 t_{s2}^2 + (U - U_1)^2]})}+ \nonumber \\
- \frac{(c_{E1} c_{E3} \sqrt{[
    8 t_{s2}^2 + 2 (U + \sqrt{[4 t_{s2}^2 + (U - U_1)^2]} - U_1)^2]} \sqrt{[
    8 t_{s2}^2 + 2 (-U + \sqrt{[4 t_{s2}^2 + (U - U1)^2]} + U_1)^2]}
     Cos[\phi_{E1} -
      \phi_{E3} + (-E_1 + E_3) t])}{(8 (4 t_{s2}^2 + (U -
        U_1)^2))} \nonumber \\
 - \frac{(c_{E2} c_{E3} \sqrt{[
    8 t_{s2}^2 + 2 (U + \sqrt{[4 t_{s2}^2 + (U - U_1)^2]} - U_1)^2]} \sqrt{[
    8 t_{s2}^2 + 2 (-U + \sqrt{[4 t_{s2}^2 + (U - U_1)^2]} + U_1)^2]}
     Cos[\phi_{E2} - \phi_{E3} + (-E_2 + E_3) t])}{(8 (4 t_{s2}^2 + (U -
        U_1)^2))} \nonumber \\
- \frac{( c_{E1} c_{E4} \sqrt{[
    t_{s2}^2 (4 + (U + \sqrt{[4 t_{s2}^2 + (U - U_1)^2]} - U_1)^2)]} \sqrt{[
    8 t_{s2}^2 + 2 (-U + \sqrt{[4 t_{s2}^2 + (U - U_1)^2]} + U_1)^2]}
     Cos[\phi_{E1} - \phi_{E4} + (-E_1 + E_4) t])} {(4 \sqrt{
    2} (4 t_{s2}^2 + (U - U_1)^2))} \nonumber \\
- \frac{(c_{E2} c_{E4} \sqrt{[
    t_{s2}^2 (4 + (U + \sqrt{[4 t_{s2}^2 + (U - U_1)^2]} - U_1)^2)]} \sqrt{[
    8 t_{s2}^2 + 2 (-U + \sqrt{[4 t_{s2}^2 + (U - U_1)^2]} + U_1)^2]}
     Cos[\phi_{E2} - \phi_{E4} + (-E_2 + E_4) t])}{(4 \sqrt{
    2} (4 t_{s2}^2 + (U - U_1)^2))} \nonumber \\
+ \frac{(c_{E3} c_{E4} \sqrt{[
    t_{s2}^2 (4 + (U + \sqrt{[4 t_{s2}^2 + (U - U_1)^2]} - U_1)^2)]} \sqrt{[
    8 t_{s2}^2 + 2 (U + \sqrt{[4 t_{s2}^2 + (U - U_1)^2]} - U_1)^2]}
     Cos[ \phi_{E3} - \phi_{E4} + (-E_3 + E_4) t])}{(4 \sqrt{
    2} (4 t_{s2}^2 + (U - U1)^2) )} \nonumber \\
\end{eqnarray*}
and probability of finding electron A at 2 node and electron B at node 2' as
\small
\begin{eqnarray}
\rho(4,4,t)=p(2,2',t)=\frac{(2 c_{E4}^2)}{(4 + (U + \sqrt{[4 t_{s2}^2 + (U - U1)^2]} - U_1)^2)} + \nonumber \\
+ \frac{(
 4 c_{E3}^2 t_{s2}^2)}{(
 8 t_{s2}^2 +
  2 (U + \sqrt{[4 t_{s2}^2 + (U - U_1)^2]} - U_1)^2)}+ \nonumber \\
+ \frac{((c_{E1}^2 +
    c_{E2}^2) t_{s2}^2)}{(
 4 t_{s2}^2 - (U - U_1) (-U + \sqrt{4 t_{s2}^2 + (U - U_1)^2} + U_1))}+ \nonumber \\
  + \frac{(
 2 c_{E1}c_{E2} t_{s2}^2 Cos[\phi_{E1} - \phi_{E2} + (-E_1 + E_2) t])}{(
 4 t_{s2}^2 - (U - U_1) (-U + \sqrt{[4 t_{s2}^2 + (U - U_1)^2]} +
     U_1))}+ \nonumber \\
     + \frac{(2 c_{E1}c_{E3} t_{s2}^2 Cos[\phi_{E1} - \phi_{E3} + (-E_1 + E_3) t])}{(\sqrt{[
    4 t_{s2}^2 + (U - U_1) (U + \sqrt{[4 t_{s2}^2 + (U - U_1)^2]} - U_1)]} \sqrt{[
    4 t_{s2}^2 - (U - U_1) (-U + \sqrt{[4 t_{s2}^2 + (U - U_1)^2]} +
        U_1)]})}+ \nonumber \\
          + \frac{(2 c_{E2} c_{E3} t_{s2}^2 Cos[
     \phi_{E2} - \phi_{E3} + (-E_2 + E_3) t])}{(\sqrt{[
    4 t_{s2}^2 + (U - U1) (U + \sqrt{[4 t_{s2}^2 + (U - U_1)^2]} - U_1)]} \sqrt{[
    4 t_{s2}^2 - (U - U_1) (-U + \sqrt{[4 t_{s2}^2 + (U - U_1)^2]} +
        U1)}])}+ \nonumber \\
         + \frac{(2 c_{E1}c_{E4} t_{s2}^2 Cos[
     \phi_{E1} - \phi_{E4} + (-E_1 + E_4) t])}{(\sqrt{[
    t_{s2}^2 (2 +
       2 t_{s2}^2 + (U - U_1) (U + \sqrt{[4 t_{s2}^2 + (U - U_1)^2]} - U_1))]}
     \sqrt{[4 t_{s2}^2 - (U - U_1) (-U + \sqrt{[4 ts2^2 + (U - U_1)^2]} +
        U1)]})} + \nonumber \\
        +\frac{(2 c_{E2} c_{E4} t_{s2}^2 Cos[
     \phi_{E2} - \phi_{E4} + (-E_2 + E_4) t])}{(\sqrt{[
    t_{s2}^2 (2 +
       2 t_{s2}^2 + (U - U_1) (U + \sqrt{[4 ts2^2 + (U - U1)^2]} - U_1))]}
     \sqrt{[4 t_{s2}^2 - (U - U_1) (-U + \sqrt{[4 t_{s2}^2 + (U - U_1)^2]} +
        U_1)]})} + \nonumber \\
        +\frac{(2 c_{E3} c_{E4} t_{s2}^2 Cos[
     \phi_{E3} - \phi_{E4} + (-E_3 + E_4) t])}{(\sqrt{[
    t_{s2}^2 (2 +
       2 t_{s2}^2 + (U - U_1) (U + \sqrt{[4t_{s2}^2 + (U - U_1)^2]} - U_1))]}
     \sqrt{[4 t_{s2}^2 + (U - U_1) (U + \sqrt{[4t_{s2}^2 + (U - U_1)^2]} - U_1)]})} \nonumber \\
     \end{eqnarray}.
\normalsize

\section{Quasiclassical approach towards tunable swap gate}
\subsection{Quasiclassical approach towards symmetric swap gate with the same localizing potentials}
We assume that one particle is in the field of another particle that is given by integro-differential equations as often used in quantum chemistry.
In addition we will assume that kinetic energy of particles A and B (electrons in qubit A and B) is very small so $t_{s12} \rightarrow 0$ and $t_{s1'2'} \rightarrow 0$ however still $t_{s12} \neq 0$ and $t_{s1'2'} \neq 0$ so after certain long time particles have chance to find proper configuration. We assume that $p_{A1}$ is the probability of finding particle A at node 1 and that $p_{B1'}$ is the probability of finding particle B at node 1'.
We have effective Hamiltonian omitting kinetic terms in the form
\begin{eqnarray}
\label{HphSymmetricSwap}
H(p_{A1},p_{B1'},E_{p1},E_{p2},E_{p1'},E_{p2'})=p_{A1} E_{p_1}+(1-p_{A1})E_{p_2}+p_{B1'} E_{p_{1'}}+(1-p_{B1'})E_{p_{2'}}+ \nonumber \\
+p_{A1}p_{B1'} \frac{q^2}{d}+(1-p_{A1})(1-p_{B1'}) \frac{q^2}{d} +p_{A1}(1-p_{B1'}) \frac{q^2}{\sqrt{d^2+(a+b)^2}} +p_{B1'}(1-p_{A1'})\frac{q^2}{\sqrt{d^2+(a+b)^2}}. 
\end{eqnarray}
Well designed classical Swap and Q-Swap gate (inverting state 0 to 1 and 1 to 0) will have the property that function $H(p_{A1},p_{B1'},E_{p1},E_{p2},E_{p1'},E_{p2'})$ will reach its minima at $(p_{A1}=0,p_{B1'}=1)$ and at $(p_{A1}=1,p_{B1'}=0)$ and that $H(1,0,E_{p1},E_{p2},E_{p1'},E_{p2'})=H(0,1,E_{p1},E_{p2},E_{p1'},E_{p2'})$. Consequently we assume \newline $H(1,1,E_{p1},E_{p2},E_{p1'},E_{p2'})>H(0,1,E_{p1},E_{p2},E_{p1'},E_{p2'})$ and \newline
 $H(0,0,E_{p1},E_{p2},E_{p1'},E_{p2'})>H(0,1,E_{p1},E_{p2},E_{p1'},E_{p2'})$ 
 \newline
 what implies that logical states that are not allowed have higher energy.  
Further simplification can be done by setting for example $E_{p1}$ and $E_{p1'}$ to constant value (as $E_{p1}=const_1$ and $E_{p1'}=const_{1'}$) for example 0 or 1 or any other fixed real number. It shall be underlined that we can chose the one among 4 possible combinations $(E_{p1},E_{p1'})$, $(E_{p2},E_{p2'})$, $(E_{p1},E_{p2'})$, $(E_{p2},E_{p1'})$ whose values needs to be fixed. We set $q=1$, $d=1$, $a+b=0.2$, $E_{p2}=E_{p2'}=1$.
We obtain
\begin{eqnarray}
H(p_{A1},p_{B1'},E_{p1},1,E_{p1'},1)=p_{A1} E_{p_1}+(1-p_{A1})+p_{B1'} E_{p_{1'}}+(1-p_{B1'})+ \nonumber \\
+p_{A1}p_{B1'}+(1-p_{A1})(1-p_{B1'})+p_{A1}(1-p_{B1'}) \frac{1}{\sqrt{1.01}} +p_{B1'}(1-p_{A1'})\frac{1}{\sqrt{1.01}}. 
\end{eqnarray}
Now we need to trace the numerical behaviour of 4 functions
\begin{eqnarray}
H(1,0,E_{p1},1,E_{p1'},1)= E_{p_1}+1 
+ \frac{1}{\sqrt{1.01}} . 
\end{eqnarray}
\begin{eqnarray}
H(0,1,E_{p1},1,E_{p1'},1)=+1+ E_{p_{1'}} 
 +\frac{1}{\sqrt{1.01}}. 
\end{eqnarray}
\begin{eqnarray}
H(1,1,E_{p1},1,E_{p1'},1)=E_{p_1}+ E_{p_{1'}}+1 . 
\end{eqnarray}
\begin{eqnarray}
H(0,0,E_{p1},1,E_{p1'},1)=+1+1+1=3 . 
\end{eqnarray}
Since we impose $H(1,0,E_{p1},1,E_{p1'},1)=H(0,1,E_{p1},1,E_{p1'},1)$ we obtain $E_{p1'}=E_{p1}$.
Imposing $H(0,0,E_{p1},1,E_{p1'},1)=H(1,1,E_{p1},1,E_{p1'},1)$ we finally obtain $E_{p1'}=E_{p1}=1$.
We observe that $H(1,0,E_{p1},1,E_{p1'},1)<H(0,0,E_{p1},1,E_{p1'},1)$
\subsection{Quasi-classical approach towards transition from anticorrelated symmetric swap gate to the correlated swap gate}
We have effective Hamiltonian omitting kinetic terms for any angle $\alpha$ in the form
\begin{eqnarray}
H(p_{A1},p_{B1'},E_{p1},E_{p2},E_{p1'},E_{p2'})=p_{A1} E_{p_1}+(1-p_{A1})E_{p_2}+p_{B1'} E_{p_{1'}}+(1-p_{B1'})E_{p_{2'}}+ \nonumber \\
+p_{A1}p_{B1'} \frac{q^2}{\sqrt{d^2+(a+b)^2}}+(1-p_{A1})(1-p_{B1'}) \frac{q^2}{\sqrt{(d+Cos(\alpha)(a+b))^2+(a+b)^2Sin(\alpha)^2}} + \nonumber \\
p_{A1}(1-p_{B1'}) \frac{q^2}{\sqrt{(d+(a+b)Cos(\alpha))^2+((1+Sin(\alpha))(a+b))^2}} +p_{B1'}(1-p_{A1'})\frac{q^2}{\sqrt{d^2+(a+b)^2}}. 
\end{eqnarray}
Previously considered case was for the angle $\alpha=-\Pi/2$ that corresponds to the symmetric anticorrelated swap gate as specified by formula \ref{HphSymmetricSwap}. Let us set $E_{p2}$ and $E_{p2'}$ to be 1. We obtain following $H(p_{A1},p_{B1'},E_{p1},1,E_{p1'},1)$ values as
\begin{eqnarray}
H(p_{A1}=0,p_{B1'}=0,E_{p1},E_{p2}=1,E_{p1'},E_{p2'}=1)=1+1+ \nonumber \\
+ \frac{q^2}{\sqrt{(d+Cos(\alpha)(a+b))^2+(a+b)^2Sin(\alpha)^2}}=V_1, 
\end{eqnarray}
\begin{eqnarray}
H(p_{A1}=1,p_{B1'}=1,E_{p1},E_{p2}=1,E_{p1'},E_{p2'}=1)= E_{p_1}+ E_{p_{1'}}+ \nonumber \\
+ \frac{q^2}{\sqrt{d^2+(a+b)^2}}=V_1, 
\end{eqnarray}
\begin{eqnarray}
H(p_{A1}=1,p_{B1'}=0,E_{p1},E_{p2}=1,E_{p1'},E_{p2'}=1)=E_{p_1}+1+ \nonumber \\
 \frac{q^2}{\sqrt{(d+(a+b)Cos(\alpha))^2+((1+Sin(\alpha))(a+b))^2}}=V_2, 
\end{eqnarray}
\begin{eqnarray}
H(p_{A1}=0,p_{B1'}=1,E_{p1},E_{p2}=1,E_{p1'},E_{p2'}=1)=1+ E_{p_{1'}} 
 +\frac{q^2}{\sqrt{d^2+(a+b)^2}}=V_2, 
\end{eqnarray}
with condition $V_2 < V_1$ that implies
\begin{eqnarray}
1+1+\frac{q^2}{\sqrt{(d+Cos(\alpha)(a+b))^2+(a+b)^2Sin(\alpha)^2}}-\frac{q^2}{\sqrt{d^2+(a+b)^2}}= 
E_{p_1}+ E_{p_{1'}},
\end{eqnarray}

\begin{eqnarray}
+\frac{q^2}{\sqrt{d^2+(a+b)^2}}-\frac{q^2}{\sqrt{(d+(a+b)Cos(\alpha))^2+((1+Sin(\alpha))(a+b))^2}}=E_{p_1}-E_{p_{1'}}
\end{eqnarray}
and we obtain
\begin{eqnarray}
1+\frac{1}{2}(\frac{q^2}{\sqrt{(d+Cos(\alpha)(a+b))^2+(a+b)^2Sin(\alpha)^2}}-\frac{q^2}{\sqrt{(d+(a+b)Cos(\alpha))^2+((1+Sin(\alpha))(a+b))^2}}=E_{p1},
\end{eqnarray}
\begin{eqnarray}
1+\frac{1}{2}(\frac{q^2}{\sqrt{(d+Cos(\alpha)(a+b))^2+(a+b)^2Sin(\alpha)^2}}+\frac{q^2}{\sqrt{(d+(a+b)Cos(\alpha))^2+((1+Sin(\alpha))(a+b))^2}} \nonumber \\
-\frac{q^2}{\sqrt{d^2+(a+b)^2}}=E_{p1'},
\end{eqnarray}
We can check the designing condition $V_1>V_2$ that implies
\begin{eqnarray}
H(p_{A1}=1,p_{B1'}=1,E_{p1},E_{p2}=1,E_{p1'},E_{p2'}=1)= E_{p_1}+ E_{p_{1'}} 
+ \frac{q^2}{\sqrt{d^2+(a+b)^2}}=V_1>, \nonumber \\
H(p_{A1}=1,p_{B1'}=0,E_{p1},E_{p2}=1,E_{p1'},E_{p2'}=1)=E_{p_1}+1+ 
 \frac{q^2}{\sqrt{(d+(a+b)Cos(\alpha)))^2+((1+Sin(\alpha))(a+b))^2}}=V_2, \nonumber \\ 
\end{eqnarray}
and consequently we obtain
\begin{eqnarray}
E_{p_{1'}} > 
+1+ \frac{q^2}{\sqrt{(d+(a+b)Cos(\alpha))^2+((1+Sin(\alpha))(a+b))^2}}-\frac{q^2}{\sqrt{d^2+(a+b)^2}}
\end{eqnarray}
what brings the condition
\begin{eqnarray}
\frac{q^2}{\sqrt{(d+Cos(\alpha)(a+b))^2+(a+b)^2Sin(\alpha)^2}}>\frac{q^2}{\sqrt{(d+(a+b)Cos(\alpha)))^2+((1+Sin(\alpha))(a+b))^2}}.
\end{eqnarray}
that is equivalent to the condition $Sin(\alpha)^2<(1+Sin(\alpha))^2=1+Sin^2(\alpha)+2Sin(\alpha)$ or $0<Sin(\alpha)$ that $\alpha \in (0, \Pi)$.
The conducted considerations are visualized by Fig.\ref{GeometricTransition} that describes the dependence of phenomenological electrostatic Hamiltonian on logical values $p_{A1}$ and $p_{B1}$ what gives the prescription for construction fuzzy logic swap gate or quantum swap gate implemented in single electron devices.
\section{Electrostatic quantum antiswap gate}
We can use the fact that two qubits can be in maximum correlated states (if $p_{A1}=1$ we have $p_{B1}=1-p_{A1}$).  Due to repulsive electrostatic interaction we can build the electrostatic quantum antiswap gate so it corresponds to the situation when one qubits is being copied by another qubit. It shall be underlined that the same considerations can be done for electron-hole system ($q^2=-1$) or hole-hole system ($q^2=1$).
\begin{figure}
\centering 
 \includegraphics[scale=0.6]{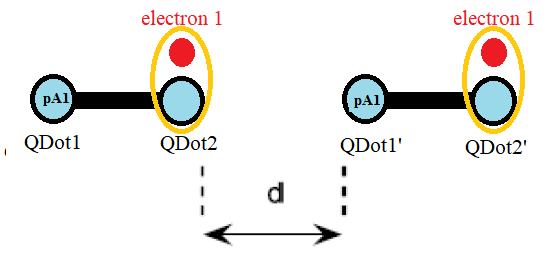}
 \caption{Scheme of electrostatic antiswap gate }
\end{figure}
We have
\begin{eqnarray}
H(p_{A1},p_{B1'},E_{p1},E_{p2},E_{p1'},E_{p2'})=p_{A1} E_{p_1}+(1-p_{A1})E_{p_2}+p_{B1'} E_{p_{1'}}+(1-p_{B1'})E_{p_{2'}}+ \nonumber \\
+p_{A1}p_{B1'} \frac{q^2}{d+a+b}+(1-p_{A1})(1-p_{B1'}) \frac{q^2}{d+a+b} +p_{A1}(1-p_{B1'}) \frac{q^2}{d+2(a+b)} +p_{B1'}(1-p_{A1'})\frac{q^2}{d}= \nonumber \\. 
\end{eqnarray}
We set $E_{p1}=1, E_{p2'}=1$ and we have
\begin{eqnarray}
H(p_{A1},p_{B1'},1,E_{p2},E_{p1'},1)=p_{A1}+(1-p_{A1})E_{p_2}+p_{B1'} E_{p_{1'}}+(1-p_{B1'})+ \nonumber \\
+p_{A1}p_{B1'} \frac{q^2}{d+a+b}+(1-p_{A1})(1-p_{B1'}) \frac{q^2}{d+a+b} +p_{A1}(1-p_{B1'}) \frac{q^2}{d+2(a+b)} +p_{B1'}(1-p_{A1'})\frac{q^2}{d}= \nonumber \\
=p_{A1}+(1-p_{A1})E_{p_2}+p_{B1'} E_{p_{1'}}+(1-p_{B1'})+ \nonumber \\
+(2p_{A1}p_{B1'}+1-(p_{A1}+p_{B1'}) )\frac{q^2}{d+a+b}+p_{A1}(1-p_{B1'}) \frac{q^2}{d+2(a+b)} +p_{B1'}(1-p_{A1'})\frac{q^2}{d}
\end{eqnarray}

In order to obtain quantum repeater we have $p_{A1}=p_{B1'}$ and such configuration is energetically favourable (minimizes Hamiltonian energy) so we have
\begin{eqnarray}
H(0,0,1,E_{p2},E_{p1'},1)=H(1,1,1,E_{p2},E_{p1'},1)=V_1, \nonumber \\
H(1,0,1,E_{p2},E_{p1'},1)=H(0,1,1,E_{p2},E_{p1'},1)=V_2, V_2<V_1.
\end{eqnarray}

We have the condition
\begin{eqnarray}
H(0,0,1,E_{p2},E_{p1'},1)=E_{p_2}+1+\frac{q^2}{d+a+b}=V_1,
\end{eqnarray}

\begin{eqnarray}
H(1,1,1,E_{p2},E_{p1'},1)
=1+ E_{p_{1'}} 
+(2+1-2 )\frac{q^2}{d+a+b}=1+ E_{p_{1'}}+\frac{q^2}{d+a+b}=V_1.
\end{eqnarray}
and it implies $E_{p_{1'}}=E_{p_2}=E_p$. We have

\begin{eqnarray}
H(1,0,1,E_{p2},E_{p1'},1)=1+1 
+ \frac{q^2}{d+2(a+b)} =V_2,
\end{eqnarray}

\begin{eqnarray}
H(0,1,1,E_{p2},E_{p1'},1)=+E_{p_2}+E_{p_{1'}}+\frac{q^2}{d}=V_2.
\end{eqnarray}
We have
\begin{eqnarray}
E_p=1+\frac{1}{2}(\frac{q^2}{d+2(a+b)}-\frac{q^2}{d})=E_{p_2}=E_{p_{1'}}.
\end{eqnarray}
and it implies
\begin{eqnarray}
V_2=2+(\frac{q^2}{d+2(a+b)}-\frac{q^2}{d})+\frac{q^2}{d}=2+\frac{q^2}{d+2(a+b)},
\end{eqnarray}
\begin{eqnarray}
V_1=2+\frac{q^2}{d+a+b}+\frac{1}{2}(\frac{q^2}{d+2(a+b)}-\frac{q^2}{d}).
\end{eqnarray}
where
\begin{eqnarray}
V_2-V_1=(\frac{q^2}{d+2(a+b)}-\frac{q^2}{d+a+b})+\frac{1}{2}(\frac{q^2}{d+2(a+b)}-\frac{q^2}{d})<0,
\end{eqnarray}
since $(\frac{q^2}{d+2(a+b)}-\frac{q^2}{d+a+b})<0$ and $\frac{1}{2}(\frac{q^2}{d+2(a+b)}-\frac{q^2}{d})<0$ so condition $V_2-V_1<0$ is fulfilled.

\begin{figure}
\centering 
 \includegraphics[scale=0.6]{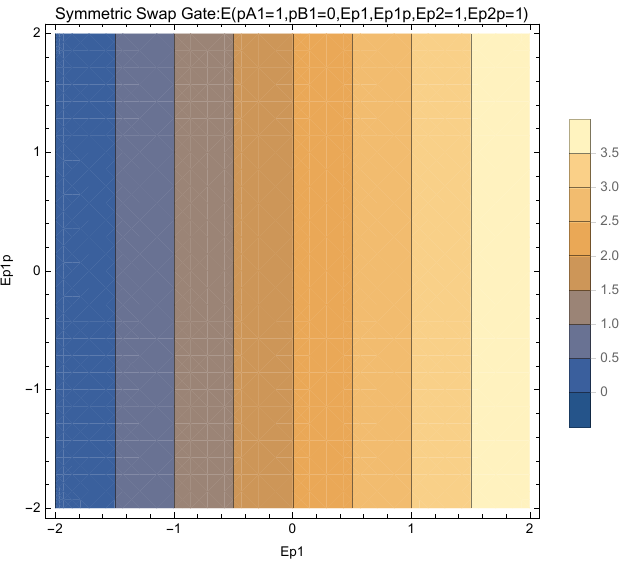}\includegraphics[scale=0.6]{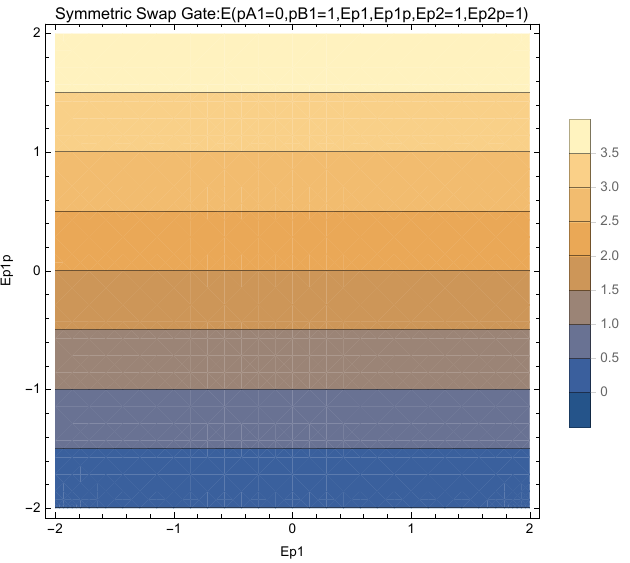}
 \includegraphics[scale=0.6]{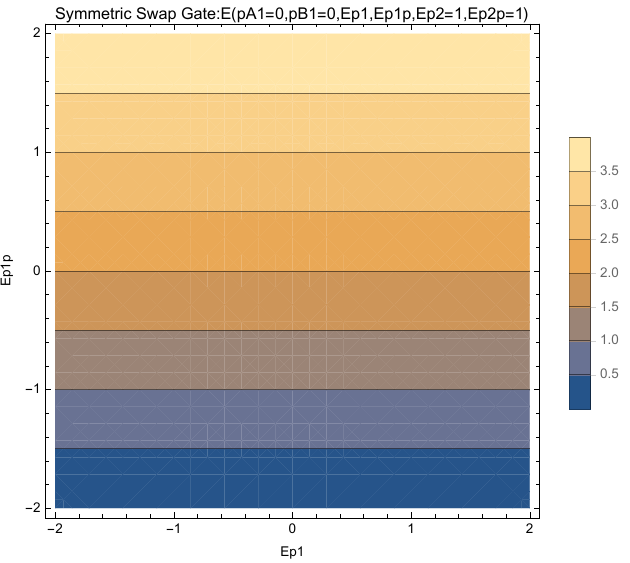}\includegraphics[scale=0.6]{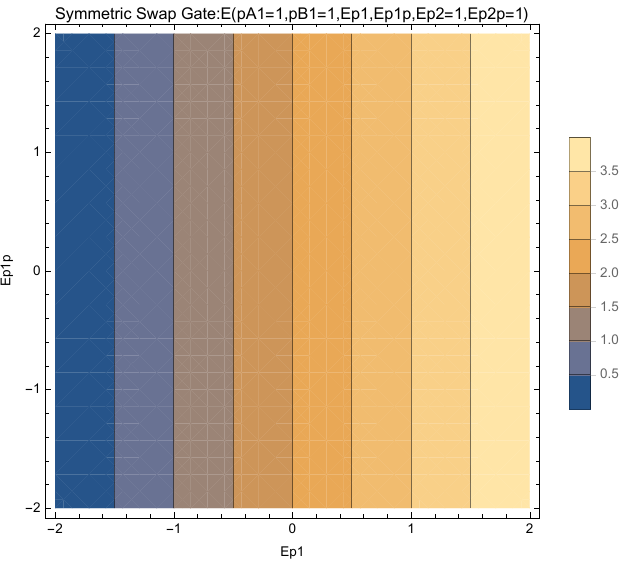}
 \caption{Dependence of quasi-classical Swap Gate Hamiltonian on polarizing potentials that are electrostatically controlled given by $E_{p1}$ and $E_{p1'}$ for fixed ($E_{p2}=1$,$E_{p2'}=1$).}
\label{qp1}
\end{figure}

\begin{figure}
\includegraphics[scale=0.6]{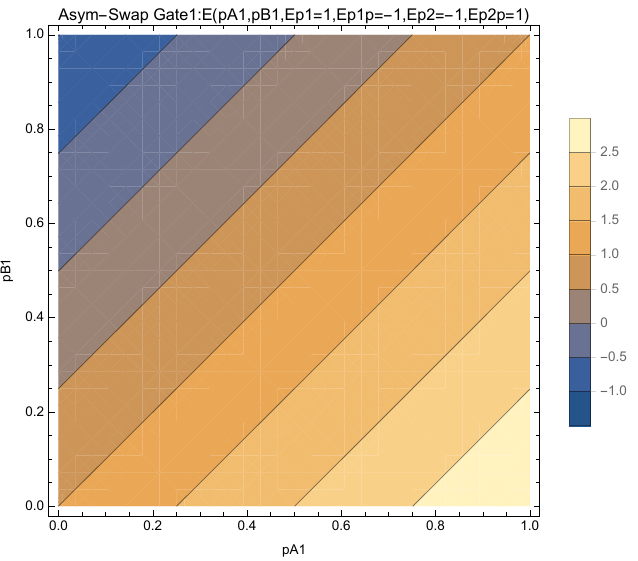}
\includegraphics[scale=0.6]{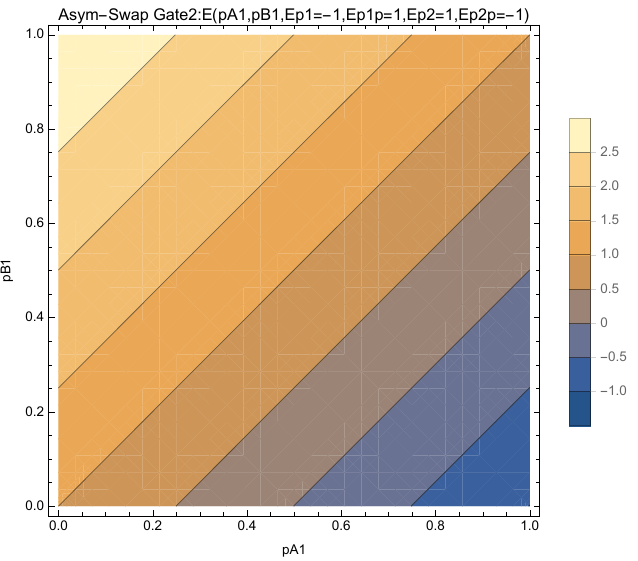}
\includegraphics[scale=0.6]{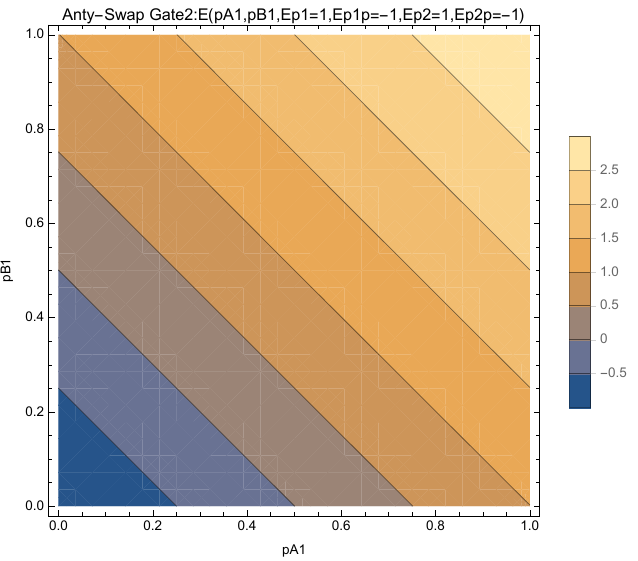}
\includegraphics[scale=0.6]{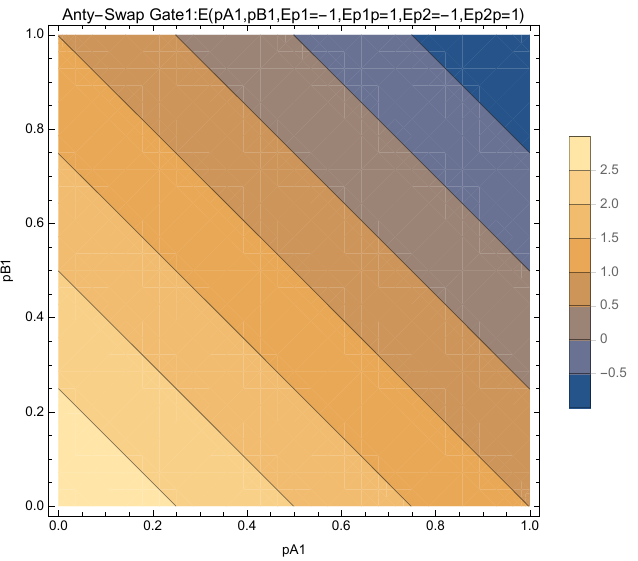}
\includegraphics[scale=0.6]{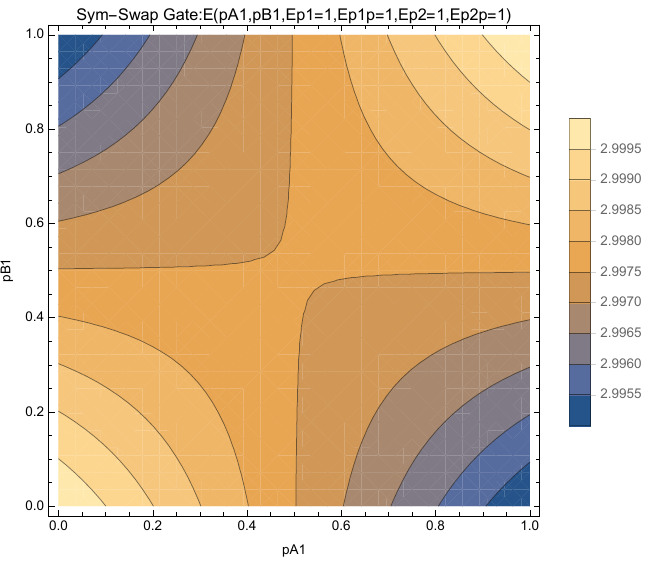} \\ $ $ \\ 
\includegraphics[scale=0.6]{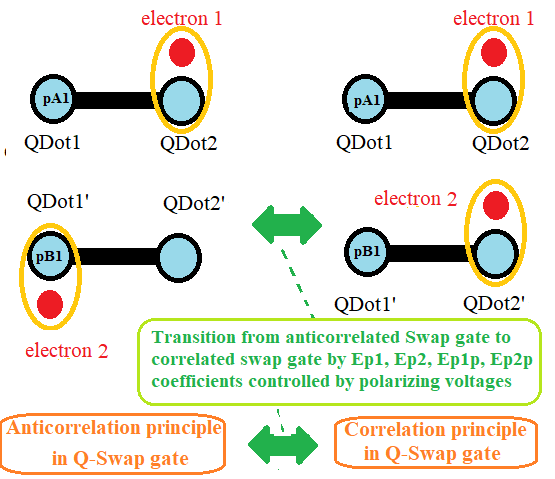}
 \caption{Role of polarizing voltages in transition from symmetric classical and quantum swap gate into asymmetric swap gate (upper middle and left) and into transition into anti-swap gate (bottom plots).}
 \label{qp2}
\end{figure}

\begin{figure}
\includegraphics[scale=0.5]{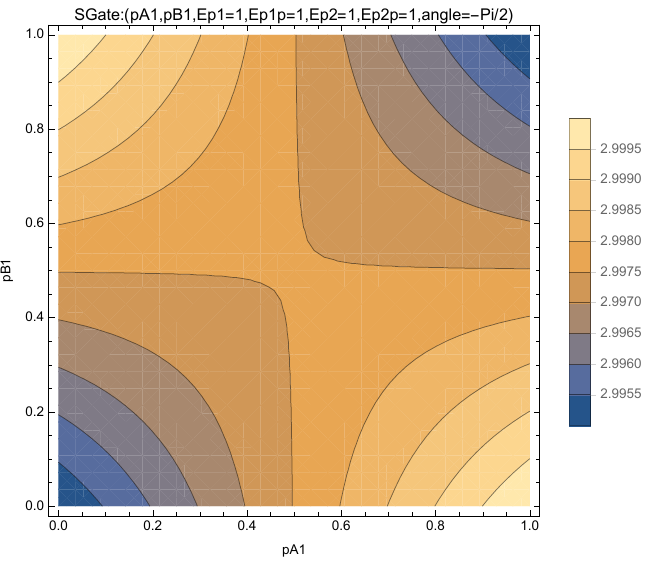}
\includegraphics[scale=0.4]{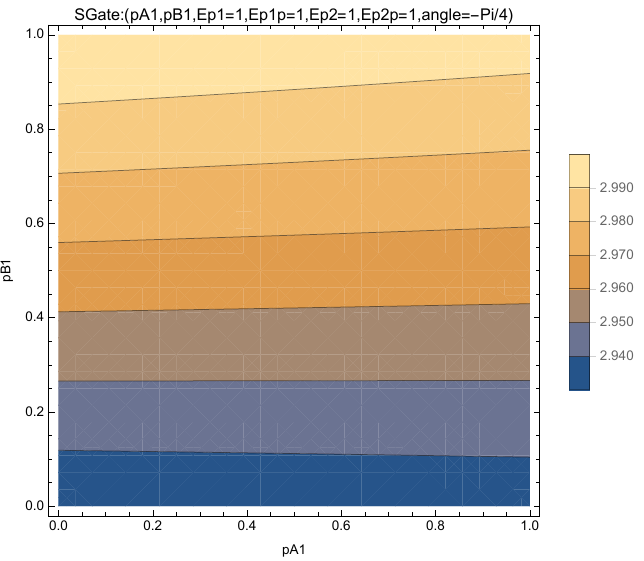}
\includegraphics[scale=0.4]{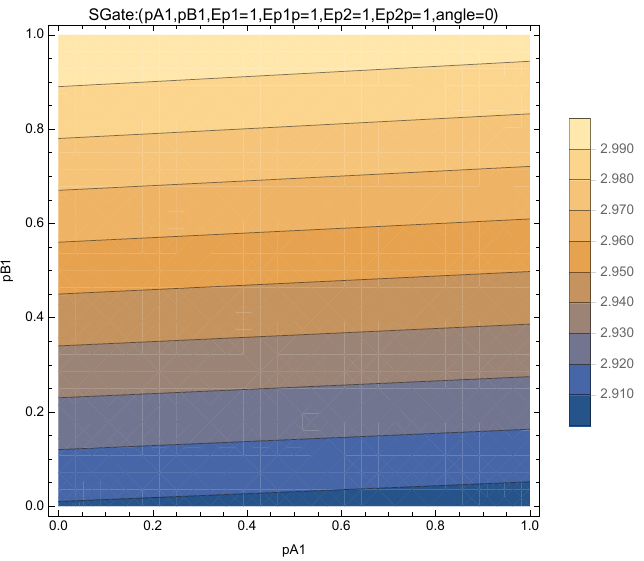}
\includegraphics[scale=0.4]{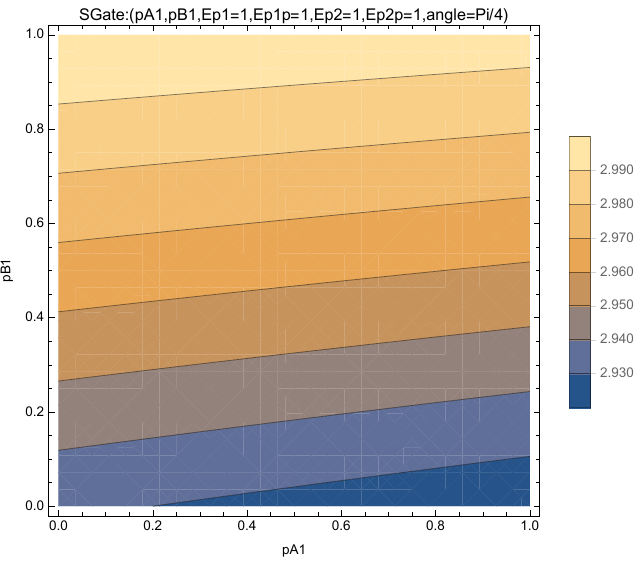}
\includegraphics[scale=0.5]{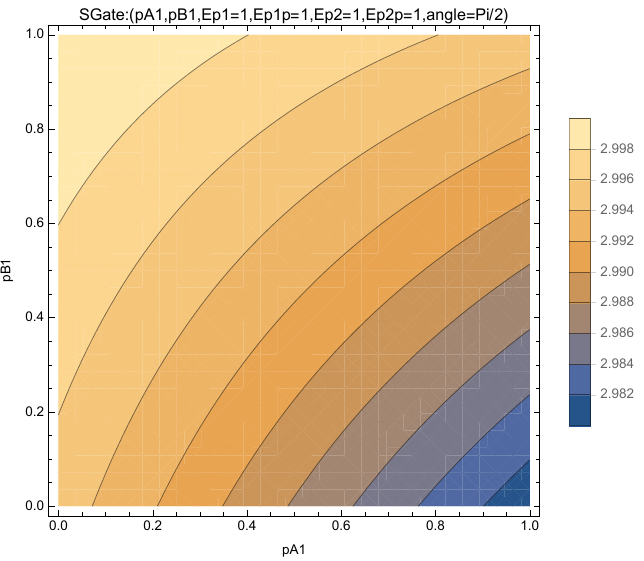}
\\
\includegraphics[scale=0.5]{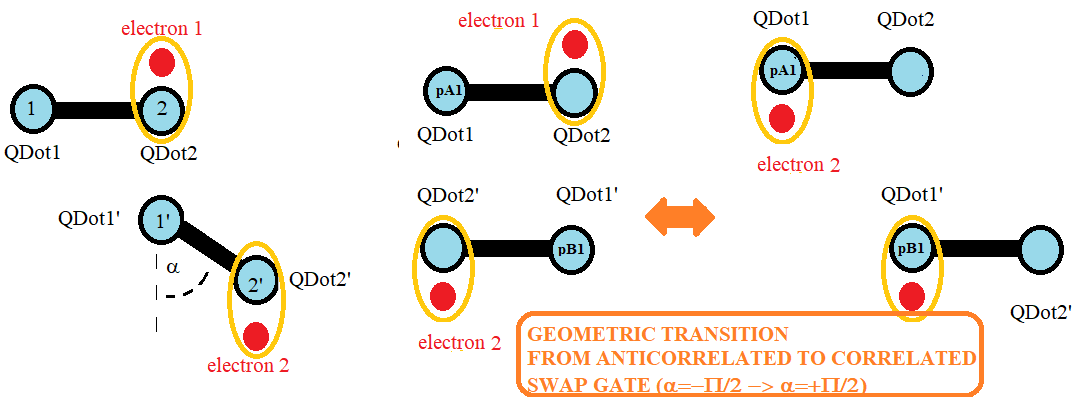}
\\
\includegraphics[scale=0.4]{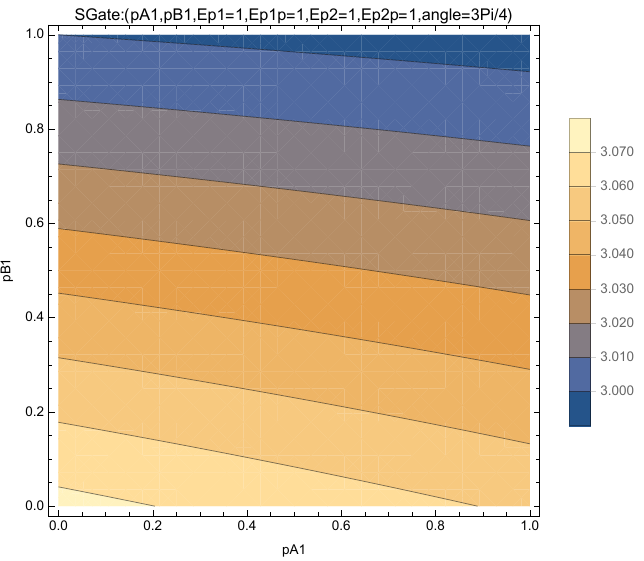}
\includegraphics[scale=0.4]{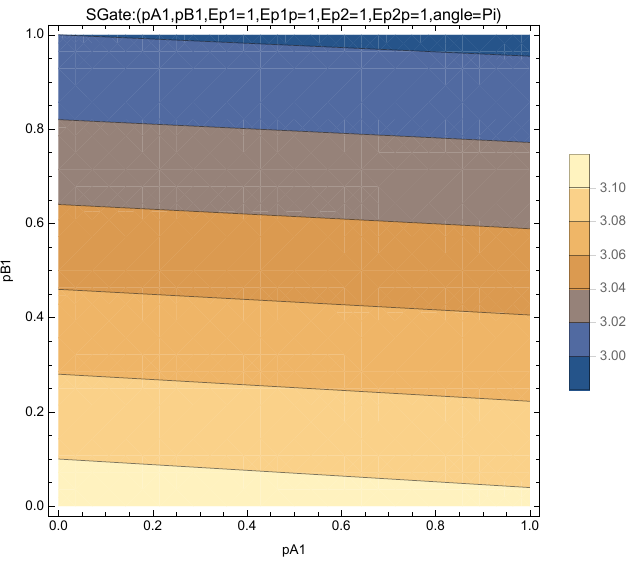}
\includegraphics[scale=0.4]{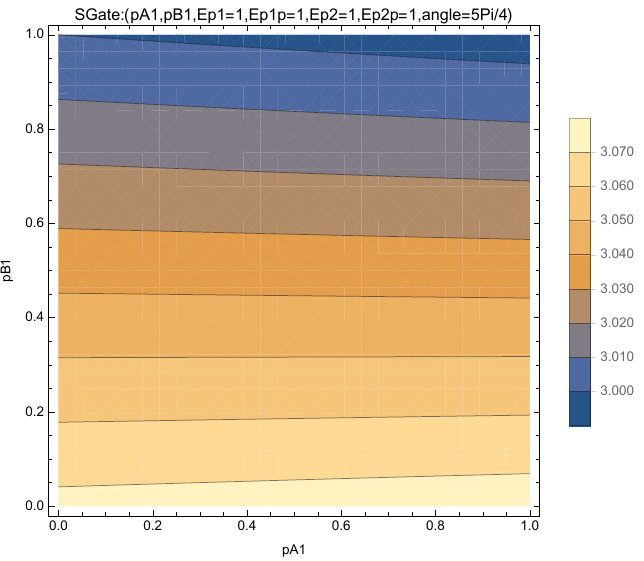}
\caption{Transition from classical/quantum swap gate into antiswap gate by changing the angle $\alpha$ from $-\Pi$ to $\Pi$. Minima in phenomenological Hamiltonian identifies the preferable logical state (states of ($p_{A1}=1,p_{B1}=0$) and ($p_{A1}=0,p_{B1}=1$) are preferred in the same way) into only one preferred state ($p_{A1}=1,p_{B1}=0$). Contour plots describing qubit-qubit Hamiltonian values in function of probabilities ($p_{A1}$,$p_{B1}$) of SWAP and ANTISWAP gates were magnified.}
\label{GeometricTransition}
\end{figure}
\begin{figure}
\centering 
 \includegraphics[scale=0.6]{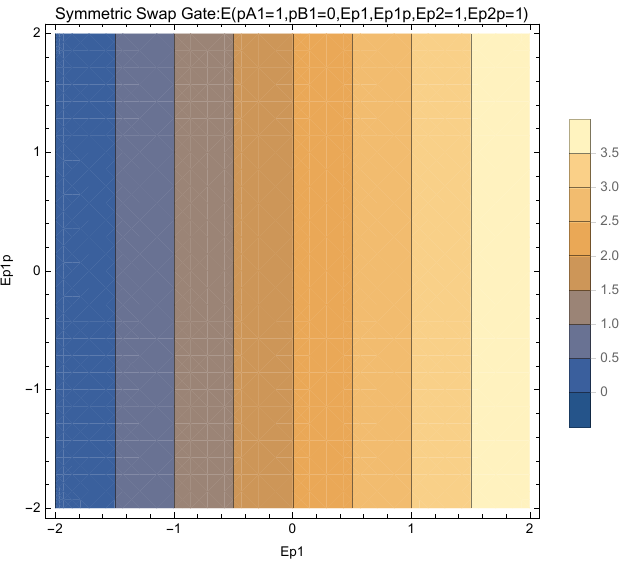}\includegraphics[scale=0.6]{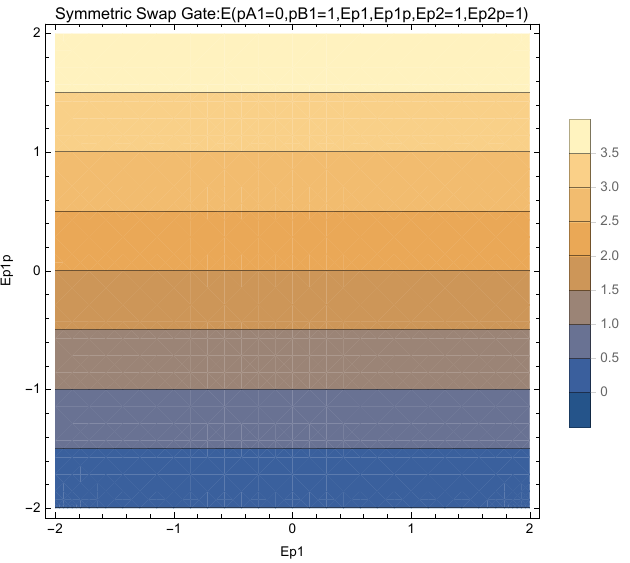}
 \includegraphics[scale=0.6]{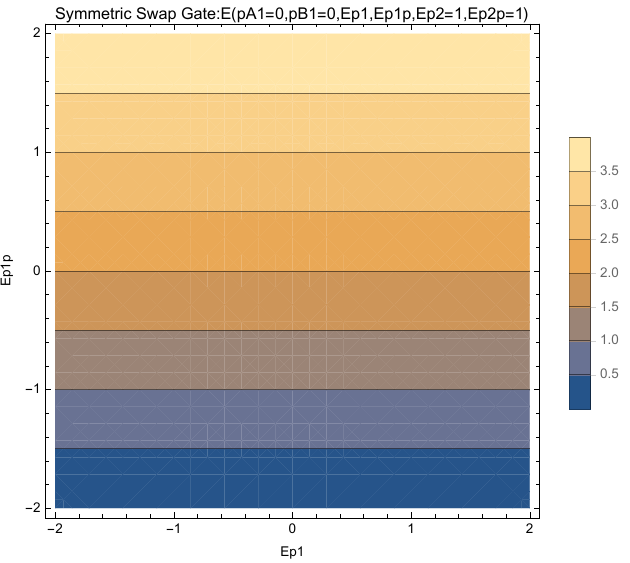}\includegraphics[scale=0.6]{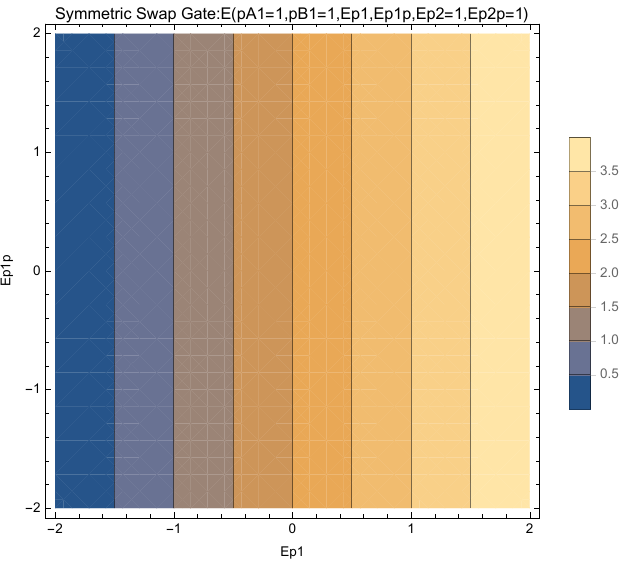}
 \caption{Dependence of quasi-classical Swap Gate Hamiltonian on polarizing potentials that are electrostatically controlled given by $E_{p1}$ and $E_{p1'}$ for fixed ($E_{p2}=1$,$E_{p2'}=1$).}
\label{qp1}
\end{figure}

\begin{figure}
\includegraphics[scale=0.6]{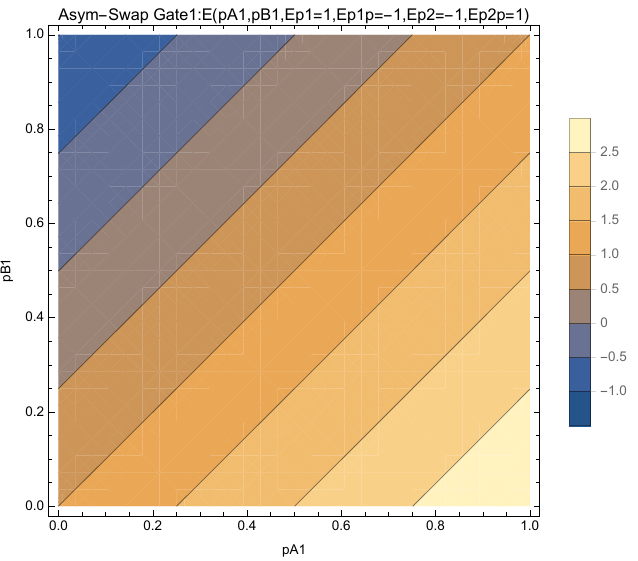}
\includegraphics[scale=0.6]{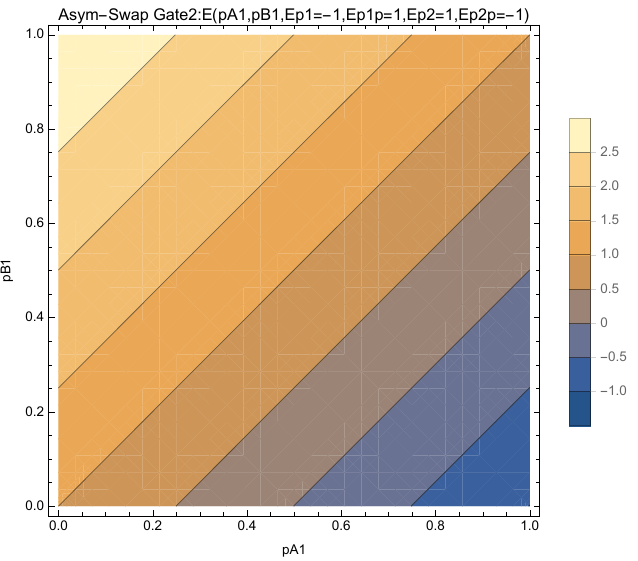}
\includegraphics[scale=0.6]{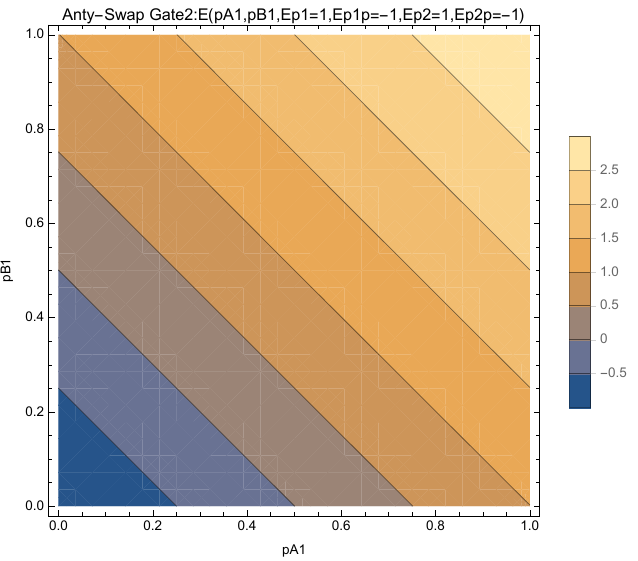}
\includegraphics[scale=0.6]{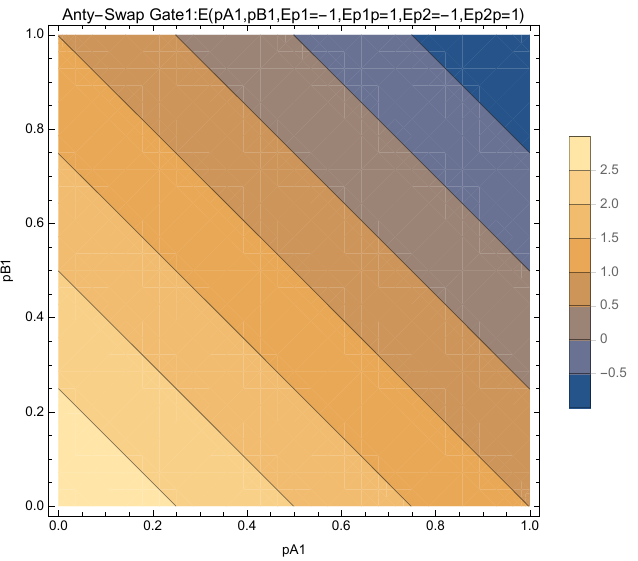}
\includegraphics[scale=0.6]{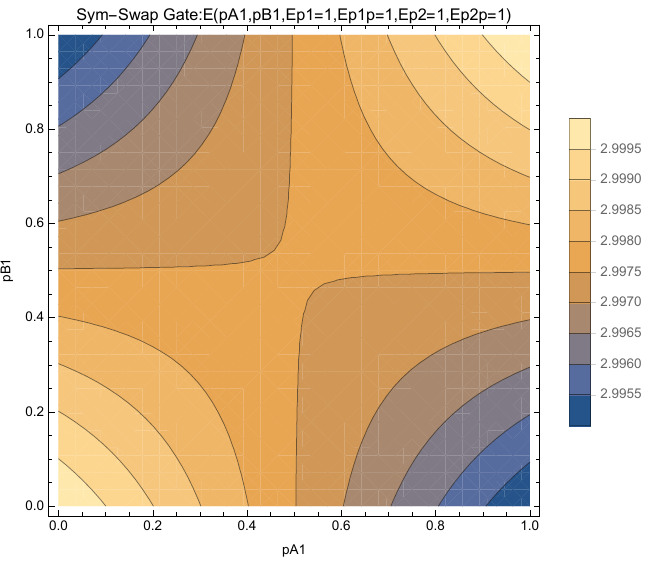} \\ $ $ \\
\includegraphics[scale=0.6]{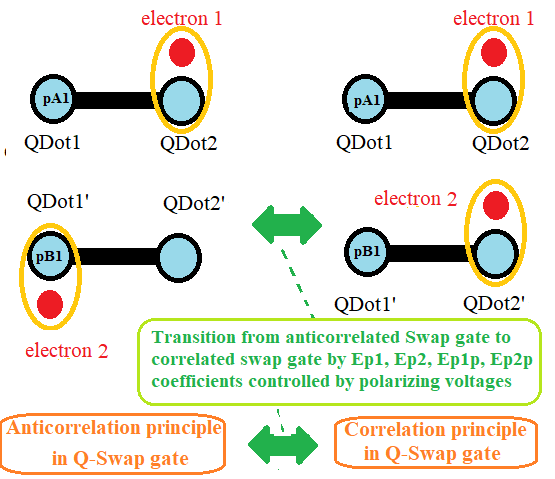}
 \caption{Role of polarizing voltages in transition from symmetric classical and quantum swap gate into asymmetric swap gate (upper middle and left) and into transition into anti-swap gate (bottom plots).}
 \label{qp2}
\end{figure}

\begin{figure}
\includegraphics[scale=0.5]{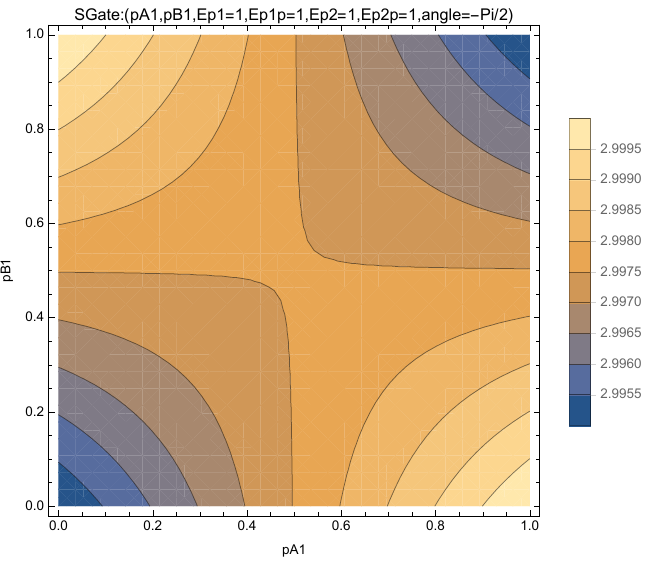}
\includegraphics[scale=0.4]{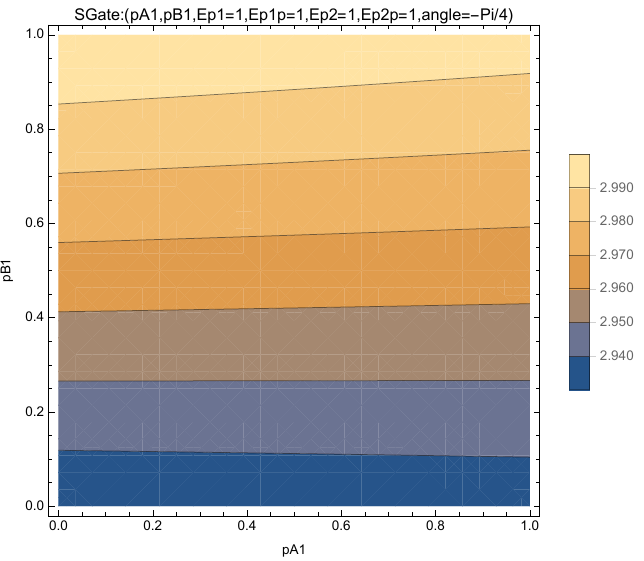}
\includegraphics[scale=0.4]{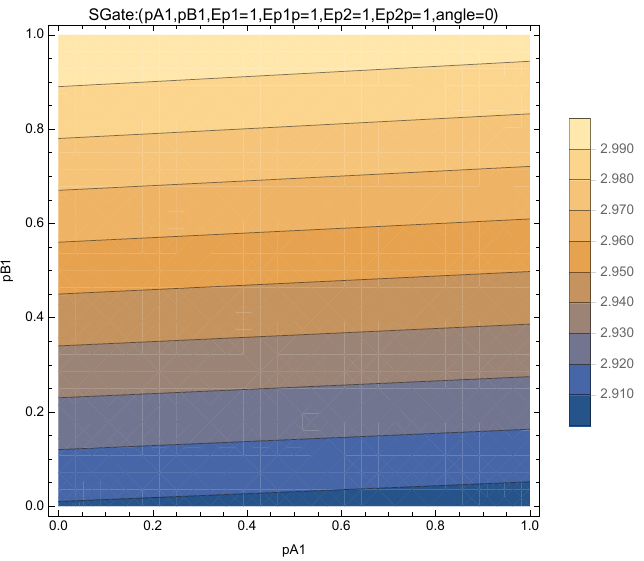}
\includegraphics[scale=0.4]{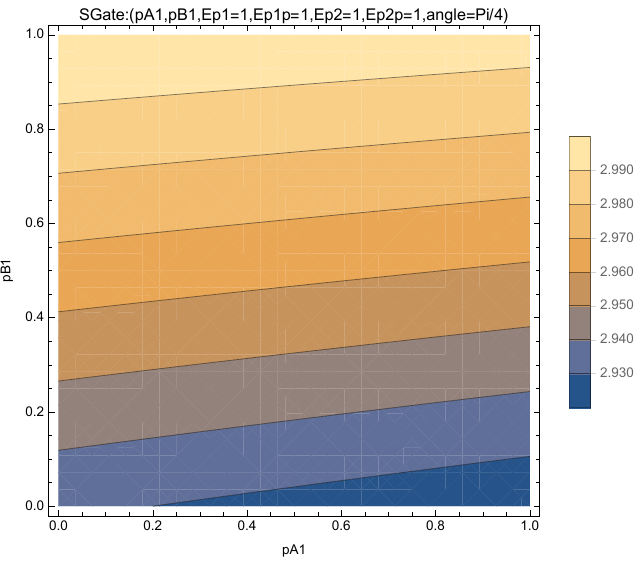}
\includegraphics[scale=0.5]{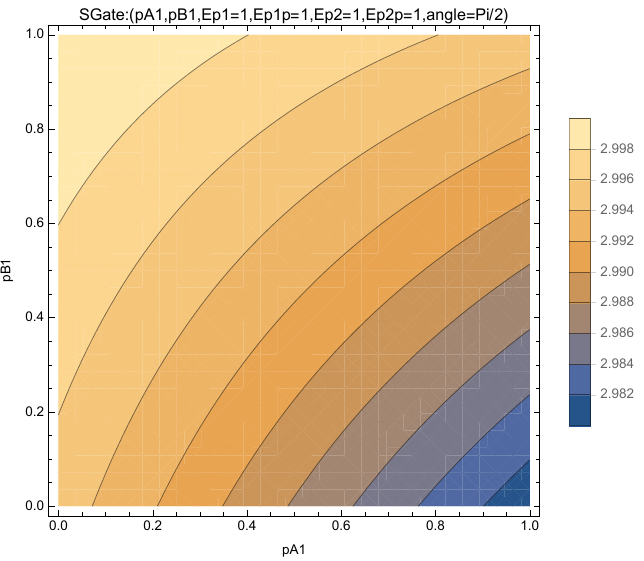} \\
\includegraphics[scale=0.5]{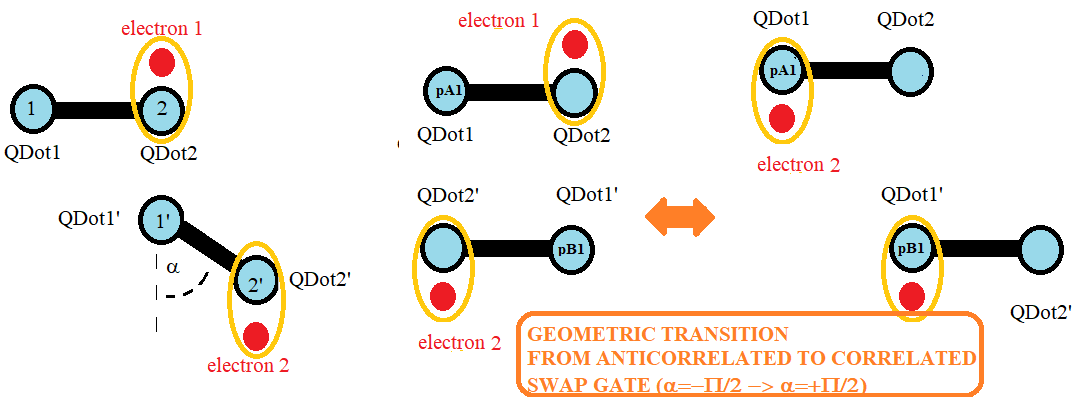} \\
\includegraphics[scale=0.4]{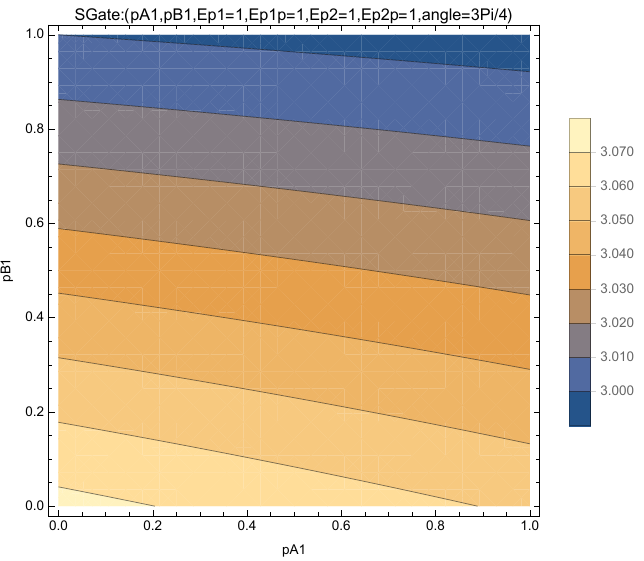}
\includegraphics[scale=0.4]{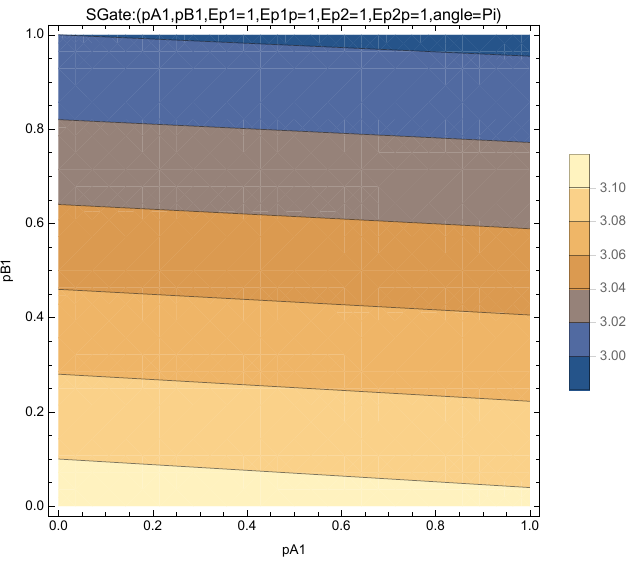}
\includegraphics[scale=0.4]{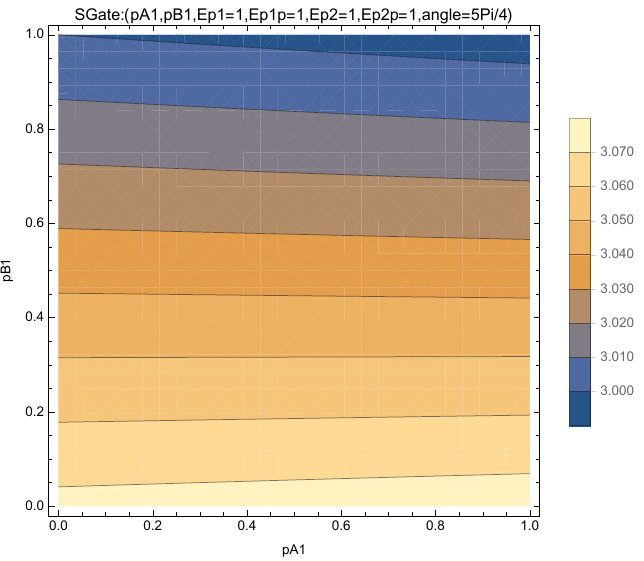}
\caption{Transition from classical/quantum swap gate into antiswap gate by changing the angle $\alpha$ from $-\Pi$ to $\Pi$. Minima in phenomenological Hamiltonian identifies the preferable logical state (states of ($p_{A1}=1,p_{B1}=0$) and ($p_{A1}=0,p_{B1}=1$) are preferred in the same way) into only one preferred state ($p_{A1}=1,p_{B1}=0$). Contour plots describing qubit-qubit Hamiltonian values in function of probabilities ($p_{A1}$,$p_{B1}$) of SWAP and ANTISWAP gates were magnified.}
\label{GeometricTransition}
\end{figure}

\begin{figure}
\centering
\includegraphics[scale=0.6]{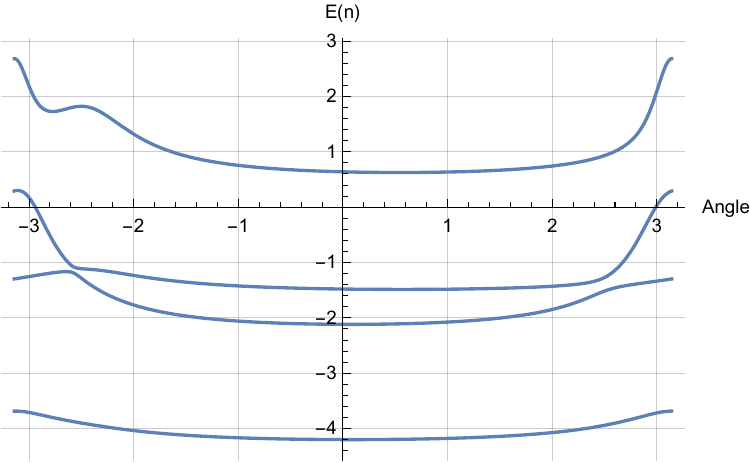}\includegraphics[scale=0.6]{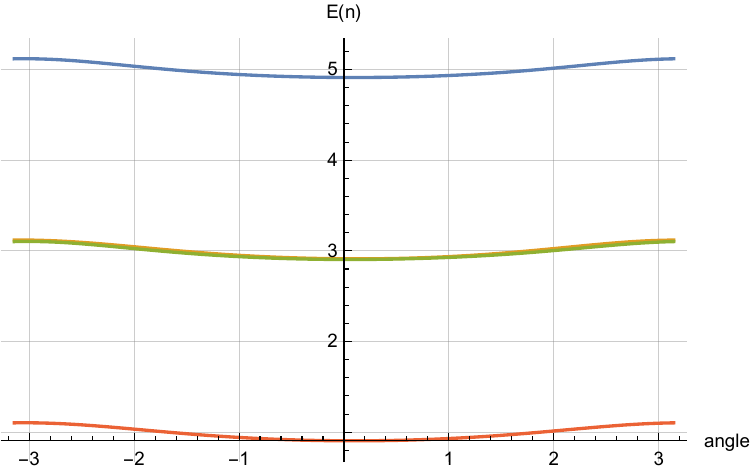}
\caption{Two main scenarios of dependence of eigenergies of electrostatically interacting qubit on angle. Left plot corresponds to parameters (d=1, a+b=0.8, q=1, $E_{p1}=1$,$E_{p2}=-1$,$E_{p1'}=-3$,$E_{p2'}= -2$, $t_{s12}=1$, $t_{s1'2'}=1$) and right plot has the same parameters except ( $E_{p1}=E_{p2}=E_{p1'}=E_{p2'}=1$, a+b=0.1). }
\label{qp4}
\end{figure}

\begin{figure}
\centering
\includegraphics[scale=0.38]{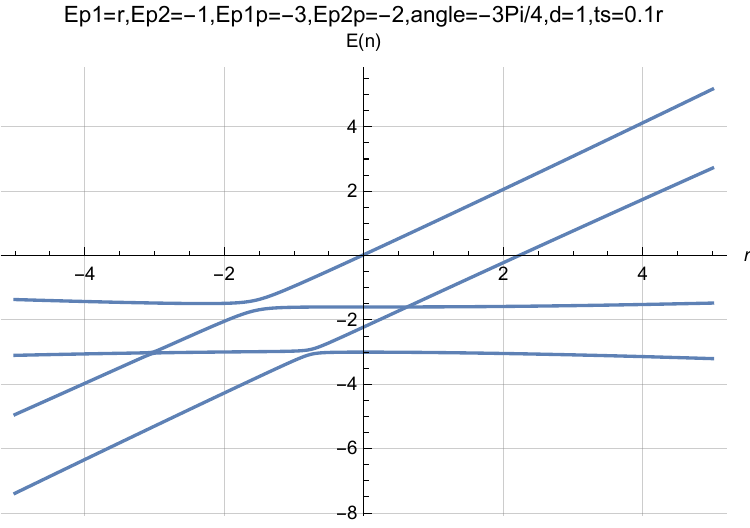}
\includegraphics[scale=0.38]{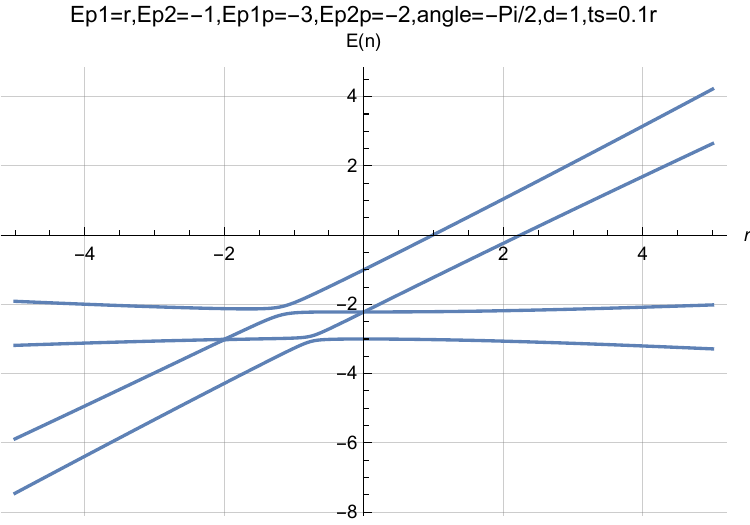}
\includegraphics[scale=0.38]{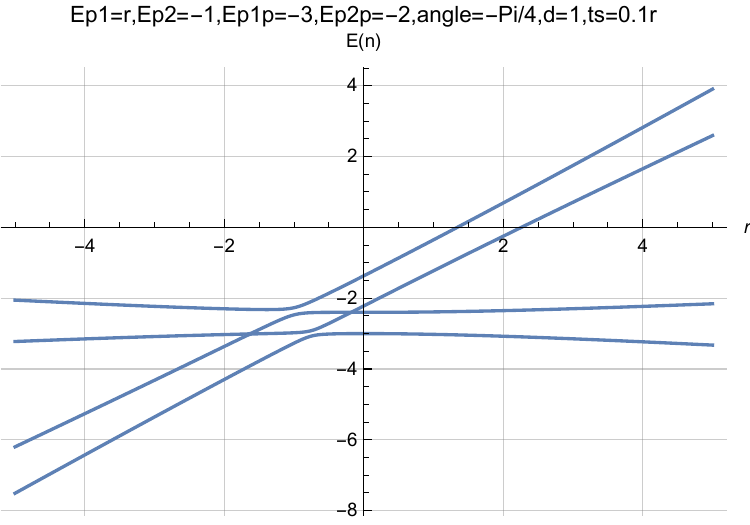}
\includegraphics[scale=0.38]{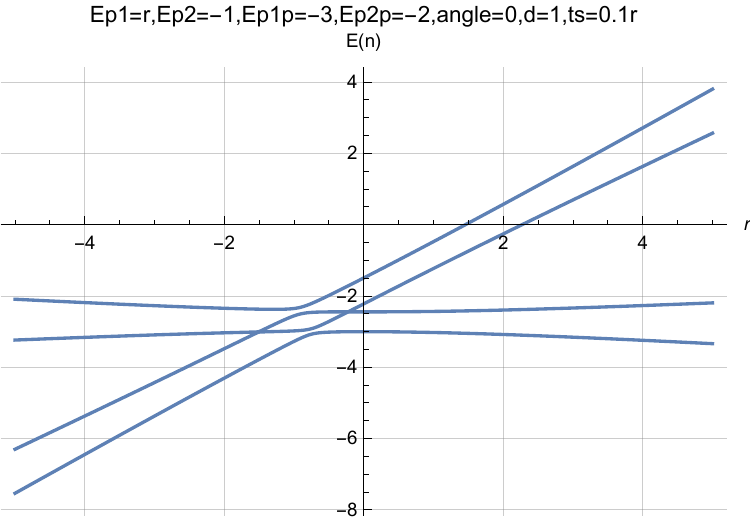}
\includegraphics[scale=0.38]{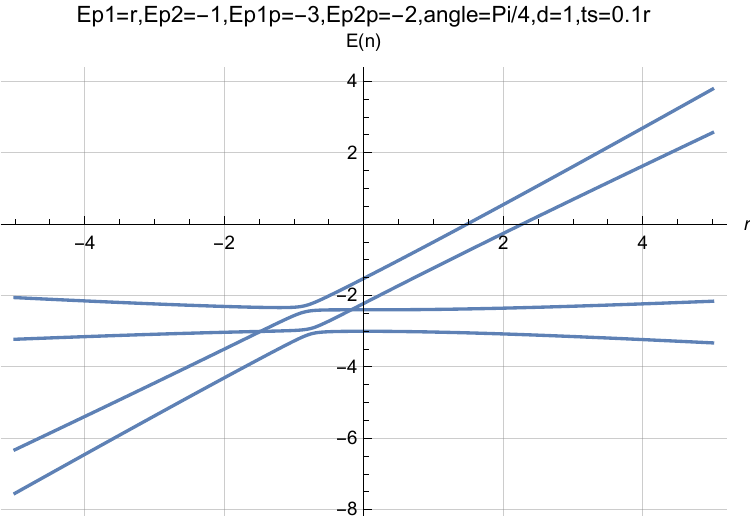}
\includegraphics[scale=0.38]{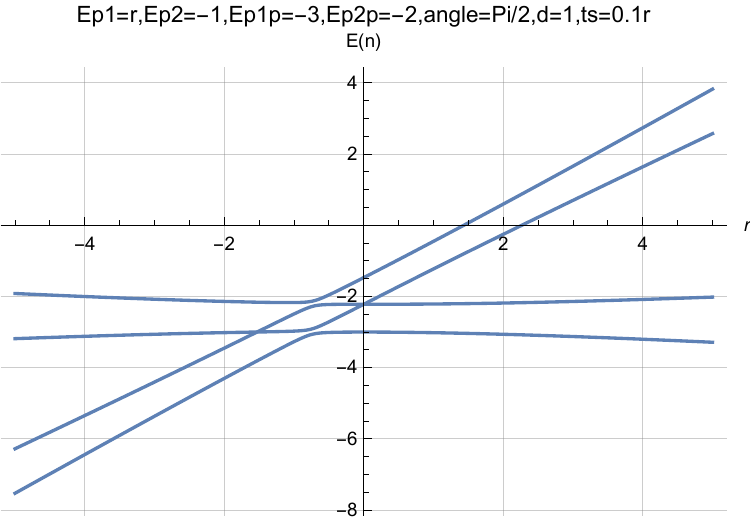}
\includegraphics[scale=0.38]{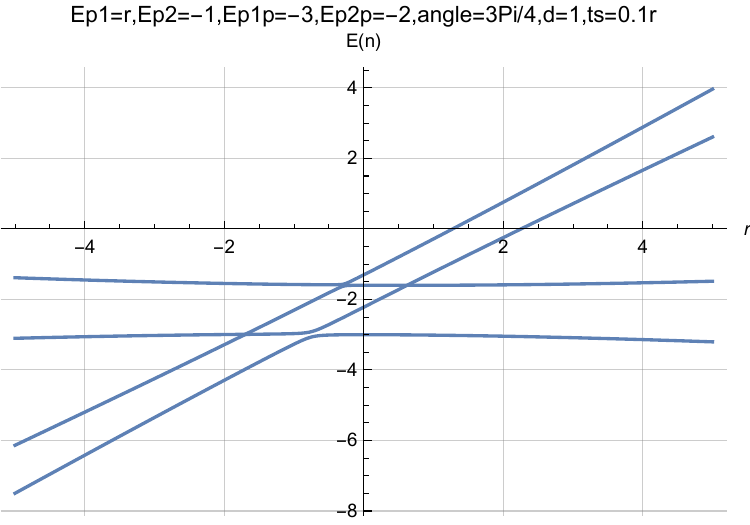}
\includegraphics[scale=0.38]{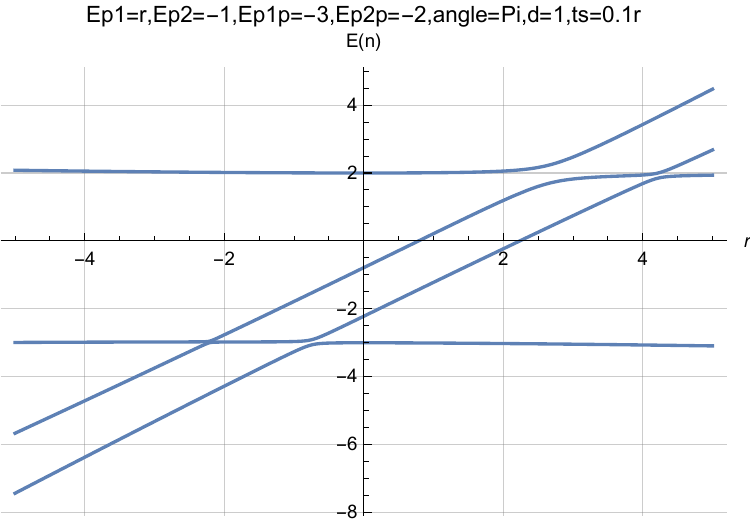}
\includegraphics[scale=0.38]{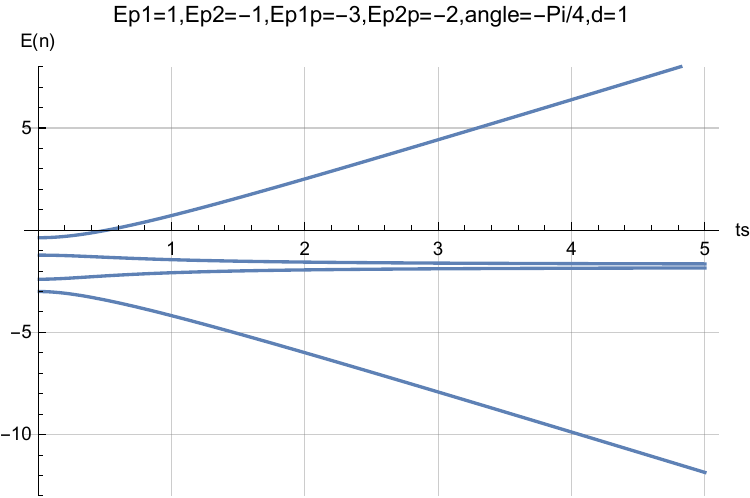}
\includegraphics[scale=0.38]{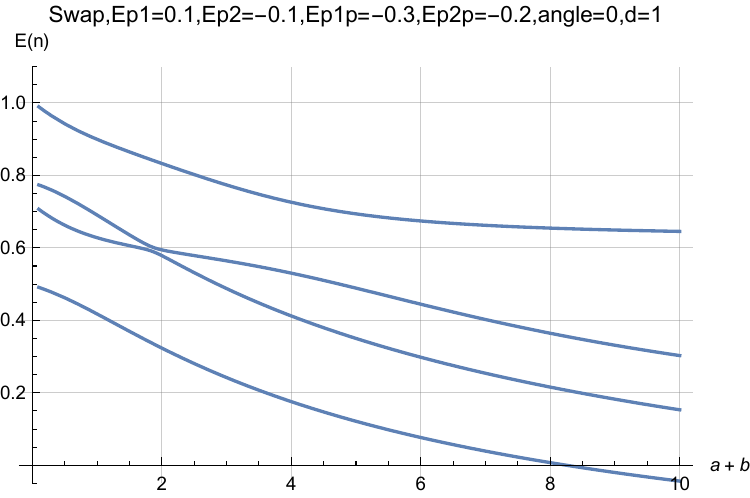}
\includegraphics[scale=0.38]{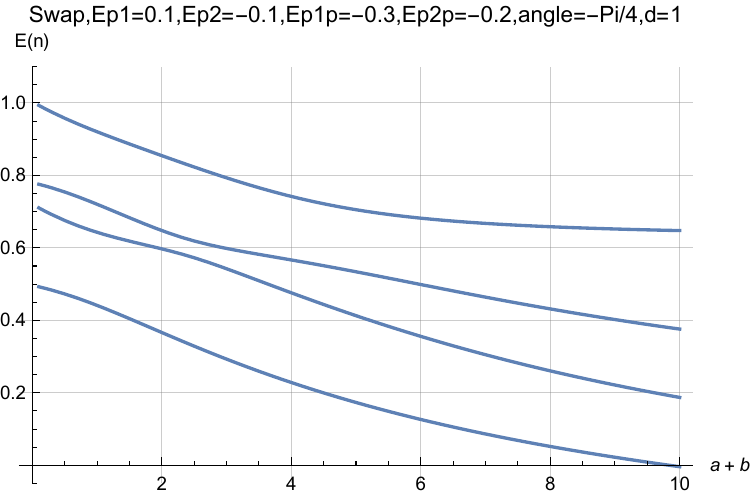}
\includegraphics[scale=0.38]{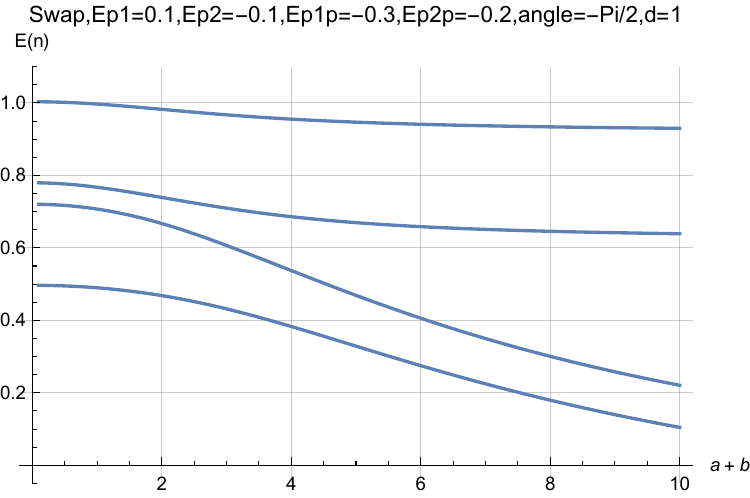}
\includegraphics[scale=0.38]{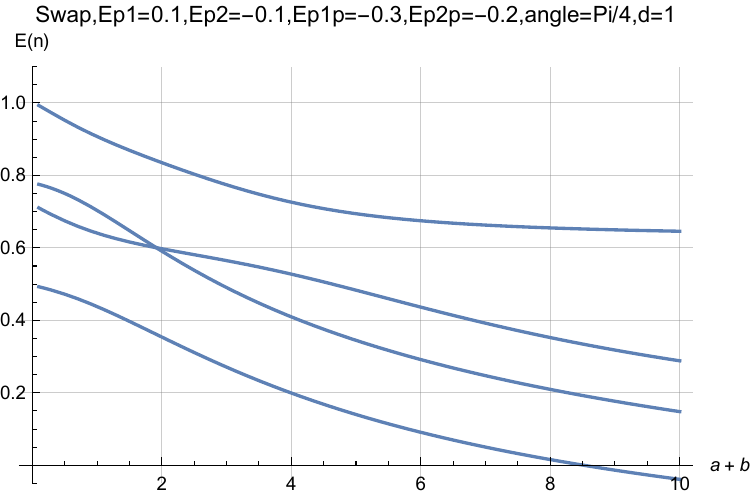}
\includegraphics[scale=0.38]{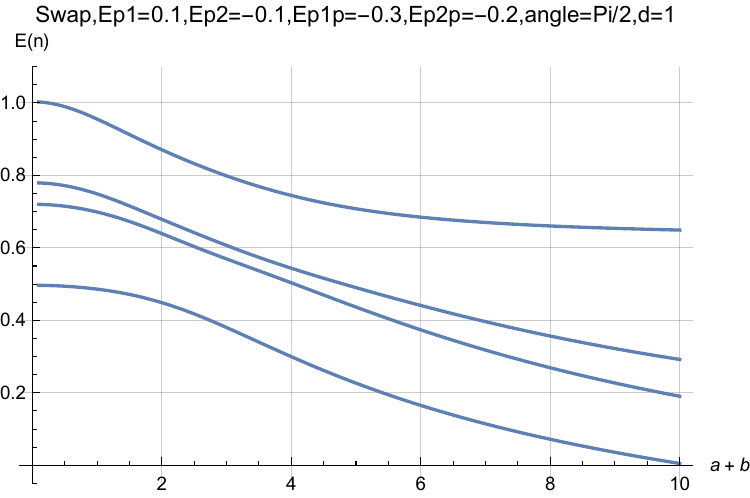}
\includegraphics[scale=0.38]{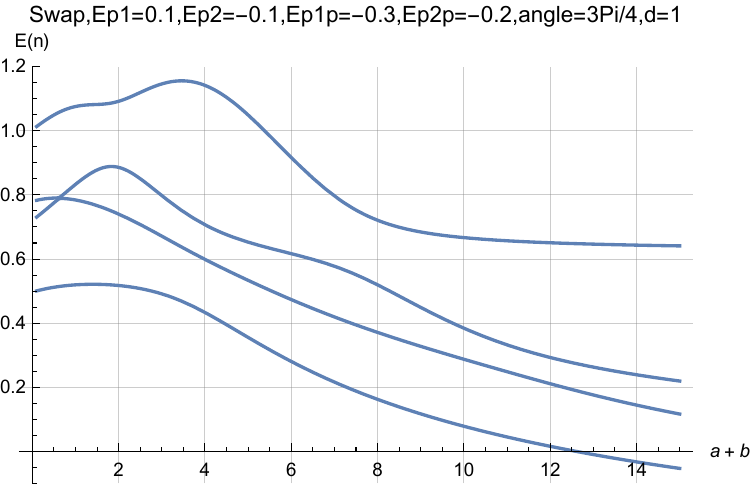}
\includegraphics[scale=0.38]{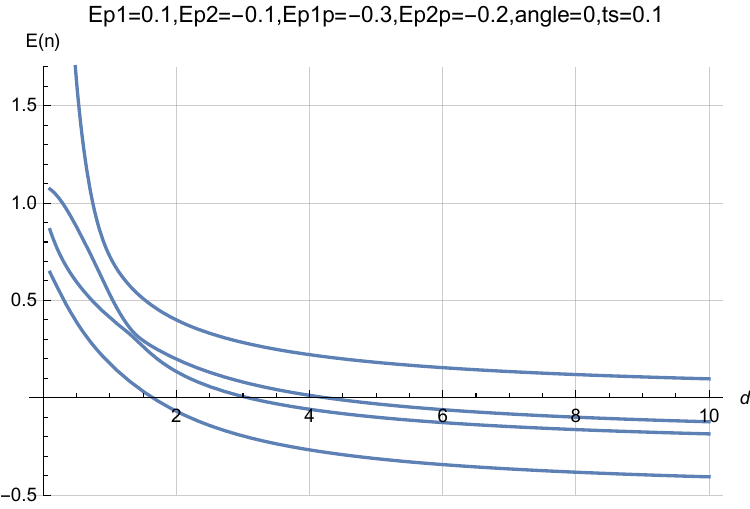}
\includegraphics[scale=0.38]{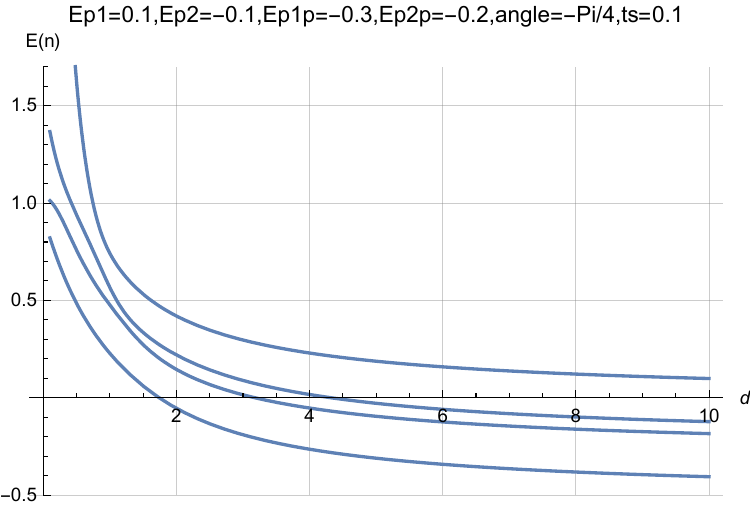}
\includegraphics[scale=0.38]{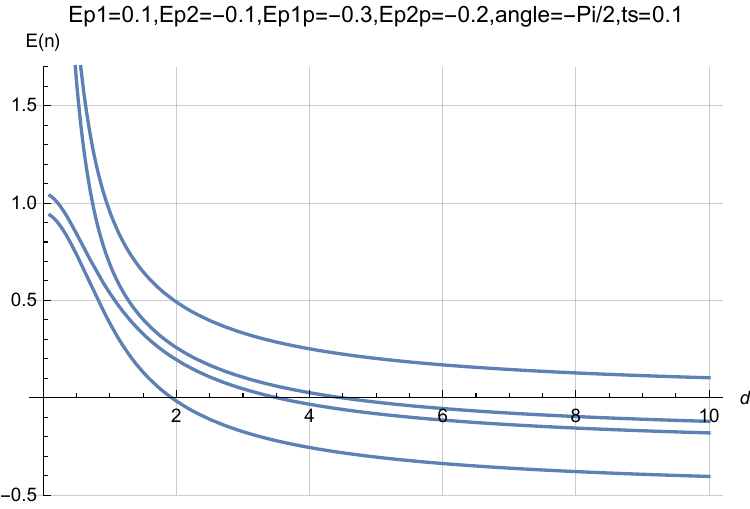}
\includegraphics[scale=0.38]{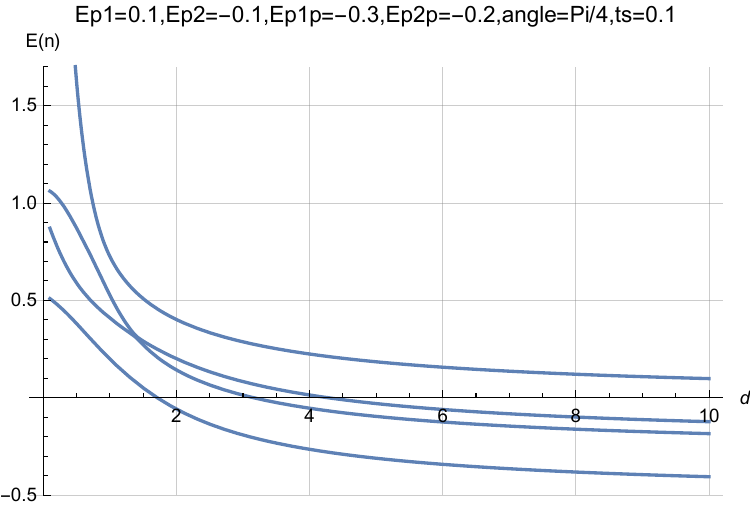}
\includegraphics[scale=0.38]{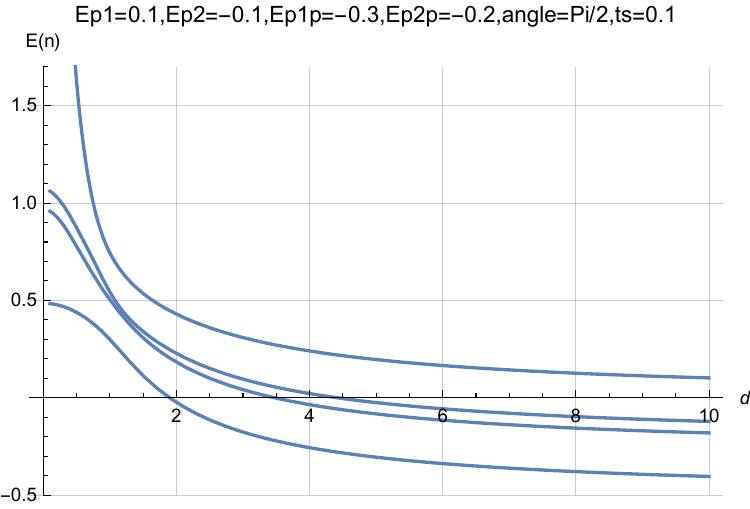}
\includegraphics[scale=0.38]{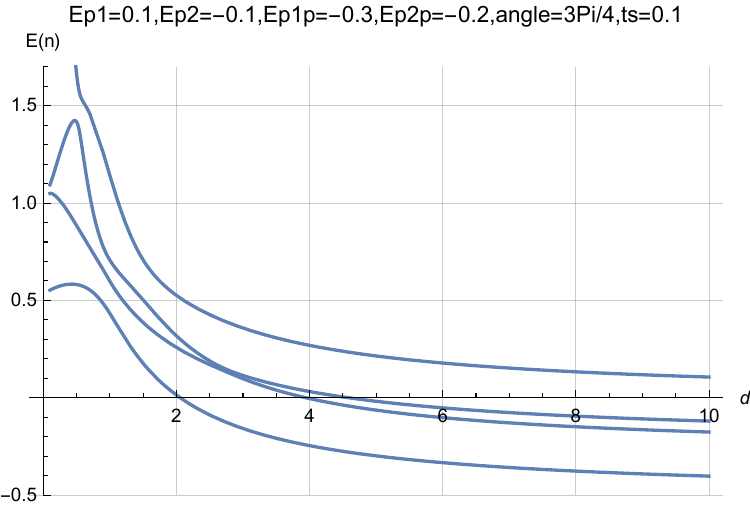}

	\caption{Energy spectrum of generalized electrostatic quantum gate in dependence on angle for fixed $E_{p1}$, $E_{p2}$, $E_{p1'}$, $E_{p2'}$, $t_{s12}=t_{s1'2'}=0.1$, $q=1$, $a+b=0.8$. No crossing of energy levels was reported.}
	\label{GQSwapGateSpectrum} 
\end{figure}

In the next section, we present a fundamental derivation of the tight-binding model from Schrödinger equation. 
\newpage 

\newpage

\section{From Eigenenergy Qubits to Position-Based Qubits}
Hamiltonian of the tight-binding model describing single-electron devices(as presented by Fujisawa \cite{Fujisawa} or Pomorski \cite{CurvedWannier}) uses Hamiltonian of position-based qubit $\hat{H}_P=\hat{R} \hat{H}_E \hat{R}^{-1}$ with rotation matrix $\hat{R}$, fictitious angle $\Theta_s$ and eigenenergy Hamiltonian $\hat{H}_E$. That in case of two energy level system can be expressed in the following form:
\begin{multline}
        \label{eq:Hp}
    \hat{H}_P = \begin{bmatrix}
        E_{P1} & t_s\cos{\theta_P}+\text{i}t_s \sin{\theta_P} \\
        t_s\cos{\theta_P}-\text{i}t_s\sin{\theta_P} & E_{P2}
    \end{bmatrix}
    =
    \begin{bmatrix}
        E_{P1} & t_s e^{+i\theta_P} \\
        t_se^{-i\theta_P} & E_{P2}
    \end{bmatrix} = \hat{R}[ E_1 \ket{E_1}\bra{E_1}+E_2 \ket{E_2}\bra{E_2} ] \hat{R}^{-1} \\ 
    =
    \begin{bmatrix}
        +\cos(\Theta_s) & +\sin(\Theta_s) \\
        -\sin(\Theta_s) & +\cos(\Theta_s)
    \end{bmatrix}
    \begin{bmatrix}
        E_{1} & 0 \\
        0 & E_{2}
    \end{bmatrix}
    \begin{bmatrix}
        +\cos(\Theta_s) & -\sin(\Theta_s) \\
        +\sin(\Theta_s) & +\cos(\Theta_s)
    \end{bmatrix}= \hat{R}\hat{H}_E \hat{R}^{-1}= \\
    =
     \begin{bmatrix}
        E_{1}\cos(\Theta_s)^2+E_{2}\sin(\Theta_s)^2 & (E_{2}-E_{1})\sin(\Theta_s)\cos(\Theta_s) \\
        +(E_{2}-E_{1})\sin(\Theta_s)\cos(\Theta_s)  & E_{2}\cos(\Theta_s)^2+E_{1}\sin(\Theta_s)^2
    \end{bmatrix}
    =
    \begin{bmatrix}
        E_{1}+(E_{2}-E_{1})\sin(\Theta_s)^2 & \frac{1}{2}(E_{2}-E_{1})\sin(2\Theta_s) \\
        +\frac{1}{2}(E_{2}-E_{1})\sin(2\Theta_s) & (E_{2}-E_{1})\cos(\Theta_s)^2+E_{1}
    \end{bmatrix}=   \\
    =
      \begin{bmatrix}
        \frac{1}{2}(E_{2}+E_{1}) & 0 \\
        0 & \frac{1}{2}(E_{2}+E_{1})
    \end{bmatrix} +
      \begin{bmatrix}
         E_{1}\cos(\Theta_s)^2+E_{2}\sin(\Theta_s)^2-\frac{E_1+E_2}{2} & \frac{1}{2}(E_{2}-E_{1})\sin(2\Theta_s) \\
        +\frac{1}{2}(E_{2}-E_{1})\sin(2\Theta_s) &   E_{1}\sin(\Theta_s)^2+E_{2}\cos(\Theta_s)^2-\frac{E_1+E_2}{2}
    \end{bmatrix}=
     \\
     \frac{E_1+E_2}{2}\hat{I}
     +
     \begin{bmatrix}
         E_{1}(1-\sin(\Theta_s)^2)+E_{2}\sin(\Theta_s)^2-\frac{E_1+E_2}{2} & \frac{1}{2}(E_{2}-E_{1})\sin(2\Theta_s) \\
        +\frac{1}{2}(E_{2}-E_{1})\sin(2\Theta_s) &   E_{1}\sin(\Theta_s)^2+E_{2}(1-\sin(\Theta_s))^2-\frac{E_1+E_2}{2}
    \end{bmatrix}        .
\end{multline}
We continue elementary mathematical steps 
\begin{multline}
\hat{H}_P=
 \begin{bmatrix}
         E_{1}(1-\sin(\Theta_s)^2)+E_{2}\sin(\Theta_s)^2-\frac{E_1+E_2}{2} & \frac{1}{2}(E_{2}-E_{1})\sin(2\Theta_s) \\
        +\frac{1}{2}(E_{2}-E_{1})\sin(2\Theta_s) &   E_{1}\sin(\Theta_s)^2+E_{2}(1-\sin(\Theta_s))^2-\frac{E_1+E_2}{2}
    \end{bmatrix} + \frac{1}{2}(E_{2}+E_{1})\hat{I} = \nonumber \\ =
    \frac{1}{2}(E_{2}+E_{1})\hat{I}\hat{\sigma}_0 
    + \frac{1}{2}(E_{2}-E_{1})\sin(2\Theta_s)
    \hat{\sigma}_1 + \Big[ (E_{2}-E_{1})\sin(\Theta_s)^2-\frac{E_2-E_1}{2} \Big] \hat{\sigma}_3= 
     \\
     =
    \begin{bmatrix}
         +[(E_{2}-E_{1})\sin(\Theta_s)^2-\frac{E_2-E_1}{2}] & \frac{1}{2}(E_{2}-E_{1})\sin(2\Theta_s) \\
        +\frac{1}{2}(E_{2}-E_{1})\sin(2\Theta_s) &   -[(E_{2}-E_{1})\sin(\Theta_s)^2-\frac{E_2-E_1}{2}]
    \end{bmatrix}+ \frac{1}{2}(E_{2}+E_{1})\hat{I} = \nonumber \\ 
    =(E_{2}-E_{1})
     \begin{bmatrix}
         +[\sin(\Theta_s)^2-\frac{1}{2}] & \frac{1}{2}\sin(2\Theta_s) \\
        +\frac{1}{2}\sin(2\Theta_s) &   -[\sin(\Theta_s)^2-\frac{1}{2}]
    \end{bmatrix}+ \frac{1}{2}(E_{2}+E_{1})\hat{I}= \nonumber \\
   = (E_{2}-E_{1})
     \begin{bmatrix}
         +[\sin(\Theta_s)^2-\frac{1}{2}] & \pm \sin(\Theta_s)\sqrt{1-\sin(\Theta_s)^2} \\
        \pm \sin(\Theta_s)\sqrt{1-\sin(\Theta_s)^2} &   -[\sin(\Theta_s)^2-\frac{1}{2}]
    \end{bmatrix}+ \frac{1}{2}(E_{2}+E_{1})\hat{I} = \nonumber \\
    = (E_{2}-E_{1})\sin(\Theta_s)^2
     \begin{bmatrix}
         +[1-\frac{1}{2\sin(\Theta_s)^2}] & \pm \sqrt{\frac{1}{\sin(\Theta_s)^2}-1} \\
        \pm \sqrt{\frac{1}{\sin(\Theta_s)^2}-1} &   -[1-\frac{1}{2\sin(\Theta_s)^2}]
    \end{bmatrix}+ \frac{1}{2}(E_{2}+E_{1})\hat{I} 
    = \nonumber \\
    = \frac{(E_{2}-E_{1})}{2}\frac{1}{w}
     \begin{bmatrix}
         +[1-w] & \pm \sqrt{2}\sqrt{w-\frac{1}{2}} \\
        \pm \sqrt{2}\sqrt{w-\frac{1}{2}} &   -[1-w]
    \end{bmatrix}+ \frac{1}{2}(E_{2}+E_{1})\hat{I},
    w=\frac{1}{2 \sin(\Theta_s)^2}, 
    \nonumber \\
\end{multline}
where we have used the property of maximization of
the presence of the left Wannier wavefunction $w_L(x)=\cos(\Theta_s)\psi_{E1}(x)+\sin(\Theta_s)\psi_{E2}(x)$ in the region from $-\infty$ to $x_0$ encoded in minimization of functional $\int_{-\infty}^{x_0}|w_L(x)|^2dx=F[\Theta_s]$, so it is achieved that $\frac{d}{d\Theta_s}F[\Theta_s]=0$. 
Having the logical state zero assigned from $-\infty$ to $x_0$ (medium distance between 2 quantum dots)
we obtain the formula for angle $\Theta_s$ as given by \cite{CurvedWannier} and the expression
\begin{eqnarray}
      \Theta_s=\frac{1}{2}\arctan{\Bigg[ \frac{\int_{-\infty}^{+x_0}\frac{1}{2}[\psi_{E1}(x)\psi_{E2}^{*}(x)+\psi_{E1}^{*}(x)\psi_{E2}(x)]dx}{\int_{-\infty}^{+x_0}dx[|\psi_{E2}(x)|^2-|\psi_{E1}(x)|^2]} \Bigg]}.
  \end{eqnarray}
  It is useful to parameterize $\hat{H}_P$ by the introduced quantity $w=\frac{1}{2 \sin(\Theta_s)^2}$ and two eigenenergies $E_1$ and $E_2$. It is proper to underline that 
 \begin{eqnarray}
    \hat{H}_E=
     \begin{bmatrix}
        [ - \frac{\hbar^2}{2m}\frac{d^2}{dx^2} + V_{eff}(x)  ] & 0 \\
        0 & [ - \frac{\hbar^2}{2m}\frac{d^2}{dx^2} + V_{eff}(x)  ]
    \end{bmatrix}
    =
     \begin{bmatrix}
        E_{1} & 0 \\
        0 & E_{2}
    \end{bmatrix}=E_1\ket{E_1}\bra{E_1}+E_2\ket{E_2}\bra{E_2}, \nonumber \\
      \hat{H}_E \ket{\psi_{E}}=
     \hat{H}_E 
     \begin{bmatrix}
       \sqrt{p_{E1}(t)}e^{i \gamma_1(t)}\psi_{E1}(x) \\
         \sqrt{p_{E2}(t)}e^{i \gamma_2(t)}
         \psi_{E2}(x)
    \end{bmatrix} = i \hbar \frac{d}{dt}
    \begin{bmatrix}
       \sqrt{p_{E1}(t)}e^{i \gamma_1(t)}\psi_{E1}(x) \\
         \sqrt{p_{E2}(t)}e^{i \gamma_2(t)}\psi_{E2}(x)
    \end{bmatrix} = i \hbar \frac{d}{dt}\ket{\psi_{E}}= \nonumber \\
    =
    \begin{bmatrix}
        +\cos(\Theta_s) & -\sin(\Theta_s) \\
        +\sin(\Theta_s) & +\cos(\Theta_s)
    \end{bmatrix}
     \begin{bmatrix}
        +\cos(\Theta_s) & +\sin(\Theta_s) \\
        -\sin(\Theta_s) & +\cos(\Theta_s)
    \end{bmatrix}i \hbar \frac{d}{dt}
    \begin{bmatrix}
       \sqrt{p_{E1}(t)}e^{i \gamma_1(t)}\psi_{E1}(x) \\
         \sqrt{p_{E2}(t)}e^{i \gamma_2(t)}\psi_{E2}(x)
    \end{bmatrix}, \nonumber \\
    \ket{\psi}_P=
    \begin{bmatrix}
       w_L(x) \\
       w_R(x)
    \end{bmatrix}=
    \begin{bmatrix}
        +\cos(\Theta_s) & +\sin(\Theta_s) \\
        -\sin(\Theta_s) & +\cos(\Theta_s)
    \end{bmatrix}
    \begin{bmatrix}
      \psi_{E1}(x) \\
      \psi_{E2}(x)
    \end{bmatrix}, 
     \ket{\psi}_P=
    \begin{bmatrix}
       \psi_{E1}(x) \\
       \psi_{E2}(x)
    \end{bmatrix}=
    \begin{bmatrix}
        +\cos(\Theta_s) & -\sin(\Theta_s) \\
        +\sin(\Theta_s) & +\cos(\Theta_s)
    \end{bmatrix}
    \begin{bmatrix}
      w_L(x) \\
      w_R(x)
    \end{bmatrix}, \nonumber \\
     \hat{H}_E \ket{\psi}_E=
    \hat{H}_E 
     \begin{bmatrix}
       \sqrt{p_{E1}(t)}e^{i \gamma_1(t)}\psi_{E1}(x) \\
         \sqrt{p_{E2}(t)}e^{i \gamma_2(t)}
         \psi_{E2}(x)
    \end{bmatrix} =  \hat{H}_E 
    \begin{bmatrix}
        \sqrt{p_{E1}(t)}e^{i \gamma_1(t)} & 0 \\
        0 & \sqrt{p_{E2}(t)}e^{i \gamma_2(t)}
    \end{bmatrix}
      \begin{bmatrix}
        +\cos(\Theta_s) & -\sin(\Theta_s) \\
        +\sin(\Theta_s) & +\cos(\Theta_s)
    \end{bmatrix}
    \begin{bmatrix}
      w_L(x) \\
      w_R(x)
    \end{bmatrix}=
    \end{eqnarray}
    \begin{eqnarray*}
    \nonumber \\
    i \hbar \frac{d}{dt} \Bigg[
    \sin(\Theta_s)[ \sqrt{p_{E2}(t)}e^{i \gamma_2(t)}-\sqrt{p_{E1}(t)}e^{i \gamma_1(t)} ]
     \begin{bmatrix}
       0 &  1\\
       1 & 0
    \end{bmatrix}
    + \nonumber \\
    +
    \begin{bmatrix}
        \cos(\Theta_s) & -\sin(\Theta_s) \\
        \sin(\Theta_s) & +\cos(\Theta_s)
    \end{bmatrix}
    \begin{bmatrix}
        \sqrt{p_{E1}(t)}e^{i \gamma_1(t)} & 0 \\
        0 & \sqrt{p_{E2}(t)}e^{i \gamma_2(t)}
    \end{bmatrix} \Bigg]
    \begin{bmatrix}
      w_L(x) \\
      w_R(x)
    \end{bmatrix}.
\end{eqnarray*}
Consequently using the fact that $\Theta_s$ is time independent we obtain the following chain of equivalent expressions as giving the transformation from energy base to the position base
\begin{eqnarray}
 \hat{H}_E 
  \begin{bmatrix}
        +\cos(\Theta_s) & -\sin(\Theta_s) \\
        +\sin(\Theta_s) & +\cos(\Theta_s)
    \end{bmatrix}
    \begin{bmatrix}
        +\cos(\Theta_s) & +\sin(\Theta_s) \\
        +\sin(\Theta_s) & -\cos(\Theta_s)
    \end{bmatrix} 
    \times \nonumber \\
    \times 
    \begin{bmatrix}
        \sqrt{p_{E1}(t)}e^{i \gamma_1(t)} & 0 \\
        0 & \sqrt{p_{E2}(t)}e^{i \gamma_2(t)}
    \end{bmatrix}
      \begin{bmatrix}
        +\cos(\Theta_s) & -\sin(\Theta_s) \\
        +\sin(\Theta_s) & +\cos(\Theta_s)
    \end{bmatrix}
    \begin{bmatrix}
      w_L(x) \\
      w_R(x)
    \end{bmatrix}=
    \nonumber \\
    =
    \begin{bmatrix}
        +\cos(\Theta_s) & -\sin(\Theta_s) \\
        +\sin(\Theta_s) & +\cos(\Theta_s)
    \end{bmatrix} 
    \times \nonumber \\
    \times
    i\hbar \frac{d}{dt}    \Bigg[
     \begin{bmatrix}
        +\cos(\Theta_s) & +\sin(\Theta_s) \\
        +\sin(\Theta_s) & -\cos(\Theta_s)
    \end{bmatrix}
    \begin{bmatrix}
    \sqrt{p_{E1}(t)}e^{i \gamma_1(t)} &   0                                \\
    0                                 &  \sqrt{p_{E2}(t)}e^{i \gamma_2(t)}
    \end{bmatrix} 
   \begin{bmatrix}
        \cos(\Theta_s) & -\sin(\Theta_s) \\
        \sin(\Theta_s) & +\cos(\Theta_s)
    \end{bmatrix}
    \Bigg]
    \begin{bmatrix}
      w_L(x) \\
      w_R(x)
    \end{bmatrix}, \nonumber \\
    \Bigg[
    \begin{bmatrix}
        +\cos(\Theta_s) & +\sin(\Theta_s) \\
        -\sin(\Theta_s) & +\cos(\Theta_s)
    \end{bmatrix}
    \hat{H}_E 
  \begin{bmatrix}
        +\cos(\Theta_s) & -\sin(\Theta_s) \\
        +\sin(\Theta_s) & +\cos(\Theta_s)
    \end{bmatrix}  \Bigg] \times \nonumber \\ \times
    \Bigg[
     \begin{bmatrix}
        +\cos(\Theta_s) & +\sin(\Theta_s) \\
        +\sin(\Theta_s) & -\cos(\Theta_s)
    \end{bmatrix}
    \begin{bmatrix}
        \sqrt{p_{E1}(t)}e^{i \gamma_1(t)} & 0 \\
        0 & \sqrt{p_{E2}(t)}e^{i \gamma_2(t)}
    \end{bmatrix}
      \begin{bmatrix}
        +\cos(\Theta_s) & -\sin(\Theta_s) \\
        +\sin(\Theta_s) & +\cos(\Theta_s)
    \end{bmatrix} \Bigg]
    \begin{bmatrix}
      w_L(x) \\
      w_R(x)
    \end{bmatrix}=
    \nonumber \\
  i\hbar \frac{d}{dt}    \Bigg[
     \begin{bmatrix}
        +\cos(\Theta_s) & +\sin(\Theta_s) \\
        +\sin(\Theta_s) & -\cos(\Theta_s)
    \end{bmatrix}
    \begin{bmatrix}
    \sqrt{p_{E1}(t)}e^{i \gamma_1(t)} & 0 \\
        0 &  \sqrt{p_{E2}(t)}e^{i \gamma_2(t)}
    \end{bmatrix} 
    \begin{bmatrix}
        \cos(\Theta_s) & -\sin(\Theta_s) \\
        \sin(\Theta_s) & +\cos(\Theta_s)
    \end{bmatrix}
    \Bigg]
    \begin{bmatrix}
      w_L(x) \\
      w_R(x)
    \end{bmatrix}= \nonumber \\
    i\hbar \frac{d}{dt}    
    \begin{bmatrix}
    \sqrt{p_{E1}(t)}e^{i \gamma_1(t)}\cos(\Theta_s)^2+\sqrt{p_{E2}(t)}e^{i \gamma_2(t)}\sin(\Theta_s)^2 & [ \sqrt{p_{E2}(t)}e^{i \gamma_2(t)} - \sqrt{p_{E1}(t)}e^{i \gamma_1(t)}]\sin(\Theta_s)\cos(\Theta_s) \\
         [ \sqrt{p_{E2}(t)}e^{i \gamma_2(t)} - \sqrt{p_{E1}(t)}e^{i \gamma_1(t)}]\sin(\Theta_s)\cos(\Theta_s) &  \sqrt{p_{E1}(t)}e^{i \gamma_1(t)}\sin(\Theta_s)^2+\sqrt{p_{E2}(t)}e^{i \gamma_2(t)}\cos(\Theta_s)^2
    \end{bmatrix} 
    \begin{bmatrix}
      w_L(x) \\
      w_R(x)
    \end{bmatrix}= \nonumber \\
    = i \hbar  \frac{d}{dt}\ket{\psi}_P=\hat{H}_P\ket{\psi}_P= \nonumber \\
      i\hbar \frac{d}{dt}    
     \begin{bmatrix}
        +\cos(\Theta_s) & +\sin(\Theta_s) \\
        -\sin(\Theta_s) & +\cos(\Theta_s)
    \end{bmatrix}
    \begin{bmatrix}
    \sqrt{p_{E1}(t)}e^{i \gamma_1(t)} & 0 \\
        0 &  \sqrt{p_{E2}(t)}e^{i \gamma_2(t)}
    \end{bmatrix} 
    \begin{bmatrix}
      \psi_{E1}(x) \\
      \psi_{E2}(x)
    \end{bmatrix}, \nonumber \\
    i \hbar \frac{d}{dt} \psi_p(x,t)=
   \begin{bmatrix}
      1, & 1 \\
    \end{bmatrix}
    i \hbar \frac{d}{dt}
      \ket{\psi}_P=   \begin{bmatrix}
      1, & 1 \\
    \end{bmatrix} 
    \hat{H}_P
      \ket{\psi}_P= \nonumber \\
      =
    \begin{bmatrix}
      1, & 1 \\
    \end{bmatrix} 
     \Bigg[
    \begin{bmatrix}
        +\cos(\Theta_s) & +\sin(\Theta_s) \\
        -\sin(\Theta_s) & +\cos(\Theta_s)
    \end{bmatrix}
    \hat{H}_E 
  \begin{bmatrix}
        +\cos(\Theta_s) & -\sin(\Theta_s) \\
        +\sin(\Theta_s) & +\cos(\Theta_s)
    \end{bmatrix}  \Bigg] \times \nonumber \\ \times
    \Bigg[
     \begin{bmatrix}
        +\cos(\Theta_s) & +\sin(\Theta_s) \\
        +\sin(\Theta_s) & -\cos(\Theta_s)
    \end{bmatrix}
    \begin{bmatrix}
        \sqrt{p_{E1}(t)}e^{i \gamma_1(t)} & 0 \\
        0 & \sqrt{p_{E2}(t)}e^{i \gamma_2(t)}
    \end{bmatrix}
      \begin{bmatrix}
        +\cos(\Theta_s) & -\sin(\Theta_s) \\
        +\sin(\Theta_s) & +\cos(\Theta_s)
    \end{bmatrix} \Bigg]
    \begin{bmatrix}
      w_L(x) \\
      w_R(x)
    \end{bmatrix}= \nonumber \\
     =
    \begin{bmatrix}
      1, & 1 \\
    \end{bmatrix} 
     \Bigg[
    \begin{bmatrix}
        +\cos(\Theta_s) & +\sin(\Theta_s) \\
        -\sin(\Theta_s) & +\cos(\Theta_s)
    \end{bmatrix} \nonumber \times \\
    \times
     \begin{bmatrix}
        \int_{-\infty}^{+\infty}dx_1\psi_{E1}^{*}(x_1)[-\frac{\hbar^2}{2m}\frac{d^2}{dx_1^2}+V(x_1,t)]\psi_{E1}(x_1) & \int_{-\infty}^{+\infty}dx_1\psi_{E1}(x_1)[-\frac{\hbar^2}{2m}\frac{d^2}{dx_1^2}+V(x_1,t)]\psi_{E2}^{*}(x_1) \\
        \int_{-\infty}^{+\infty}dx_1\psi_{E2}(x_1)[-\frac{\hbar^2}{2m}\frac{d^2}{dx_1^2}+V(x_1,t)]\psi_{E1}^{*}(x_1)  &  \int_{-\infty}^{+\infty}dx_1\psi_{E2}^{*}(x_1)[-\frac{\hbar^2}{2m}\frac{d^2}{dx^2}+V(x,t)]\psi_{E2}(x_1)
    \end{bmatrix} \times \nonumber \\
    \times
  \begin{bmatrix}
        +\cos(\Theta_s) & -\sin(\Theta_s) \\
        +\sin(\Theta_s) & +\cos(\Theta_s)
    \end{bmatrix}  \Bigg] \times \nonumber \\ \times
    \Bigg[
     \begin{bmatrix}
        +\cos(\Theta_s) & +\sin(\Theta_s) \\
        +\sin(\Theta_s) & -\cos(\Theta_s)
    \end{bmatrix}
    \begin{bmatrix}
        \sqrt{p_{E1}(t)}e^{i \gamma_1(t)} & 0 \\
        0 & \sqrt{p_{E2}(t)}e^{i \gamma_2(t)}
    \end{bmatrix}
\Bigg]
    \begin{bmatrix}
      \psi_{E1}(x) \\
      \psi_{E2}(x)
    \end{bmatrix}, \nonumber 
\end{eqnarray}
where we have introduced the complex value function $\psi_P(x)$ expressed as a linear combination of maximal localized Wannier functions localized on the left or on the right side ($w_L(x)$ and $w_R(x)$) or as linear combination of 2 lowest energy eigenstates $\psi_{E1}(x)$ and $\psi_{E2}(x)$ with effective potential $V(x,t)$ describing a position-based qubit in the single electron regime as depicted in Fig. 4 and 5.
We assume that this potential is only weakly time-dependent. 
\begin{eqnarray}
\hat{H}_p=
\left(
\begin{array}{cc}
 \sin (2\Theta_s)  \frac{E_{12}+E_{21}}{2}+E_{1} \cos ^2(\Theta_s)+E_{2} \sin ^2(\Theta_s) & E_{12} \cos ^2(\Theta_s)+\sin
   (2\Theta_s) \frac{E_{2}-E_{1}}{2}-E_{21} \sin ^2(\Theta_s) \\
 -E_{12} \sin ^2(\Theta_s)+\sin(2\Theta_s)  \frac{E_2-E_1}{2}+E_{21} \cos ^2(\Theta_s) & -\sin (2\Theta_s)
\frac{E_{12}+E_{21}}{2}+E_1 \sin ^2(\Theta_s)+E_2 \cos ^2(\Theta_s) \\
\end{array}
\right), \nonumber \\
E_{21}(t)=E_{12}^{*}(t)=\int_{-\infty}^{-\infty}dx_1
\psi_{E2}^{*}(x_1,t)[-\frac{\hbar^2}{2m}\frac{d^2}{dx^2}+V(x_1,t)]\psi_{E2}(x_1,t), \nonumber \\
E_{ss}(t)=E_{s}(t)=\int_{-\infty}^{-\infty}dx_1
\psi_{Es}^{*}(x_1,t)[-\frac{\hbar^2}{2m}\frac{d^2}{dx^2}+V(x_1,t)]\psi_{Es}(x_1,t).
\end{eqnarray}

We observe that a given quantum system expressing single-electron qubits can be described by eigenenergy Hamiltonian $H_{E}$ and this picture can be transferred into position-based Hamiltonian $H_{P}$ or reversely.

\section{Higher order effects and picture beyond tight-binding model}
One particle is moving in the electric field of another particle that corresponds to the case of Hamiltonian having more prominent time-independent part and small time-dependent part. 

We have given Schroedinger equation for the system of 2 interacting particles in the regime of small particle-particle interaction subjected to time-dependent electric biasing of metallic gates and thus having the form as 
\begin{eqnarray}
i\hbar\frac{d}{dt}\ket{\psi}=\hat{H}\ket{\psi}= [-\frac{\hbar^2}{2m_a}\frac{d^2}{dx_a^2}-\frac{\hbar^2}{2m_b}\frac{d^2}{dx_b^2}+ V_A(x_A,t)+ V_A(x_B,t)+\frac{e^2}{4\pi \epsilon_0}\frac{1}{\sqrt{d^2+(x_1-x_2)^2}}]\ket{\psi}, \nonumber \\
\ket{\psi}=\gamma_1(t)\ket{E_{1a}}\ket{E_{1b}}+\gamma_2(t)\ket{E_{1a}}\ket{E_{2b}}+\gamma_3(t)\ket{E_{2a}}\ket{E_{1b}}+\gamma_4(t)\ket{E_{2a}}\ket{E_{2b}}, \nonumber \\
\ket{E_{s,a}}\ket{E_{k,b}}=\int_{-\infty}^{+\infty} \int_{-\infty}^{+\infty} dx_a dx_b \psi_{E_{s,a}}(x_{s,A})\psi_{E_{s,b}}(x_{k,b})\ket{x_{s,a}}\ket{x_{k,b}}.
\end{eqnarray}

At first we refrain to situation of toy model of two qubits having occupancy of two energy levels each and with situation of weak qubit interaction.  
We can state that the identity operators can be written as $\hat{I_a}=\ket{E_{1a}}\bra{E_{1a}}+\ket{E_{2a}}\bra{E_{2a}}$ and $\hat{I_b}=\ket{E_{1b}}\bra{E_{1b}}+\ket{E_{2b}}\bra{E_{2b}}$. System Hamiltonian in spectral form can be written as
\begin{eqnarray}
\hat{H}= [ f_1(t) \ket{E_{1a}}\bra{E_{2a}} \times \hat{I_b}+f_2(t) \ket{E_{2a}}\bra{E_{1a}} \times \hat{I_b} ] + \nonumber \\
+[ \hat{I_a} \times  g_1(t) \ket{E_{1b}}\bra{E_{2b}} + \hat{I_a} \times g_2(t) \ket{E_{2b}}\bra{E_{1b}}  ] + \nonumber \\
+[ E_{1a}(t) \ket{E_{1a}}\bra{E_{1a}} ] \times \hat{I_b} + [ E_{2a}(t) \ket{E_{2a}}\bra{E_{2a}} ] \times \hat{I_b} + \nonumber \\
+  \hat{I_a} \times  [ E_{1b}(t) \ket{E_{1b}}\bra{E_{1b}} ]  +  \hat{I_a} \times  [ E_{2b}(t) \ket{E_{2b}}\bra{E_{2b}} ] + \nonumber \\
+ \textcolor{red}{ [ h_1 [ E_{1a} \rightarrow E_{1a}, E_{1b} \rightarrow E_{1b} ](t) \ket{E_{1a}}\bra{E_{1a}} \times \ket{E_{1b}}\bra{E_{1b}}  ][11,11]}  + \nonumber \\
+ \textcolor{red}{ [ h_2 [ E_{1a} \rightarrow E_{1a}, E_{2b} \rightarrow E_{2b} ](t) \ket{E_{1a}}\bra{E_{1a}} \times \ket{E_{2b}}\bra{E_{2b}}  ][11,22]}  + \nonumber \\
+ \textcolor{red}{ [ h_3 [ E_{2a} \rightarrow E_{2a}, E_{1b} \rightarrow E_{1b} ](t) \ket{E_{2a}}\bra{E_{2a}} \times \ket{E_{1b}}\bra{E_{1b}}  ][22,11]} + \nonumber \\
+ \textcolor{red}{ [ h_4 [ E_{2a} \rightarrow E_{2a}, E_{2b} \rightarrow E_{2b} ](t) \ket{E_{2a}}\bra{E_{2a}} \times \ket{E_{2b}}\bra{E_{2b}}  ][22,22]}  + \nonumber \\
+ \textcolor{green}{ [ h_5 [ E_{1a} \rightarrow E_{2a}, E_{2b} \rightarrow E_{1b} ](t) \ket{E_{2a}}\bra{E_{1a}} \times \ket{E_{2b}}\bra{E_{1b}} ][12,21] }  + \nonumber \\
+\textcolor{green}{  [ h_6 [ E_{2a} \rightarrow E_{1a}, E_{1b} \rightarrow E_{2b} ](t) \ket{E_{1a}}\bra{E_{2a}} \times \ket{E_{2b}}\bra{E_{1b}} ][21,12]}  + \nonumber \\
+\textcolor{orange}{[ h_7 [ E_{1a} \rightarrow E_{2a}, E_{1b} \rightarrow E_{2b} ](t) \ket{E_{2a}}\bra{E_{1a}} \times \ket{E_{2b}}\bra{E_{1b}} ][12,12]}  + \nonumber \\
+\textcolor{blue}{[ h_8 [ E_{2a} \rightarrow E_{1a}, E_{2b} \rightarrow E_{1b} ](t) \ket{E_{1a}}\bra{E_{2a}} \times \ket{E_{1b}}\bra{E_{2b}} ][21,21]}  + \nonumber \\
+\textcolor{brown}{[ h_9 [ E_{1a} \rightarrow E_{1a}, E_{1b} \rightarrow E_{2b} ](t) \ket{E_{1a}}\bra{E_{1a}} \times \ket{E_{2b}}\bra{E_{1b}} ][11,12]}  + \nonumber \\
+\textcolor{brown}{[ h_{10} [ E_{2a} \rightarrow E_{2a}, E_{1b} \rightarrow E_{2b} ](t) \ket{E_{2a}}\bra{E_{2a}} \times \ket{E_{2b}}\bra{E_{1b}} ][22,12]}  + \nonumber \\
+\textcolor{brown}{[ h_{11} [ E_{1a} \rightarrow E_{2a}, E_{1b} \rightarrow E_{1b} ](t) \ket{E_{2a}}\bra{E_{1a}} \times \ket{E_{1b}}\bra{E_{1b}} ][12,11]}  + \nonumber \\
+\textcolor{brown}{[ h_{12} [ E_{1a} \rightarrow E_{2a}, E_{2b} \rightarrow E_{2b} ](t) \ket{E_{2a}}\bra{E_{1a}} \times \ket{E_{2b}}\bra{E_{2b}} ][12,22]}  + \nonumber \\
+\textcolor{pink}{ [ h_{13} [ E_{1a} \rightarrow E_{1a}, E_{2b} \rightarrow E_{1b} ](t) \ket{E_{1a}}\bra{E_{1a}} \times \ket{E_{2b}}\bra{E_{1b}} ][11,21]}  + \nonumber \\
+\textcolor{pink}{ [ h_{14} [ E_{2a} \rightarrow E_{2a}, E_{2b} \rightarrow E_{1b} ](t) \ket{E_{2a}}\bra{E_{2a}} \times \ket{E_{2b}}\bra{E_{1b}} ][22,21]}  + \nonumber \\
+\textcolor{pink}{[ h_{15} [ E_{2a} \rightarrow E_{1a}, E_{1b} \rightarrow E_{1b} ](t) \ket{E_{2a}}\bra{E_{1a}} \times \ket{E_{1b}}\bra{E_{1b}} ][21,11]}  + \nonumber \\
+\textcolor{pink}{[ h_{16} [ E_{2a} \rightarrow E_{1a}, E_{2b} \rightarrow E_{2b} ](t) \ket{E_{2a}}\bra{E_{1a}} \times \ket{E_{2b}}\bra{E_{2b}} ][21,22]} . \nonumber \\
\end{eqnarray}

We have introduced the following colour notation in regard to Coulomb interaction term:
\begin{itemize}
  \item red when both two qubits are not changing the energy
  \item blue when both two qubits de-excite
  \item orange when both qubits transfer into higher energy
  \item brown when both qubits transfer into higher energy
  \item pink when we have emission of photons by one of qubit, while second state is unchanged. 
\end{itemize}


Single–particle energies in the two lines are defined as expectation values of the corresponding time–dependent Hamiltonians,
\begin{align}
\varepsilon^{(A)}_s(t)&=\!\int\!dx_A\;\psi^{(A)\!*}_s(x_A,t)\Big[-\tfrac{\hbar^2}{2m_A}\partial_{x_A}^2+V_A(x_A,t)\Big]\psi^{(A)}_s(x_A,t),\\
\varepsilon^{(B)}_s(t)&=\!\int\!dx_B\;\psi^{(B)\!*}_s(x_B,t)\Big[-\tfrac{\hbar^2}{2m_B}\partial_{x_B}^2+V_B(x_B,t)\Big]\psi^{(B)}_s(x_B,t).
\end{align}
We also use short–hand transition matrix elements within each line,
\begin{align}
F_A(t)&=\!\int\!dx_A\;\psi^{(A)\!*}_2(x_A,t)\Big[-\tfrac{\hbar^2}{2m_A}\partial_{x_A}^2+V_A(x_A,t)\Big]\psi^{(A)}_1(x_A,t),\\
G_B(t)&=\!\int\!dx_B\;\psi^{(B)\!*}_2(x_B,t)\Big[-\tfrac{\hbar^2}{2m_B}\partial_{x_B}^2+V_B(x_B,t)\Big]\psi^{(B)}_1(x_B,t).
\end{align}

We have transition amplitude from different energy occupancy levels given in the form as 
\begin{eqnarray}
 f_1(t)= \int dx_A \, \psi_{E_A(2)}^*(x_A,t) \left( -\frac{\hbar^2}{2m_A} \frac{d^2}{dx_A^2} + V_A(x_A,t) \right) \psi_{E_A(1)}(x_A,t),
\end{eqnarray}
and 
\begin{eqnarray}
 g_1(t)= \int dx_B \, \psi_{E_B(2)}^*(x_B,t) \left( -\frac{\hbar^2}{2m_B} \frac{d^2}{dx_B^2} + V_B(x_B,t) \right) \psi_{E_B(1)}(x_B,t).
\end{eqnarray}

We represent the two–body coupling in the spectral (single–particle eigenstate) basis as
\begin{align}
h_{k s r p}
=\!\!\int\! dx_A\,dx_B\,
\psi^{(A)\!*}_{k}(x_A)\,\psi^{(B)\!*}_{s}(x_B)\,
\Big[-\frac{\hbar^2}{2m_A}\partial_{x_A}^2 + V_A(x_A)
-\frac{\hbar^2}{2m_B}\partial_{x_B}^2 + V_B(x_B) + V_{\mathrm{int}}(x_A,x_B)\Big]\,
\psi^{(A)}_{r}(x_A)\,\psi^{(B)}_{p}(x_B).
\end{align}
For the discretized operators we use the standard finite–difference stencil for the second derivative (a tridiagonal Toeplitz operator). As a result, computing $\psi^{(A)}$ and $\psi^{(B)}$ reduces to an eigenproblem determined by the gate–defined effective potential; algebraically this resembles emitter–field models (e.g., Jaynes–Cummings), though here the structure is constrained by device geometry.

\subsection{Entangled two-photon emission}
Situation of emission of each photon by each qubit is accounted by mathematical term \newline 
$\textcolor{blue}{[ h_8 [ E_{2a} \rightarrow E_{1a}, E_{2b} \rightarrow E_{1b} ](t) \ket{E_{1a}}\bra{E_{2a}} \times \ket{E_{1b}}\bra{E_{2b}} ][21,21]} $
and describes the physical situation when the photon state is given as $\ket{Photon-pair}=\ket{E_{ph1,A}}\ket{E_{ph1,B}}$, when $E_{ph1,A}=E_{2a}-E_{1a}$ and $E_{ph2,B}=E_{2b}-E_{1b}$. 

Two–photon entanglement is conveniently expressed in the Bell basis,
\begin{align}
\ket{\Phi^\pm}&=\tfrac{1}{\sqrt{2}}\!\left(\ket{\omega^{(A)}_1}\ket{\omega^{(B)}_1}\pm\ket{\omega^{(A)}_2}\ket{\omega^{(B)}_2}\right),\\
\ket{\Psi^\pm}&=\tfrac{1}{\sqrt{2}}\!\left(\ket{\omega^{(A)}_1}\ket{\omega^{(B)}_2}\pm\ket{\omega^{(A)}_2}\ket{\omega^{(B)}_1}\right),
\end{align}
provided that each emitter supports at least a three–level ladder enabling distinct radiative channels.
Instead of enumerating non–Hermitian coefficients $s_{1,2}$ and $s_{1a,b},s_{2a,b}$, we model losses and correlated radiative processes via an effective generator
\begin{equation}
\hat H_{\mathrm{eff}}(t)=\sum_{i,j\in\{0,x,y,z\}} C_{ij}(t)\,\sigma_i^{(A)}\!\otimes\!\sigma_j^{(B)}
-\frac{i}{2}\sum_\mu \gamma_\mu\,\hat L_\mu^\dagger \hat L_\mu,
\end{equation}
with jump operators $\hat L_\mu\propto \sigma^{(A)}_-\!\otimes\!\sigma^{(B)}_-$ capturing pairwise radiative relaxation.
This formulation keeps control–relevant couplings in the coherent part $C_{ij}(t)$ while relegating emission physics to the Lindblad term.

One can expect the effective system Hamiltonian to be as following 
\begin{eqnarray}
\hat{H}=(s_1(t)\ket{E_{2a}}\bra{E_{1a}} \times \ket{E_{2b}}\bra{E_{1b}}+
s_2(t)\ket{E_{3a}}\bra{E_{1a}}\times \ket{E_{3b}}\bra{E_{1b}}), \nonumber \\
\end{eqnarray}
Since aforementioned Hamiltonian is non-Hermitian one can suggest the maximum intensification of 2 entangled photons takes place for Hamiltonian as
\begin{eqnarray}
\hat{H}=(s_{1a}(t)\ket{E_{2a}}\bra{E_{1a}} \times \ket{E_{2b}}\bra{E_{1b}}+s_{1b}(t)\ket{E_{1a}}\bra{E_{2a}} \times \ket{E_{1b}}\bra{E_{2b}}) + \nonumber \\
+(s_{2a}(t)\ket{E_{3a}}\bra{E_{1a}}\times \ket{E_{3b}}\bra{E_{1b}}+s_{2b}(t)\ket{E_{1a}}\bra{E_{3a}}\times \ket{E_{1b}}\bra{E_{3b}} ), \nonumber \\
\end{eqnarray}
Therefore, coefficients $s_{1a}(t)$, $s_{1b}(t)$, $s_{2a}(t)$, $s_{2b}(t)$  are pronounced and other less pronounced. This is thus optimization problem from the point of view of
electrodes geometries and function of polarization voltage pattern applied to 6 subsequent metallic gates.

Therefore, coefficients $s_{1a}(t)$, $s_{1b}(t)$, $s_{2a}(t)$, $s_{2b}(t)$  are pronounced and other less pronounced. This is thus optimization problem from the point of view of
electrodes geometries and function of polarization voltage pattern applied to 6 subsequent metallic gates.   

\begin{figure}
\label{fig2}
\centering
\includegraphics[scale=0.5]{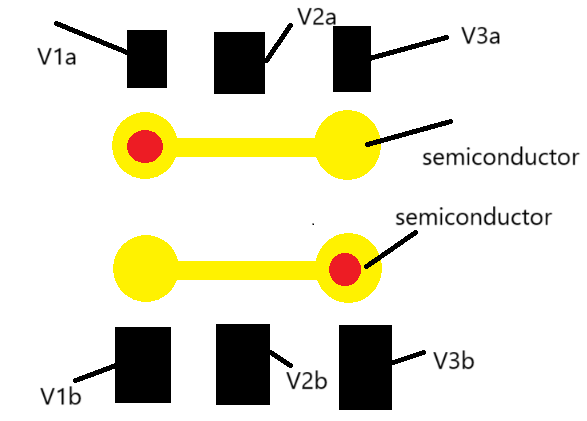}
\caption{Geometry of 2 interacting Wannier qubits to be parameterized for effective 2 entangled photon emission.}
\end{figure}
\begin{figure}
\centering
\includegraphics[scale=0.45]{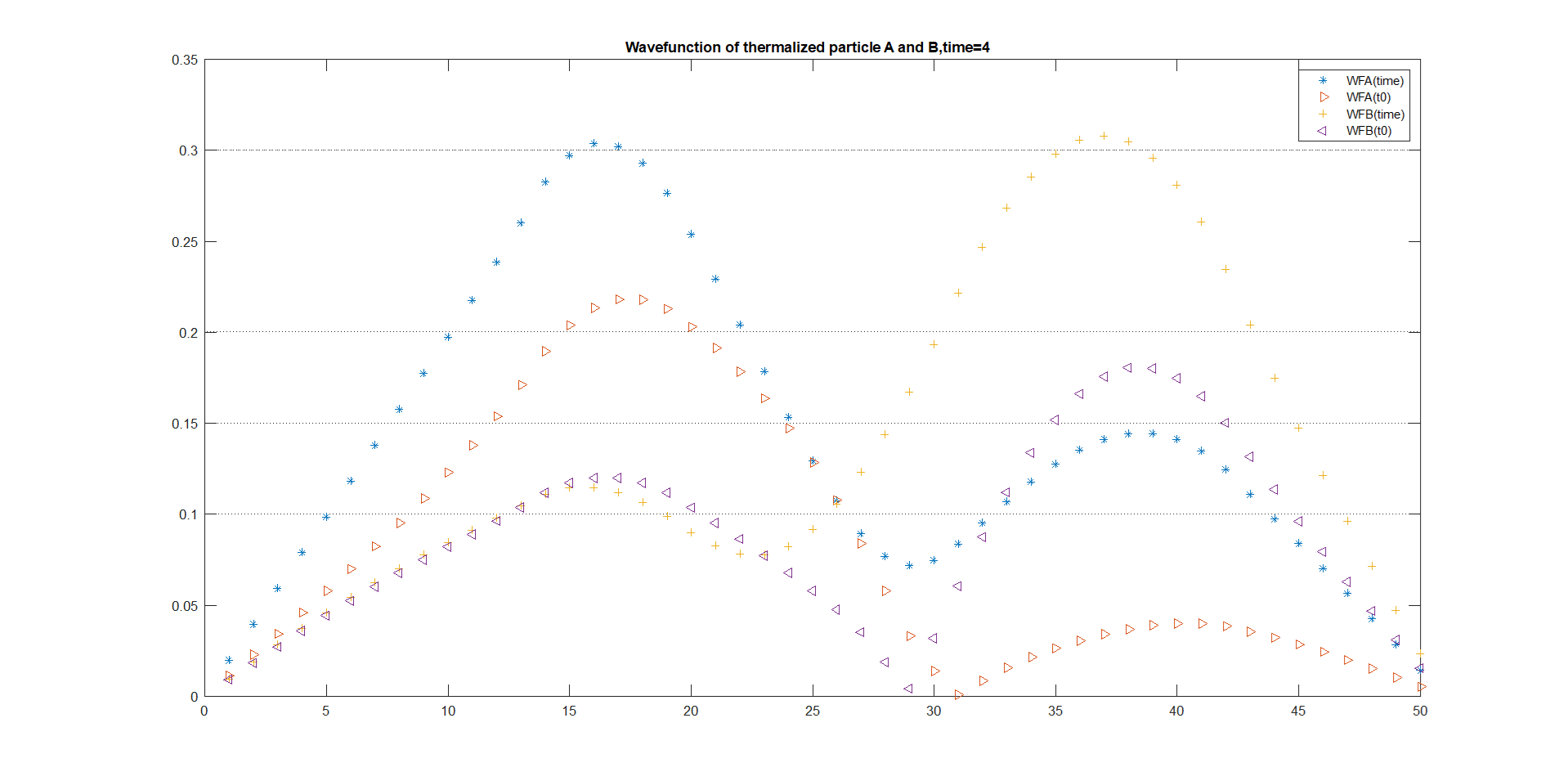} 
\caption{Wavefunctions absolute value of A and B interacting particles placed at two different single-electron lines with initial non-entangled state with the case of 2 energy occupancy for each particle at time $t_0$ and after some time t. Oscillations are never fully periodic as the system heats up and each particle starts to occupy more and more energy levels what is directly the breakdown of tight-binding model in longer time scales. One shall refer to next Figure for more details. }
\end{figure}


By proper numerical experiments one can establish proper $V_A(x_A,t)$ and $V_B(x_A,t)$ structure that maximizes the emission of entangled photons among all emitted photons that can be later translated into voltage signal pattern
$V_{1a}(t), V_{2a}(t), V_{3a}(t)$ and $V_{1b}(t), V_{2b}(t), V_{3b}(t)$ as with respect to the physical situation from Fig.17. 
The results obtained by two-body Schrödinger formalism support the validity of tight-binding model to certain degree. 
More optimal two entangled-photon emission can be optimized with reference to specified by Fig. \ref{qp1}, \ref{qp2}, \ref{GeometricTransition} , \ref{qp4}, \ref{GQSwapGateSpectrum}.

\subsection{Case of thermalization of systems of 2 position based qubits and failure of tight-binding model }
We can set 
\begin{eqnarray}
\psi_{A,B}(x_A,x_B,t_0)=(e^{i \gamma1A}\sqrt{p(E1,A)}\psi_{A,E1}(x_A,t_0) + 
\sqrt{p(E2,A)}e^{i \gamma2A}\psi_{A,E2}(x_A,t_0) \times  \nonumber \\ 
\times (\sqrt{p(E1,B)}e^{i \gamma1B}\psi_{B,E1}(x_B,t_0) + \sqrt{p(E2,B)}\psi_{B,E2}(x_B,t_0)e^{i \gamma1B}) 
\end{eqnarray} gives evolution of wavefunction modulus of each particle  as in Fig.18 when perturbative Coulomb interaction is turned on. It is interesting to observe how non-entangled system is becoming more entangled with time as indicated by Fig.19.
Dynamics of two electrons confined by two local potentials is determined for various cases both in analytical and in numerical way in tight binding model [6]. In such way the system of two coupled electrostatic position based qubits can be used for the implementation of quantum swap or antiswap gate.
Broader picture is drawn by work [6,13,15].
The obtained results have its meaning in designing the proper operation of the quantum gates implemented in chain of coupled semiconductor quantum dots that are electrostatically controlled. It opens the path for implementation of CMOS quantum computer that is only controlled by voltages applied to CMOS transistors with no need of usage of magnetic field. It is nice alternative for implementation of quantum electronics other than by the usage of Josephson junctions [20,21]. The best way of detection of entanglement present between electrostatically interacting qubits is by measure of the correlation-anticorrelation function that is achievable in experimental way. Both formulas for von-Neumann entanglement entropy and anticorrelation functions are given in this work in analytical form.
In the next section we describe elementary derivation of tight-binding model from Schroedinger formalism.

\begin{figure}
\centering
\includegraphics[scale=0.3]{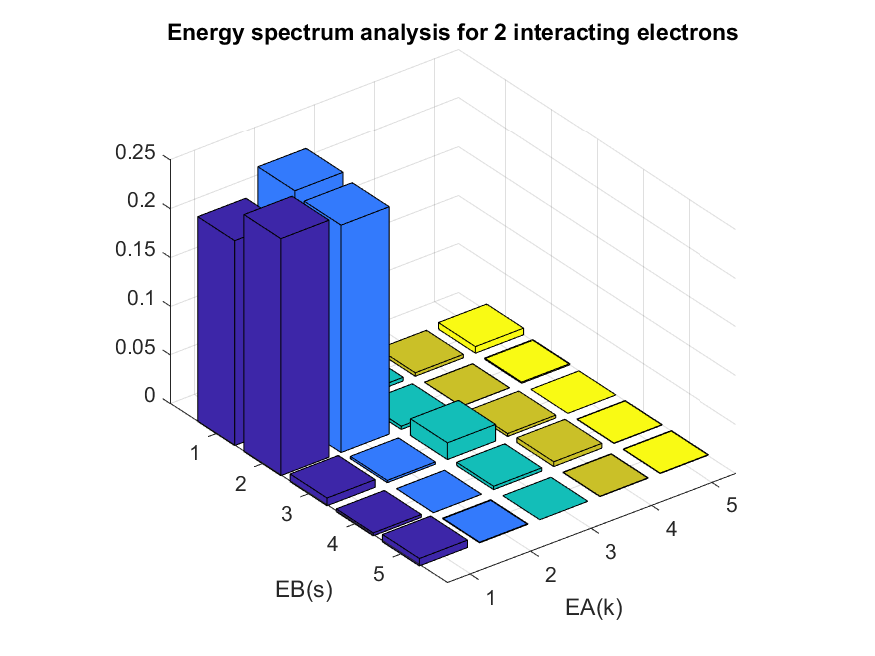}\includegraphics[scale=0.3]{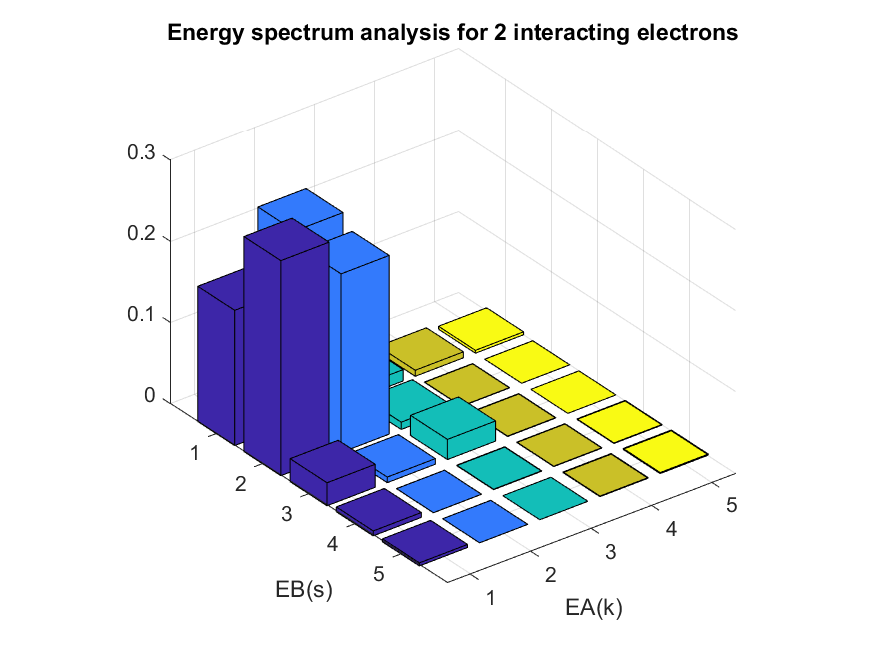}\includegraphics[scale=0.3]{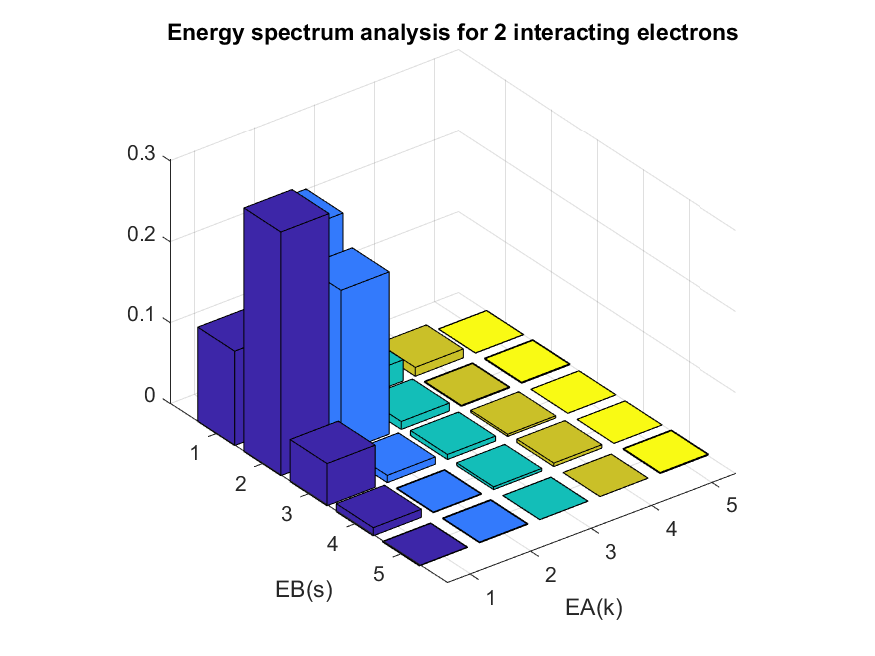}
\includegraphics[scale=0.3]{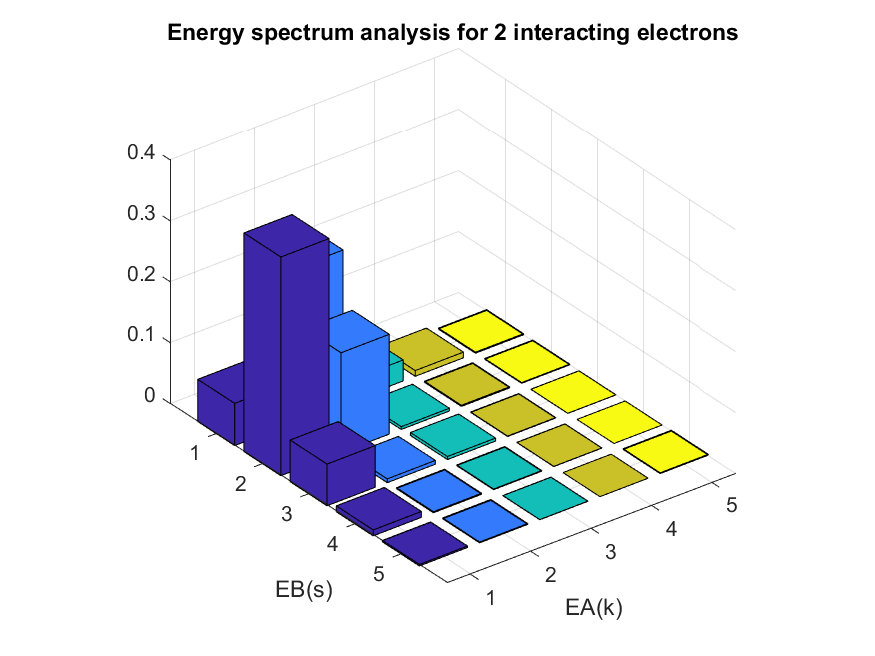}\includegraphics[scale=0.3]{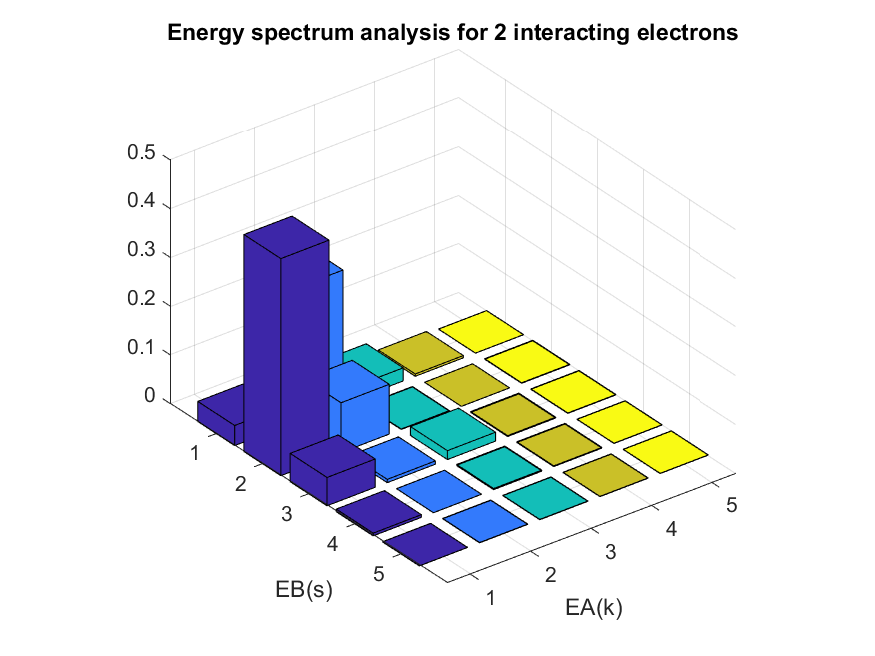}\includegraphics[scale=0.3]{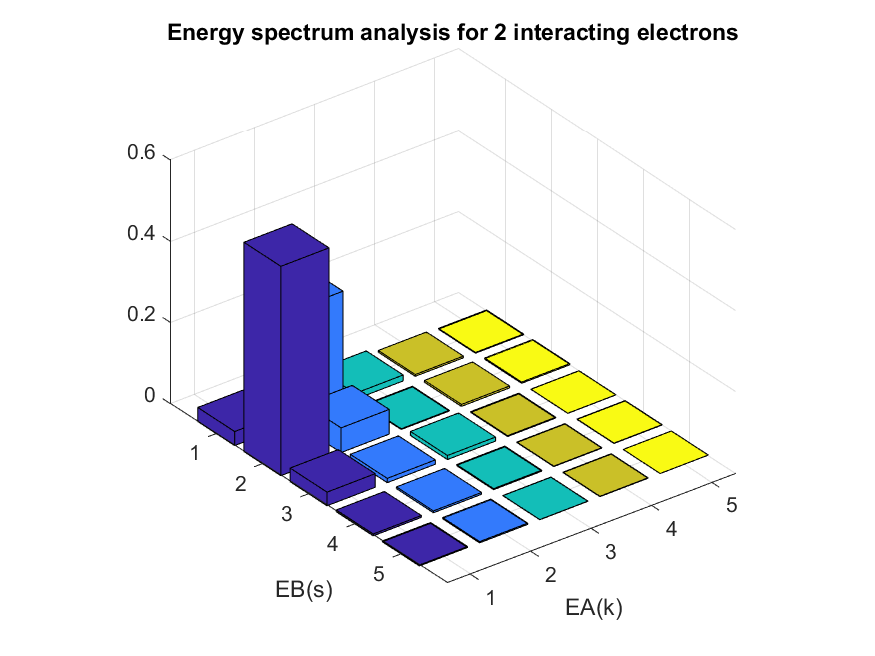}
\includegraphics[scale=0.3]{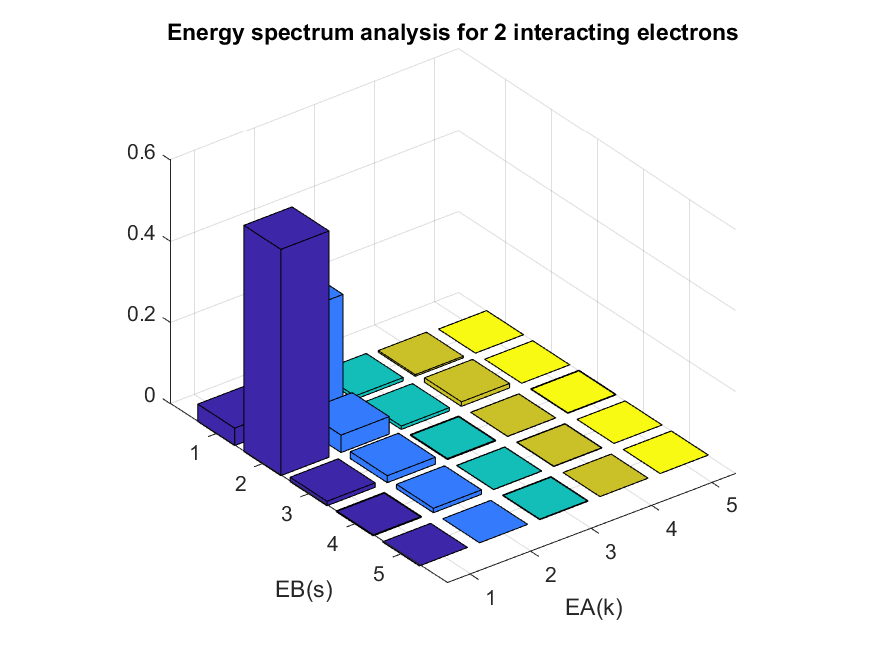}\includegraphics[scale=0.3]{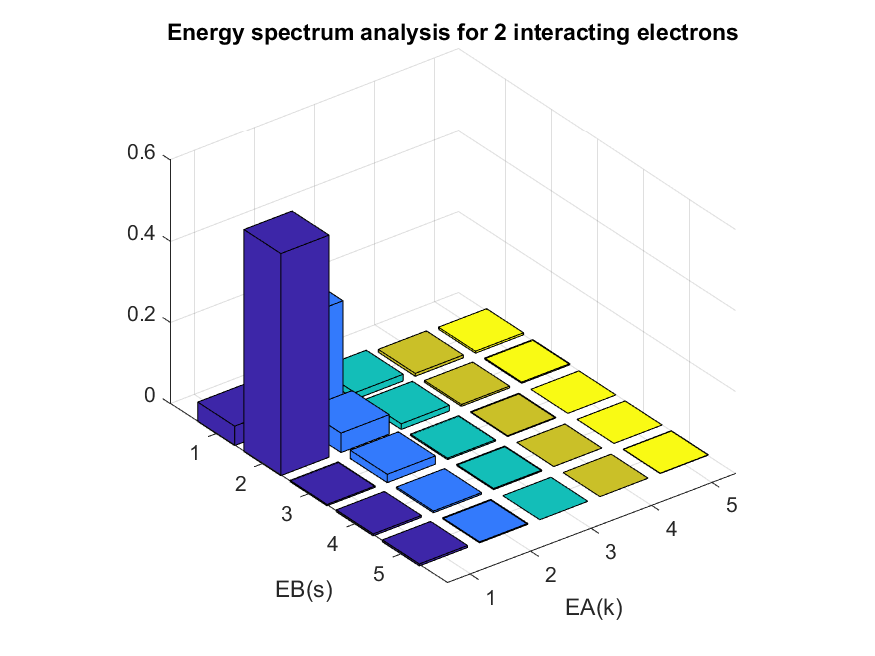}\includegraphics[scale=0.3]{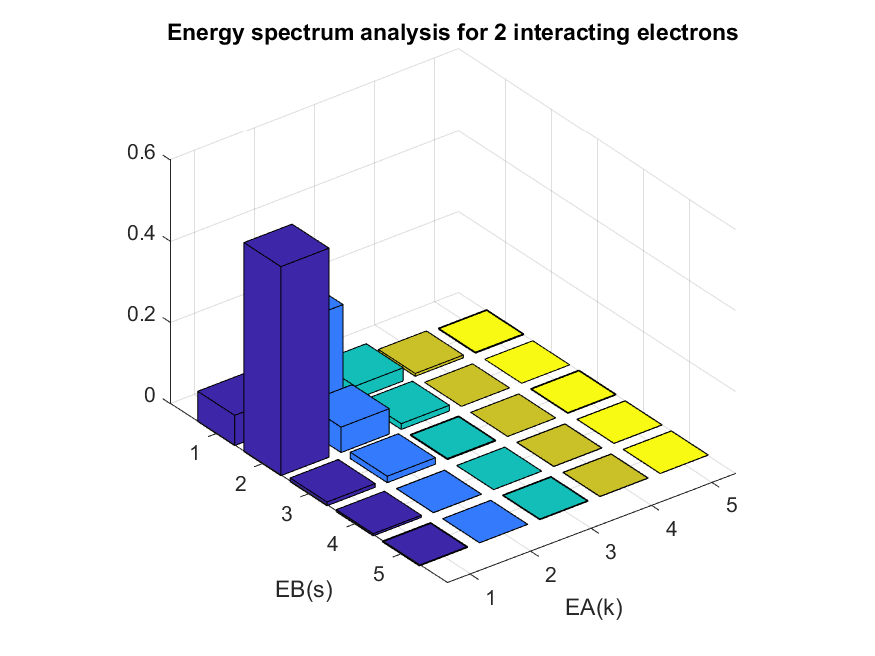}
\includegraphics[scale=0.3]{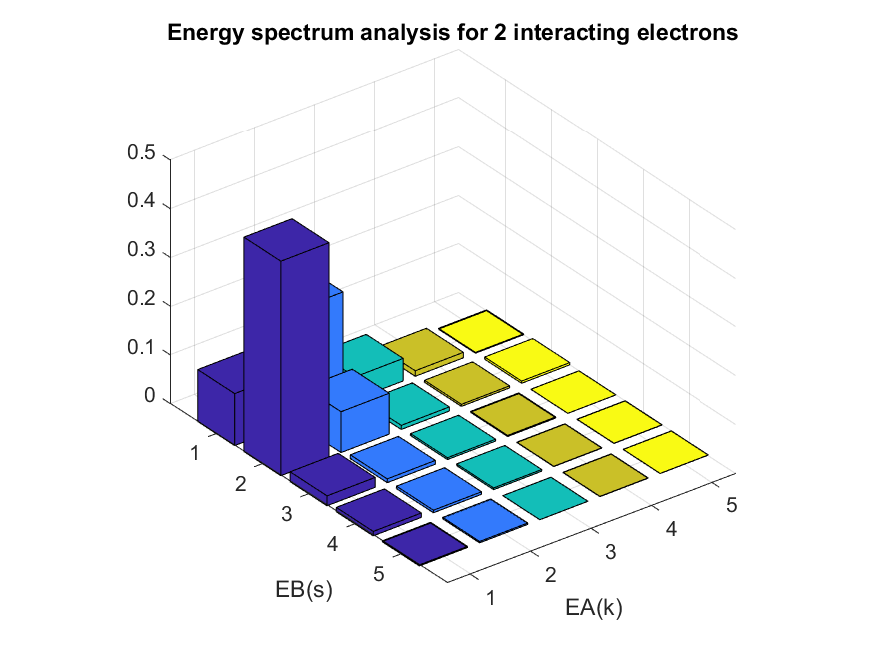}\includegraphics[scale=0.3]{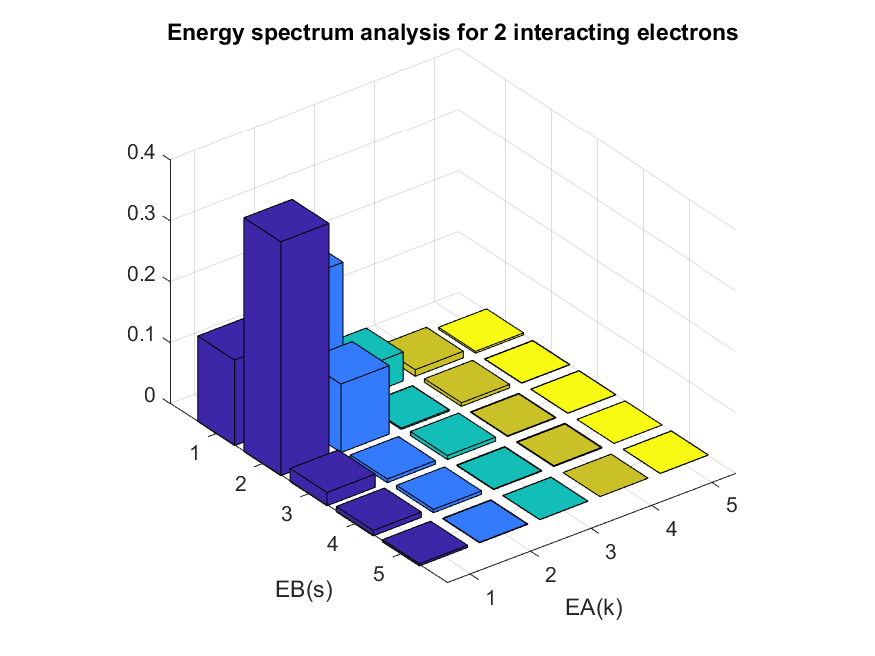}\includegraphics[scale=0.3]{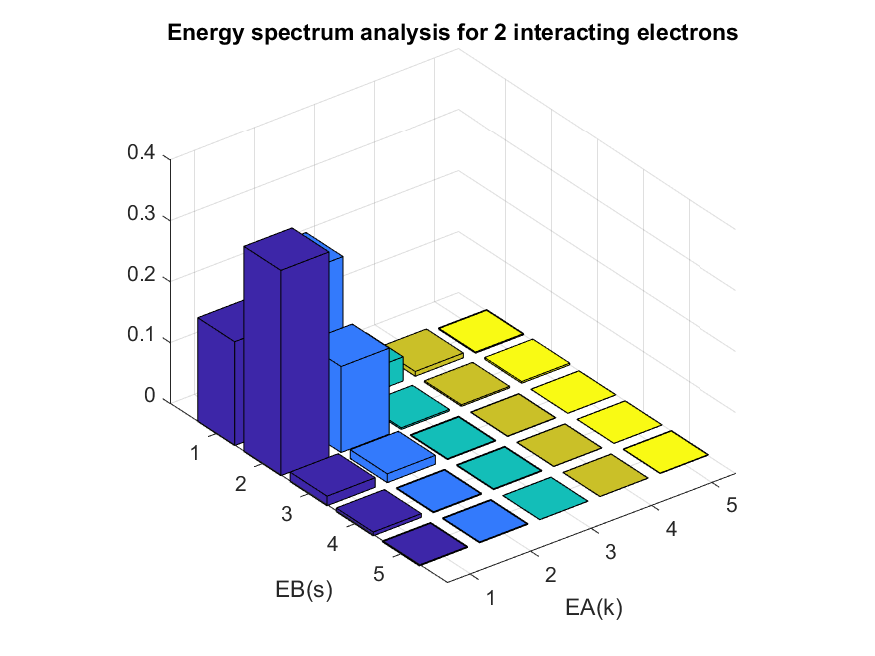}
\includegraphics[scale=0.3]{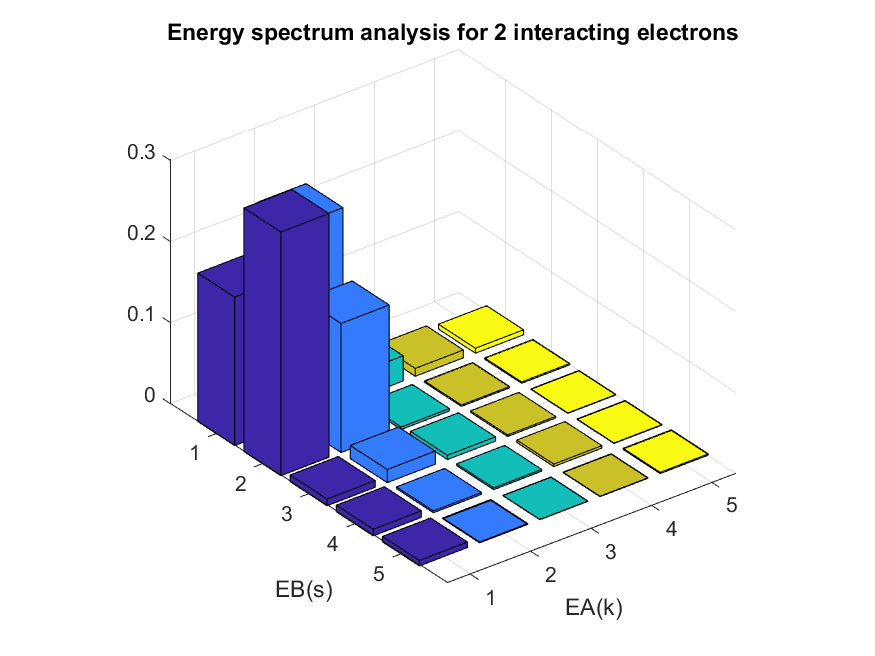}\includegraphics[scale=0.3]{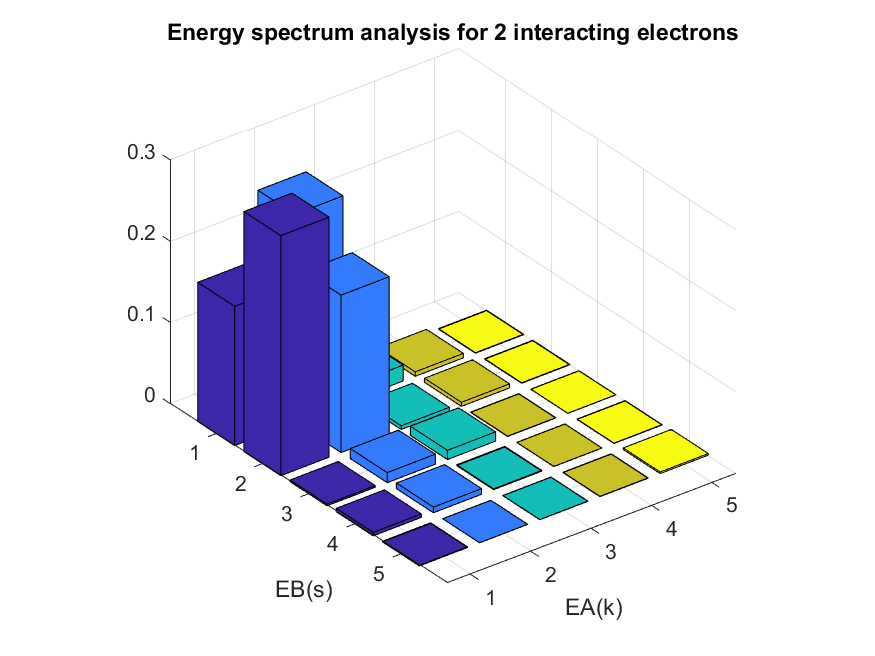}\includegraphics[scale=0.3]{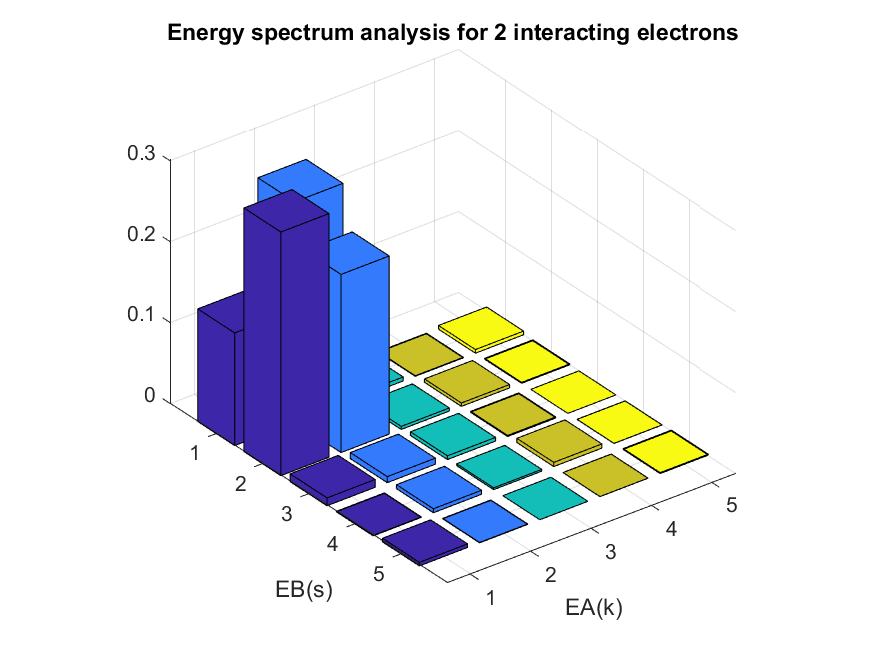}
\caption{Stages of thermalization of quantum state and evolution of occupancy of energy levels with time when time arrows goes from left to right and every lower lines represents proceeding time step. Initially A and B weakly interacting particles are placed at two different single-electron lines with initial non-entangled state being tensor product  of two qubits quantum states and each of them having the occupancy of 2 energies. System of 2 qubits is starting to occupy more and more energy levels and breakdown of tight-binding model is observed. In meantime the emergence of quantum entanglement is observed from initial non-entangled state. }
\end{figure}

\section{Fundamental derivation of tight-binding model from Schroedinger equation}

Being motivated by work \cite{p1} , \cite{f1} and starting from eigenergy qubit as given by IBM Q-Experience or many other technologies ($|\gamma_{E1}(t)|^2+|\gamma_{E2}(t)|^2=1$) we arrive into 

\begin{eqnarray}
\hat{H}(t)_E \ket{\psi(t)}_{E}=
\begin{pmatrix}
E_{1 \rightarrow 1}(t) & E_{2 \rightarrow 1}(t) \\
E_{1 \rightarrow 2}(t)  & E_{2 \rightarrow 2}(t)
\end{pmatrix}_E
\begin{pmatrix}
\gamma_{E1}(t) \\
\gamma_{E2}(t)
\end{pmatrix}_E = \nonumber \\
=i\hbar \frac{d}{dt}
\begin{pmatrix}
\gamma_{E1}(t) \\
\gamma_{E2}(t)
\end{pmatrix}_E= i\hbar \frac{d}{dt} ( \ket{\psi(t)})_E=  \nonumber \\
=i\hbar \frac{d}{dt} ( \gamma_{E1}(t) \ket{E_1(t)}+\gamma_{E2}(t) \ket{E_2(t)})=  \nonumber \\
=\begin{pmatrix}
E_{1 \rightarrow 1}(t) & E_{2 \rightarrow 1}(t) \\
E_{1 \rightarrow 2}(t)  & E_{2 \rightarrow 2}(t)
\end{pmatrix}_E
\begin{pmatrix}
+\cos(\Theta) & -\sin(\Theta) \\
+\sin(\Theta) & +\cos(\Theta)
\end{pmatrix} \times \nonumber \\
\times
\begin{pmatrix}
+\cos(\Theta) & +\sin(\Theta) \\
-\sin(\Theta) & +\cos(\Theta)
\end{pmatrix}
\begin{pmatrix}
\gamma_{E1}(t) \\
\gamma_{E2}(t)
\end{pmatrix}_E
= \nonumber \\
i \hbar \frac{d}{dt} \Bigg[
\begin{pmatrix}
+\cos(\Theta) & -\sin(\Theta) \\
+\sin(\Theta) & +\cos(\Theta)
\end{pmatrix} \times \nonumber \\
\times
\begin{pmatrix}
+\cos(\Theta) & +\sin(\Theta) \\
-\sin(\Theta) & +\cos(\Theta)
\end{pmatrix}
\begin{pmatrix}
\gamma_{E1}(t) \\
\gamma_{E2}(t)
\end{pmatrix}_E
\Bigg]
\label{eqn:centerus}
\end{eqnarray}
where we have $E_{s \rightarrow k}(t)=\bra{E_k(t)}\hat{H}(t)\ket{E_s(t)}$ since we assume $\hat{H}(t)=E_{1 \rightarrow 1}(t)\ket{E_{1}(t)}\bra{E_1(t)}+E_{2 \rightarrow 2}(t)\ket{E_2(t)}\bra{E_2(t)}+E_{1 \rightarrow 2}(t)\ket{E_2(t)}\bra{E_1(t)}+E_{2 \rightarrow 1}(t)\ket{E_1(t)}\bra{E_2(t)}$. Here $|\gamma_{E1}(t)|^2$ and $|\gamma_{E2}(t)|^2$ accounts for probability of occupancy of $E_1$ and $E_2$ energetic states.
Maintenance of quantum information is about keeping quantum information encoded in Bloch sphere by two parameters ($\Theta(t)$, $\phi(t)$) and given in reference to quantum state by
 $\ket{\psi}_E=\gamma_{E1}(t)\ket{E_1(t)}+\gamma_{E2}(t)\ket{E_2(t)}= \nonumber \\
 =e^{i \phi_0(t)}(cos(\Theta(t))\ket{E_1(t)}+e^{i \phi(t)} sin(\Theta(t))\ket{E_2(t)})$.
Due to the fact of compact size of semiconductor quantum dots as depicted in Fig.\ref{fig:swap_concept}. it is vital to express position based qubit by stating its Hamiltonian
in so called tight-binding model, where quantum state is linear superposition of two Wannier wave functions denoting maximal localization on the left or right quantum dot ($|\alpha_q(t)|^2+|\beta_q(t)|^2=1$):
\begin{eqnarray}
\hat{H}(t)_W \ket{\psi(t)}_W=
\begin{pmatrix}
E_{p1}(t) & t_{s12}(t) \\
t_{s12}^{*}(t) & E_{p2}(t)
\end{pmatrix}_W
\begin{pmatrix}
\alpha_q(t) \\
\beta_q(t)
\end{pmatrix}_W = \nonumber \\
i\hbar \frac{d}{dt}   
(\alpha_q(t)\ket{w_{L}(t)}+\beta_q(t)\ket{w_{R}(t)})_W=i\hbar \frac{d}{dt}\ket{\psi(t)}_W, 
\label{eqn:fundamentalbasic} 
\end{eqnarray}
where $\ket{w_{L}(t)}$ and $\ket{w_{R}(t)}$ are maximum localized Wannier wave functions ($\bra{w_L}\ket{w_R}=0$,$\bra{w_L}\ket{w_L}=1=\bra{w_R}\ket{w_R}$ ) respectively on left and right side, while $|\alpha_q(t)|^2$ and $|\beta_q(t)|^2$ gives the probabilities for particle (single electron or hole) to stay on the left and right quantum dot. Encoding quantum information by electron or hole position brings compactification of quantum chips. Quantum state in Wannier representation $\ket{\psi(t)}_W$ can be mapped to quantum state in energetic representation given as $\ket{\psi(t)}_E$ or reversely and those two ways of description are fully equivalent.

Following \cite{CurvedWannier}, \cite{b3} it will be useful to introduce parameter
\begin{eqnarray}
\label{special}
r=\frac{\int_{-\infty}^{0}dx (\psi_{E1}(x)\psi_{E2}^{\dag}(x)+\psi_{E1}(x)^{\dag}\psi_{E2}(x))}{\int_{-\infty}^{0}dx (|\psi_{E1}(x)|^2-|\psi_{E2}(x)|^2)}.
\end{eqnarray}
The precondition for occurrence of position-based qubit is occupancy at least two energetic levels, so neither $|\gamma_{E1}|^2$ nor $|\gamma_{E2}|^2$ is zero. 
Finally in case of static Hamiltonian from equation \ref{eqn:centerus} also given by \cite{CurvedWannier} the tight-binding Hamiltonian \cite{b3} will appear to be
\begin{eqnarray}
\label{MainFormula}
\begin{pmatrix}
+\cos(\Theta) & \sin(\Theta) \\
-\sin(\Theta) & \cos(\Theta)
\end{pmatrix}
\begin{pmatrix}
E_{1} & 0 \\
0  & E_{2}
\end{pmatrix}_E
\begin{pmatrix}
+\cos(\Theta) & -\sin(\Theta) \\
-\sin(\Theta) & +\cos(\Theta)
\end{pmatrix} 
=
\begin{pmatrix}
E_{p1} & t_{s21}  \\
t_{s12} & E_{p2}
\end{pmatrix}_W= \nonumber \\
=
E_{p1} \ket{w_{L}}\bra{w_{L}}+E_{p2} \ket{w_{R}}\bra{w_{R}}+ t_{s21} \ket{w_{R}}\bra{w_{L}}+ t_{s12} \ket{w_{L}}\bra{w_{R}}=\hat{H}_W=
\nonumber \\
\begin{pmatrix}
E_1+|sin(\frac{1}{2}ArcTan(r))|^2(E_2-E_1) & (E_2-E_1)\frac{1}{2}sin(ArcTan(r))  \\
(E_2-E_1)\frac{1}{2}sin(ArcTan(r)) & E_1+|cos(\frac{1}{2}ArcTan(r))|^2(E_2-E_1)
\end{pmatrix}, 
\end{eqnarray}

We propose maximum localized orthonormal Wannier functions (so occupancy of $w_L$ is maximized from minus infinity to zero) of the form
\begin{eqnarray}
\Theta=\frac{1}{2}ArcTan(r),
\ket{\psi}_W=
\begin{pmatrix}
w_L(x) \\
w_R(x)
\end{pmatrix}_W
=
\begin{pmatrix}
+\cos(\Theta) & \sin(\Theta) \\
-\sin(\Theta) & \cos(\Theta)
\end{pmatrix}
\begin{pmatrix}
\psi_L(x) \\
\psi_R(x)
\end{pmatrix}_E
=  
\begin{pmatrix}
+\cos(\Theta) & \sin(\Theta) \\
-\sin(\Theta) & \cos(\Theta)
\end{pmatrix}
\ket{\psi}_E
,
\nonumber \\
\begin{pmatrix}
+cos(\frac{1}{2}ArcTan [\frac{\int_{-\infty}^{0}dx (\psi_{E1}(x)\psi_{E2}^{\dag}(x)+\psi_{E1}(x)^{\dag}\psi_{E2}(x))}{\int_{-\infty}^{0}dx (|\psi_{E1}(x)|^2-|\psi_{E2}(x)|^2)}]) & sin(\frac{1}{2}ArcTan [\frac{\int_{-\infty}^{0}dx (\psi_{E1}(x)\psi_{E2}^{\dag}(x)+\psi_{E1}(x)^{\dag}\psi_{E2}(x))}{\int_{-\infty}^{0}dx (|\psi_{E1}(x)|^2-|\psi_{E2}(x)|^2)}]) \\
-sin(\frac{1}{2}ArcTan [ \frac{\int_{-\infty}^{0}dx (\psi_{E1}(x)\psi_{E2}^{\dag}(x)+\psi_{E1}(x)^{\dag}\psi_{E2}(x))}{\int_{-\infty}^{0}dx (|\psi_{E1}(x)|^2-|\psi_{E2}(x)|^2)}]) & cos(\frac{1}{2}ArcTan [ \frac{\int_{-\infty}^{0}dx (\psi_{E1}(x)\psi_{E2}^{\dag}(x)+\psi_{E1}(x)^{\dag}\psi_{E2}(x))}{\int_{-\infty}^{0}dx (|\psi_{E1}(x)|^2-|\psi_{E2}(x)|^2)}])
\end{pmatrix} 
\begin{pmatrix}
\psi_{E1}(x) \\
\psi_{E2}(x)
\end{pmatrix}_E. \nonumber \\
\end{eqnarray}
\subsection{Two interacting single-electron lines described in tight-binding model validated by perturbative Schroedinger model}

We consider the system of two position-based qubits interacting perturbatively.
Two electrons on two separated single-electron lines tends to be in anti-correlated positions due to repulsion force as depicted in Fig.3 ,5, 7. 
In such a way we encounter the situation of two electrons placed on two different left (right) quantum dots with probabilities $|\xi_1(t)|^2$ ($|\xi_4(t)|^2$) or the case left (right) occupancy of first dot and right (left) occupancy of second dot with probability $|\xi_2(t)|^2$ ($|\xi_3(t)|^2$) that specifies normalized quantum state ($|\xi_1(t)|^2+..+|\xi_4(t)|^2=1$) to be in the form as given by \cite{CurvedWannier}:
\begin{eqnarray}
\ket{\psi}=\xi_1(t)\ket{0}_A\ket{0}_B+\xi_2(t)\ket{0}_A\ket{1}_B+\xi_3(t)\ket{1}_A\ket{0}_B+\xi_4(t)\ket{1}_A\ket{1}_B.
\end{eqnarray}
In case of 2 Single-Electron Lines we can write the effective Hamiltonian to be of the form
\begin{eqnarray}
\hat{H}_{A-B}= \hat{H}_{A} \times \hat{I}_B +  \hat{I}_A \times \hat{H}_{B} + \hat{H}_{Coulomb: A-B}= \nonumber   \\
=(E_{p1A}\ket{w_{L,A}}\bra{w_{L,A}}+E_{p2A}\ket{w_{R,A}}\bra{w_{R,A}}+t_{s12A}\ket{w_{R,A}}\bra{w_{L,A}}+t_{s21A}\ket{w_{L,A}}\bra{w_{R,A}}) \times \hat{I}_B +  \nonumber   \\
+\hat{I}_A \times (E_{p1B}\ket{w_{L,B}}\bra{w_{L,B}}+E_{p2B}\ket{w_{R,B}}\bra{w_{R,B}}+t_{s12B}\ket{w_{R,B}}\bra{w_{L,B}}+t_{s21B}\ket{w_{L,B}}\bra{w_{R,B}})  +\nonumber   \\
+ \Bigg[ q_1 \ket{w_{L,A}}\ket{w_{L,B}}\bra{w_{L,A}}\bra{w_{L,B}}+ q_2 \ket{w_{L,A}}\ket{w_{L,B}}\bra{w_{R,A}}\bra{w_{R,B}} + q_3 \ket{w_{L,A}}\ket{w_{L,B}}\bra{w_{R,A}}\bra{w_{R,B}}+ \nonumber   \\
+ q_4 \ket{w_{R,A}}\ket{w_{R,B}}\bra{w_{R,A}}\bra{w_{R,B}} \Bigg]_{Coulomb}
\end{eqnarray}
Finally, we obtain validation of the tight-binding model in terms of Schroedinger equation, so we end up with the following effective Hamiltonian.
\small
\begin{eqnarray}
\hat{H}= \nonumber \\
\begin{pmatrix}
E_{1a}cos(\Theta_A)^2+E_{2a}sin(\Theta_A)^2 & (E_{2b}-E_{1b})\frac{1}{2}sin(2\Theta_B) & (E_{2a}-E_{1a})\frac{1}{2}sin(2\Theta_A) & 0  \\
(E_{2b}-E_{1b})\frac{1}{2}sin(2\Theta_B) & E_{1a}cos(\Theta_A)^2+E_{2a}sin(\Theta_A)^2  & 0 & (E_{2a}-E_{1a})\frac{1}{2}sin(2\Theta_A)   \\
(E_{2a}-E_{1a})\frac{1}{2}sin(2\Theta_A) &0 & E_{1a}sin(\Theta_A)^2+E_{2a}cos(\Theta_A)^2  & (E_{2b}-E_{1b})\frac{1}{2}sin(2\Theta_B)  \\
0 & (E_{2a}-E_{1a})\frac{1}{2}sin(2\Theta_A) & (E_{2b}-E_{1b})\frac{1}{2}sin(2\Theta_B) & E_{1a}sin(\Theta_A)^2+E_{2a}cos(\Theta_A)^2   \\
\end{pmatrix} +  \nonumber \\
\begin{pmatrix}
E_{1b}cos(\Theta_B)^2+E_{2b}sin(\Theta_B)^2 & 0 & 0 & 0 \\
0  & E_{1b}sin(\Theta_B)^2+E_{2b}cos(\Theta_B)^2  & 0 & 0 \\
0 & 0 & E_{1b}cos(\Theta_B)^2+E_{2b}sin(\Theta_B)^2 & 0 \\
0 & 0 & 0 & E_{1b}sin(\Theta_B)^2+E_{2b}cos(\Theta_B)^2
\end{pmatrix}  +  \nonumber \\
+
\begin{pmatrix}
q_1 & 0 & 0 & 0 \\
0  & q_2 & 0 & 0 \\
0 & 0 & q_3 & 0 \\
0 & 0 & 0 & q_4
\end{pmatrix}
\end{eqnarray}
\normalsize
where 4 terms are responsible for 2 qubits being at a distance d mutual Coulomb interaction (as depicted in Fig.3.) and given as

\begin{eqnarray*}
q_1=\int_{-\infty}^{+\infty}\int_{-\infty}^{+\infty}\frac{dx_Adx_B e^2}{4\pi \epsilon_0 \sqrt{d^2+(x_A-x_B)^2}}|cos(\Theta_A)\psi_{E1}(x_A)+sin(\Theta_A)\psi_{E2}(x_A)|^2 * 
|cos(\Theta_B)\psi_{E1}(x_B)+sin(\Theta_B)\psi_{E2}(x_B)|^2
\end{eqnarray*}
\begin{eqnarray*}
q_2=\int_{-\infty}^{+\infty}\int_{-\infty}^{+\infty}\frac{dx_Adx_B e^2}{4\pi \epsilon_0 \sqrt{d^2+(x_A-x_B)^2}}|cos(\Theta_A)\psi_{E1}(x_A)+sin(\Theta_A)\psi_{E2}(x_A)|^2 * 
|-sin(\Theta_B)\psi_{E1}(x_B)+cos(\Theta_B)\psi_{E2}(x_B)|^2
\end{eqnarray*}
\begin{eqnarray*}
q_3=\int_{-\infty}^{+\infty}\int_{-\infty}^{+\infty}\frac{dx_Adx_B e^2}{4\pi \epsilon_0 \sqrt{d^2+(x_A-x_B)^2}}|-sin(\Theta_A)\psi_{E1}(x_A)+cos(\Theta_A)\psi_{E2}(x_A)|^2 * 
|cos(\Theta_B)\psi_{E1}(x_B)+sin(\Theta_B)\psi_{E2}(x_B)|^2
\end{eqnarray*}
\begin{eqnarray*}
q_4=\int_{-\infty}^{+\infty}\int_{-\infty}^{+\infty}\frac{dx_Adx_B e^2}{4\pi \epsilon_0 \sqrt{d^2+(x_A-x_B)^2}}|-sin(\Theta_A)\psi_{E1}(x_A)+cos(\Theta_A)\psi_{E2}(x_A)|^2  
|-sin(\Theta_B)\psi_{E1}(x_B)+cos(\Theta_B)\psi_{E2}(x_B)|^2. \nonumber \\
\end{eqnarray*}
\normalsize


In first approximation we consider two-position based qubits as non-interacting and slightly perturbed by electrostatic single-electron to single-electron interaction
what naturally leads to first approximation of system eigenstate as tensor product of two Hilbert spaces belonging to each qubits in the form as given below:

\begin{eqnarray}
\ket{\psi}_{Energy,A-B}=
\begin{pmatrix}
\gamma_{E1A} \\
\gamma_{E2A} \\
\end{pmatrix}
\times
\begin{pmatrix}
\gamma_{E1B} \\
\gamma_{E2B} \\
\end{pmatrix}
=
\begin{pmatrix}
\gamma_{E1A}\gamma_{E1B}=cos(\xi_A)e^{i \phi_{E1A}}cos(\xi_B)e^{i \phi_{E1B}} \\
\gamma_{E1A}\gamma_{E2B}=cos(\xi_A)e^{i \phi_{E1A}}sin(\xi_B)e^{i \phi_{E2B}} \\
\gamma_{E2A}\gamma_{E1B}=sin(\xi_A)e^{i \phi_{E2A}}cos(\xi_B)e^{i \phi_{E1B}} \\
\gamma_{E2A}\gamma_{E2B}=sin(\xi_A)e^{i \phi_{E2A}}sin(\xi_B)e^{i \phi_{E2B}} \\
\end{pmatrix}.
\end{eqnarray}
Search for system ground state naturally implies minimization of Coulomb energy as criteria for lowest energy state that can be expressed by proper self-alignment of electrons in anticorrelated way and thus leading to minimization of Coulomb interaction energy. Coulomb energy is given as
\begin{eqnarray}
E_c(\xi_A,\xi_B)=
\bra{\psi}\hat{H}_c\ket{\psi}= \nonumber \\
=q_1cos(\xi_A)^2cos(\xi_B)^2+q_2cos(\xi_A)^2sin(\xi_B)^2+q_3sin(\xi_A)^2cos(\xi_B)^2+q_4 sin(\xi_A)^2 sin(\xi_B)^2.
\end{eqnarray}
Coulomb energy minimization condition brings following constrains
\begin{eqnarray}
\frac{d}{d\xi_A}E_c(\xi_A,\xi_B)=0, \frac{d}{d\xi_B}E_c(\xi_A,\xi_B)=0,
\end{eqnarray}
and hence we have
\begin{eqnarray}
0= \frac{d}{d\xi_A}E_c(\xi_A,\xi_B)=2(-sin(\xi_A)[q_1 cos(\xi_B)^2+q_2sin(\xi_B)^2]+ \nonumber \\
+cos(\xi_A)[q_3cos(\xi_B)^2+q_4sin(\xi_B)^2]),
\end{eqnarray}
that leads to
\begin{eqnarray}
tan(\xi_A)=\frac{[q_3cos(\xi_B)^2+q_4sin(\xi_B)^2]}{[q_1 cos(\xi_B)^2+q_2sin(\xi_B)^2]}=\frac{[q_3+q_4tan(\xi_B)^2]}{[q_1+q_2tan(\xi_B)^2]}.
\end{eqnarray}

Now we exercise second constrain on energy minimization what brings
\begin{eqnarray}
0= \frac{d}{d\xi_B}E_c(\xi_A,\xi_B)=2(+cos(\xi_A)^2[-q_1 sin(\xi_B)+q_2cos(\xi_B)]+ \nonumber \\
+sin(\xi_A)^2[-q_3sin(\xi_B)+q_4 cos(\xi_B)]),
\end{eqnarray}
and we obtain
\begin{eqnarray}
tan(\xi_A)^2=\frac{[-q_1 sin(\xi_B)+q_2cos(\xi_B)]}{[q_3sin(\xi_B)-q_4 cos(\xi_B)]}=\frac{[q_3cos(\xi_B)^2+q_4sin(\xi_B)^2]^2}{[q_1 cos(\xi_B)^2+q_2sin(\xi_B)^2]^2}.
\end{eqnarray}


Under assumption that qubit A is occupying 2 energy levels $E_{1A}$ and $E_{2A}$, while qubit B is occupying 2 energy levels $E_{1B}$ and $E_{2B}$ we have
\begin{eqnarray}
tan(\xi_A)^2=\frac{[-q_1 tan(\xi_B)+q_2]}{[q_3tan(\xi_B)-q_4 ]}=\frac{[q_3+q_4tan(\xi_B)^2]^2}{[q_1+q_2tan(\xi_B)^2]^2},
\end{eqnarray}
$|\gamma_{E1A}|^2=cos(\xi_A)^2=p_{E1A}$, $|\gamma_{E2A}|^2=cos(\xi_A)^2=p_{E1A}$, $|\gamma_{E1B}|^2=cos(\xi_B)^2=p_{E1B}$ and $|\gamma_{E2B}|^2=sin(\xi_B)^2=p_{E2B}$.
Geometry of two electrostatically interacting structures is encoded in $q_1$, $q_2$, $q_3$ and $q_4$ coefficients. 
We compute $(r_A$,$r_A$) in accordance with formula \ref{special} for qubits (A,B) and thus consequently we obtain $\Theta_A, \Theta_B$ coefficients. By having values of $q_1$ .. $q_4$ and consequently initial probability of occupancy eigenergies in case of separated, but perturbatively interacting qubits
$(E_{1A},E_{2A})$ and $(E_{1B},E_{2B})$ we can determine quantum dot occupancy during next time steps. Such procedure can be attempted for N interacting qubits under condition of occupancy of 2 eigenenergy levels.
\normalsize

Consiedered electrostatically interacting qubits can be further subjected to Rabi oscillations by local microwave fields acting on qubits in their direct proximity which might imply evolution of the system ground state with time.

\begin{eqnarray}
\hat{H}_{A-B}= \hat{H}_{A} \times \hat{I}_B +  \hat{I}_A \times \hat{H}_{B} + \hat{H}_{Coulomb: A-B}= \nonumber   \\
=(E_{p1A}\ket{w_{L,A}}\bra{w_{L,A}}+E_{p2A}\ket{w_{R,A}}\bra{w_{R,A}}+t_{s12A}\ket{w_{R,A}}\bra{w_{L,A}}+t_{s21A}\ket{w_{L,A}}\bra{w_{R,A}}) \times \hat{I}_B +  \nonumber   \\
+\hat{I}_A \times (E_{p1B}\ket{w_{L,B}}\bra{w_{L,B}}+E_{p2B}\ket{w_{R,B}}\bra{w_{R,B}}+t_{s12B}\ket{w_{R,B}}\bra{w_{L,B}}+t_{s21B}\ket{w_{L,B}}\bra{w_{R,B}})  +\nonumber   \\
+ \Bigg[ q_1 \ket{w_{L,A}}\ket{w_{L,B}}\bra{w_{L,A}}\bra{w_{L,B}}+ q_2 \ket{w_{L,A}}\ket{w_{L,B}}\bra{w_{R,A}}\bra{w_{R,B}} + q_3 \ket{w_{L,A}}\ket{w_{L,B}}\bra{w_{R,A}}\bra{w_{R,B}}+ \nonumber   \\
+ q_4 \ket{w_{R,A}}\ket{w_{R,B}}\bra{w_{R,A}}\bra{w_{R,B}} \Bigg]_{Coulomb}+ \nonumber   \\
+ \sum_{r,s,u,d}(\ket{w_L,A}\bra{w_L,A}+\ket{w_R,A}\bra{w_R,A})(\ket{w_L,B}\bra{w_L,b}+\ket{w_R,B}\bra{w_R,B})\nonumber   \\
 g(t,r,s,u,d)(\ket{E_{L,r}}\ket{E_{R,s}})(\bra{E_{L,u}}\bra{E_{R,d}}) (\ket{w_L,A}\bra{w_L,A}+\ket{w_R,A}\bra{w_R,A})(\ket{w_L,B}\bra{w_L,b}+\ket{w_R,B}\bra{w_R,B}).
\end{eqnarray}

It is quite straightforward to apply the presented scheme for the case of N position based qubits implementing various quantum gates \cite{b5},\cite{b6}.
Obtained results have its significance in quantum communication as well \cite{QInternet}. 
\section*{Conclusions}

Using a nearest-neighbour tight-binding description with on-site Coulomb repulsion, we derived analytical time evolutions for two electrons confined in two coupled quantum dots and validated them numerically \cite{Xu,Cryogenics,Panos}. This shows that a pair of electrostatic, position-basis qubits can implement SWAP and anti-SWAP gates via purely capacitive coupling. A broader context for semiconductor-qubit operation and device physics is reviewed in \cite{SiSpinCMOS,ChatterjeeNRP2021,BurkardRMP}.

The results guide the design of voltage-controlled gates in chains of coupled semiconductor quantum dots, opening a route to CMOS-compatible quantum information processing without magnetic fields \cite{Zwerver2022,Veldhorst2015}. This electrostatic approach provides an alternative to superconducting-circuit implementations based on Josephson junctions \cite{IBMQexp,Choi}. To quantify entanglement, we propose using the correlation/anti-correlation function, directly measurable with charge sensing; together with the von-Neumann entropy, both admit closed-form expressions within our model. The assumptions of initially localized states (e.g., maximally localized Wannier functions) and the ensuing propagation across a DQD are consistent with earlier modelling and experiments \cite{Panos,Fujisawa,Petta}.

Process- and device-level aspects within cryogenic CMOS/FDSOI flows can be explored using quantum-TCAD frameworks such as QTCAD \cite{QTCAD_APL2022,QTCAD_SSE2023}. Our formalism also extends to stochastic and dissipative dynamics through mappings between finite-state stochastic machines and (non-)dissipative tight-binding Schroedinger models \cite{b3}. Finally, analogous analytical techniques are applied to spin-based gates in quantum dots \cite{LossDiv1998}, and operation at elevated temperatures has been demonstrated in silicon devices, supporting scalable cryo-CMOS integration \cite{HotSiQubits2019}.

\subsection{Impact of noise on tight-binding model parameters}
Let us consider the effective potential for one position based qubits being given by $V(x,t)_p=V(x)_s+V_1(x,t)$ with the condition that $|V_1(x,t)|<<V(x)_s$ and the corresponding set of self-control functions for time-dependent fluctuating potential $(\psi_{E1p}(x,t)=\psi_{E1s}(x)+\delta \psi_{E1s}(x,t),\psi_{E2p}(x,t)=\psi_{E2s}(x)+\delta \psi_{E2s}(x,t))$ and non-fluctuating $(\psi_{E1s}(x),\psi_{E2s}(x))$. Based on formula \ref{special}
\begin{eqnarray}
\label{special1}
\delta r(t) = r_s-r_p(t)=\frac{\int_{-\infty}^{0}dx (\psi_{E1s}(x)\psi_{E2s}^{\dag}(x)+\psi_{E1s}(x)^{\dag}\psi_{E2s}(x))}{\int_{-\infty}^{0}dx (|\psi_{E1s}(x)|^2-|\psi_{E2s}(x)|^2)}
-\frac{\int_{-\infty}^{0}dx (\psi_{E1p}(x)\psi_{E2p}^{\dag}(x)+\psi_{E1p}(x)^{\dag}\psi_{E2p}(x))}{\int_{-\infty}^{0}dx (|\psi_{E1p}(x)|^2-|\psi_{E2p}(x)|^2)}= \nonumber \\
\frac{\int_{-\infty}^{0}dx (\psi_{E1s}(x)\psi_{E2s}^{\dag}(x)+\psi_{E1s}(x)^{\dag}\psi_{E2s}(x))}{\int_{-\infty}^{0}dx (|\psi_{E1s}(x)|^2-|\psi_{E2s}(x)|^2)}+ \nonumber \\
-\frac{\int_{-\infty}^{0}dx ((\psi_{E1s}(x)+\delta \psi_{E1s}(x,t))(\psi_{E2s}(x)+\delta \psi_{E2s}(x,t))^{\dag}+(\psi_{E1s}(x)+\delta \psi_{E1s}(x,t))^{\dag}(\psi_{E2s}(x)+\delta \psi_{E2s}(x,t)))}{\int_{-\infty}^{0}dx (|\psi_{E1s}(x)+\delta \psi_{E1s}(x,t))|^2-|(\psi_{E2s}(x)+\delta \psi_{E2s}(x,t))|^2)} \approx \nonumber \\
\frac{\int_{-\infty}^{0}dx (\delta \psi_{E1s}(x)\psi_{E2s}^{\dag}(x)+\psi_{E1s}(x)^{\dag}\delta \psi_{E2s}(x)+h.c.)}{\int_{-\infty}^{0}dx (|\psi_{E1s}(x)|^2-|\psi_{E2s}(x)|^2)},
\end{eqnarray}
where no Rabi oscillations were assumed due to the low frequency of the fluctuating effective potential component $V_1(x,t)$ with a constant with time effective potential component $V(x)_s$.
\section{Acknowledgment}
We thank Erik Staszewski (University College Dublin) for assistance with figure preparation. The first author contributed ~90 \% of the conceptual development and calculations; the second author validated the computations and provided technological context.
\twocolumn

\onecolumn
\section{Explicit simulation codes in Wolfram Mathematica software}
The system of 2 interacting position based qubits in the tight-binding model can be characterized by 4 energy eigenvalues encoded in Mathematica Wolfram by chain of commands
\begin{lstlisting}
E1qq[Ep1A_, Ep2A_, Ep1B_, Ep2B_, d_, a_, b_, tsA_, tsB_, ThetaA_, 
  ThetaB_, q_] := 
 Eigenvalues[{{Ep1A + Ep1B + (q^2/d), tsB Exp[Sqrt[-1] ThetaB], 
     tsA Exp[Sqrt[-1] ThetaA], 0}, {tsB Exp[-Sqrt[-1] ThetaB], 
     Ep1A + Ep2B + (q^2/Sqrt[d^2 + (a + b)^2]), 0, 
     tsA Exp[Sqrt[-1] ThetaA]}, {tsA Exp[-Sqrt[-1] ThetaA], 0, 
     Ep2A + Ep1B + (q^2/Sqrt[d^2 + (a + b)^2]), 
     tsB Exp[Sqrt[-1] ThetaB]}, {0, tsA Exp[-Sqrt[-1] ThetaA], 
     tsB Exp[-Sqrt[-1] ThetaB], Ep2A + Ep2B + (q^2/d)}}][[1]]
     
 E2qq[Ep1A_, Ep2A_, Ep1B_, Ep2B_, d_, a_, b_, tsA_, tsB_, ThetaA_, 
  ThetaB_, q_] := 
 Eigenvalues[{{Ep1A + Ep1B + (q^2/d), tsB Exp[Sqrt[-1] ThetaB], 
     tsA Exp[Sqrt[-1] ThetaA], 0}, {tsB Exp[-Sqrt[-1] ThetaB], 
     Ep1A + Ep2B + (q^2/Sqrt[d^2 + (a + b)^2]), 0, 
     tsA Exp[Sqrt[-1] ThetaA]}, {tsA Exp[-Sqrt[-1] ThetaA], 0, 
     Ep2A + Ep1B + (q^2/Sqrt[d^2 + (a + b)^2]), 
     tsB Exp[Sqrt[-1] ThetaB]}, {0, tsA Exp[-Sqrt[-1] ThetaA], 
     tsB Exp[-Sqrt[-1] ThetaB], Ep2A + Ep2B + (q^2/d)}}][[2]]
     
 E3qq[Ep1A_, Ep2A_, Ep1B_, Ep2B_, d_, a_, b_, tsA_, tsB_, ThetaA_, 
  ThetaB_, q_] := 
 Eigenvalues[{{Ep1A + Ep1B + (q^2/d), tsB Exp[Sqrt[-1] ThetaB], 
     tsA Exp[Sqrt[-1] ThetaA], 0}, {tsB Exp[-Sqrt[-1] ThetaB], 
     Ep1A + Ep2B + (q^2/Sqrt[d^2 + (a + b)^2]), 0, 
     tsA Exp[Sqrt[-1] ThetaA]}, {tsA Exp[-Sqrt[-1] ThetaA], 0, 
     Ep2A + Ep1B + (q^2/Sqrt[d^2 + (a + b)^2]), 
     tsB Exp[Sqrt[-1] ThetaB]}, {0, tsA Exp[-Sqrt[-1] ThetaA], 
     tsB Exp[-Sqrt[-1] ThetaB], Ep2A + Ep2B + (q^2/d)}}][[3]]
     
E4qq[Ep1A_, Ep2A_, Ep1B_, Ep2B_, d_, a_, b_, tsA_, tsB_, ThetaA_, 
  ThetaB_, q_] := 
 Eigenvalues[{{Ep1A + Ep1B + (q^2/d), tsB Exp[Sqrt[-1] ThetaB], 
     tsA Exp[Sqrt[-1] ThetaA], 0}, {tsB Exp[-Sqrt[-1] ThetaB], 
     Ep1A + Ep2B + (q^2/Sqrt[d^2 + (a + b)^2]), 0, 
     tsA Exp[Sqrt[-1] ThetaA]}, {tsA Exp[-Sqrt[-1] ThetaA], 0, 
     Ep2A + Ep1B + (q^2/Sqrt[d^2 + (a + b)^2]), 
     tsB Exp[Sqrt[-1] ThetaB]}, {0, tsA Exp[-Sqrt[-1] ThetaA], 
     tsB Exp[-Sqrt[-1] ThetaB], Ep2A + Ep2B + (q^2/d)}}][[4]]    

Plot[{E1qq[1, 1, 1, 1, d, 0.1, 0.1, 1, 1, 0, 0, 1], 
  E2qq[1, 1, 1, 1, d, 0.1, 0.1, 1, 1, 0, 0, 1], 
  E3qq[1, 1, 1, 1, d, 0.1, 0.1, 1, 1, 0, 0, 1], 
  E4qq[1, 1, 1, 1, d, 0.1, 0.1, 1, 1, 0, 0, 1]}, {d, 0.1, 1}]


\end{lstlisting}
that correspond to the following dependence of eigenergies on distances as given by Fig.20. 
\begin{figure}
\centering
\includegraphics[scale=0.95]{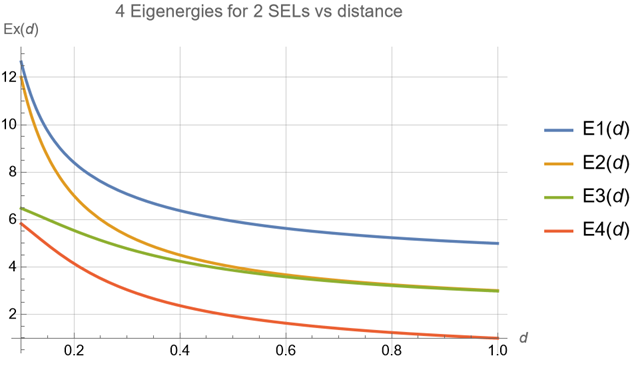}
\caption{Case of dependence of different energy eigenvalues vs distance for 2 electrostatically interacting Single Electron Position-Based Qubits. }
\end{figure}

\vfill

\end{document}